\documentclass[usenatbib]{mn2e}

\usepackage{graphicx}
\usepackage{amssymb,amsmath}
\usepackage{natbib}
\usepackage{longtable}
\usepackage[margin=1cm]{caption}
\newcommand{\hii}{H\,{\sc ii}~} 
\newcommand{\ha}{H$\alpha $~} 
\newcommand{\ntot}{82~} 
\newcommand{\dlfavg}{$-0.21$~} 
\newcommand{\dlfstd}{0.30~} 
\begin{document}

\title[LF Slopes vs. Galaxy Environment]{The Connection Between Galaxy Environment and the Luminosity Function Slopes of Star-Forming Regions}

\author[D. Cook et al.]
{David O. Cook,$^{1,2}$
Daniel A. Dale,$^1$
Janice C. Lee,$^3$
\newauthor
David Thilker,$^3$
Daniela Calzetti,$^4$
Robert C. Kennicutt,$^{5}$
\\
$^1$Department of Physics \& Astronomy, University of Wyoming, Laramie, WY 82071, USA\\
$^2$California Institute of Technology, 1200 East California Blvd, Pasadena, CA 91125, USA; dcook$@$astro.caltech.edu\\
$^3$Space Telescope Science Institute, 3700 San Martin Drive, Baltimore, MD 21218, USA\\
$^4$Department of Astronomy, University of Massachusetts, Amherst, MA 01003, USA\\
$^5$Institute of Astronomy, University of Cambridge, Cambridge CB3 0HA, UK}

\maketitle
\begin{abstract}

We present the first study of GALEX far ultra-violet (FUV) luminosity functions of individual star-forming regions within a sample of 258 nearby galaxies spanning a large range in total stellar mass and star formation properties. We identify $\sim$65,000 star-forming regions (i.e., FUV sources), measure each galaxy's luminosity function, and characterize the relationships between the luminosity function slope ($\alpha$) and several global galaxy properties. A final sample of \ntot galaxies with reliable luminosity functions are used to define these relationships and represent the largest sample of galaxies with the largest range of galaxy properties used to study the connection between luminosity function properties and galaxy environment. We find that $\alpha$ correlates with global star formation properties, where galaxies with higher star formation rates and star formation rate densities ($\Sigma_{\rm{SFR}}$) tend to have flatter luminosity function slopes. In addition, we find that neither stochastic sampling of the luminosity function in galaxies with low-number statistics nor the effects of blending due to distance can fully account for these trends. We hypothesize that the flatter slopes in high $\Sigma_{\rm{SFR}}$ galaxies is due to higher gas densities and higher star formation efficiencies which result in proportionally greater numbers of bright star-forming regions. Finally, we create a composite luminosity function composed of star-forming regions from many galaxies and find a break in the luminosity function at brighter luminosities. However, we find that this break is an artifact of varying detection limits for galaxies at different distances.

\end{abstract}

\begin{keywords}
galaxies: dwarf -- galaxies: irregular -- Local Group --  galaxies: spiral -- galaxies: star formation -- galaxies: ISM -- dust, extinction
\end{keywords}

%##################################################################################
%##################################################################################
\section{Introduction}
A galaxy's ultra-violet (UV) flux traces young, massive stars and thus is a good indicator of recent star formation \citep[i.e., far ultra-violet (FUV) traces $t < 100$~Myr;][]{kennicutt98,murphy11}. Furthermore, clustering studies of young stars and star clusters have shown that star formation occurs in a clustered environment \citep[$t < 100$~Myr;][]{lada03,gouliermis10,gouliermis15}, thus FUV sources with relatively low angular resolution (a few arcseconds) will trace groups of young, massive stars that formed at similar times and physical locations (i.e., star-forming regions).

The star formation process shows evidence of a fractal, or scale-free, structure where the distributions of stars, star clusters, and stellar complexes show similar patterns on different physical scales \citep[see review and references within][]{elmegreen10}. As a consequence of this fractal picture, the mass function (and similarly the luminosity function; LF) of these distributions can be represented as a power-law characterized by a slope of -2 \citep{elmegreen06b}. Observationally, the mass and luminosity functions of star-forming regions (i.e., star clusters and \hii regions) have shown to be adequately approximated as a power-law with a slope of -2 $\pm$ 0.2 \citep[][Adamo et al. 2016; submitted]{zhang99,larsen02,hunter03,bik03,degrijs03b,mccrady07,cook12,whitmore14,chandar15}. However, it is not clear if galaxy environment has a significant effect on $\alpha$, nor is the scatter well understood.

Many star cluster and \hii region studies that measure MF/LFs do so with small galaxy samples (usually only a few), and the methods used to identify regions and to generate MF/LFs can vary from study to study. The inhomogeneity of data and methods can add scatter to any relationship of MF/LF slope with galaxy properties and possibly mask any correlations. However, mass and luminosity functions derived for multiple galaxies with uniform data and methods have found hints of systematic trends between $\alpha$ and global galaxy properties \citep[e.g., SFR, M$_B$, galaxy type, etc. ][]{kennicutt89,elmegreen1999,youngblood99,vanzee00,thilker02,whitmore14}. These trends suggest that environment may play a role in the formation of stars and the quantification of such trends would provide clues into the star formation process.

In addition to environmental affects on the star formation process, mass and luminosity functions have shown evidence for a break in their fitted power-laws at higher masses \citep[e.g., M$_{star} \sim 10^5 - 10^6$ M$_{\odot}$][Adamo et al. 2016; submitted]{gieles06a,bastian12a} and luminosities \citep[e.g., L$_{\rm{H}\alpha}\sim$38.6 erg/s][see also Whitmore et al. 2014, Adamo et al. 2016; submitted]{kennicutt89,pleuss00,bradley06}, where the slope is steeper at higher masses/luminosities. This break has been interpreted as a truncation in the mass function due to density bounded high luminosity regions suggesting a possible upper limit on the mass of star-forming regions \citep{beckman00}. 

Since the measurements of the masses and luminosities of bright star-forming regions required to accurately characterize this break are relatively rare even in normal star-forming galaxies  \citep{larsen02,gieles06a,bastian08}, studies have constructed composite LFs that include the regions of many galaxies to increase the number statistics of bright, star-forming regions \citep{bradley06,whitmore14}. However, the galaxy samples used span a range of distance ($\sim$10s of Mpc) which may contribute to an artificial break due to different luminosity detection limits of galaxies at different distances.

This paper is the first of two papers which aim to test the universal nature of the MF/LF of star-forming regions in a sample of 258 nearby galaxies whose properties span a wide range in SFR, metallicity, luminosity, and Hubble type. The current paper will look at the FUV LFs of star-forming regions while the second paper will look at the MFs. Specifically, this paper measures the LFs, quantifies and trends between LF slope and global galaxy properties, and investigates a break in the composite LF containing star-forming regions from many galaxies.

%##################################################################################
%##################################################################################
\section{Data \& Sample \label{sec:sample}}
The local volume legacy (LVL) sample consists of 258 of our nearest galaxy neighbors reflecting a statistically complete, representative sample of the local universe. The sample selection and description are detailed in \citet{dale09}, but we provide a brief overview here. 

The LVL galaxy sample was built upon a panchromatic data set covering UV, optical, and infrared (IR) wavelengths with the aim of studying both obscured and unobscured star formation in the Local Universe. The final LVL sample consists of galaxies that appear outside the Galactic plane ($|b| > 20^{\circ}$), have a distance less than 11~Mpc ($D \leq$ 11~Mpc), span an absolute $B-$band magnitude of $-9.6 < M_{\rm{B}} < -20.7$, and span an RC3 catalog galaxy type range of $-5 < T < 10$. Although the galaxy morphology composition is diverse, the LVL sample is dominated by dwarf galaxies due to its volume-limited nature. The full LVL sample and basic properties are listed in Table~\ref{tab:genprop}.

The published data sets of LVL consist of GALEX UV \citep{lee11}, Spitzer IR and 2MASS NIR \citep{dale09}, ground-based optical \citep{cook14a}, and ground-based H$\alpha$ \citep{kennicutt08} imaging. We use the GALEX FUV images to identify star-forming regions (see \S\ref{sec:id}), and the global galaxy properties are derived from a combination of the previously published data set as a whole (see \S\ref{sec:galprop}). 

%To prepare the data for identification and photometry, contaminating sources (e.g., background galaxies and foreground stars) were identified and masked. We utilize the contaminant regions identified by \cite{cook14a}, which are available in all LVL filter bandpasses (i.e., FUV, NUV, UBVR, JHK, IRAC1,2,3,4, MIPS24). A detailed description of the contaminating regions and removal can be found in \cite{cook14a}, but we provide a brief overview here.

%The initial contaminant catalogs were generated by \cite{dale09} which were based on Spitzer Space Telescope IR colors and high resolution $HST$ images. These catalogs were then updated by \cite{cook14a} using additional high-resolution HST images via the Hubble Legacy Archive\footnote{http://hla.stsci.edu/hlaview.html}. $HST$ images were visually checked for sources with spiral structure, extended profiles, and  optical colors indicating a background galaxy, and sources with diffraction spikes indicating a foreground star. The new contaminant regions were tailored to each image and added to the modified contamination source catalog of \citet{dale09}. With an updated contaminating source catalog, each contaminant was removed through an interpolation of the surrounding local sky using the IRAF task IMEDIT. 

\begin{table*}
{General Galaxy Properties}\\
\begin{tabular}{lcccc}
\hline
\hline
Galaxy & RA        & DEC        & $D$   & $T$       \\
Name   & (J2000.0) & (J2000.0)  & (Mpc) &           \\
(1)    & (2)       & (3)        & (4)   & (5)       \\
\hline
                 WLM   & 00:01:58.16   & $-$15:27:39.3   &   ~0.92   &     ~10      \\ 
             NGC0024   & 00:09:56.54   & $-$24:57:47.3   &   ~8.13   &     ~~5      \\ 
             NGC0045   & 00:14:03.99   & $-$23:10:55.5   &   ~7.07   &     ~~8      \\ 
             NGC0055   & 00:14:53.60   & $-$39:11:47.9   &   ~2.17   &     ~~9      \\ 
             NGC0059   & 00:15:25.13   & $-$21:26:39.8   &   ~5.30   &    $-$3      \\ 
         ESO410-G005   & 00:15:31.56   & $-$32:10:47.8   &   ~1.90   &    $-$1      \\ 
        SCULPTOR-DE1   & 00:23:51.70   & $-$24:42:18.0   &   ~4.20   &     ~10      \\ 
         ESO294-G010   & 00:26:33.37   & $-$41:51:19.1   &   ~1.90   &    $-$3      \\ 
              IC1574   & 00:43:03.82   & $-$22:14:48.8   &   ~4.92   &     ~10      \\ 
             NGC0247   & 00:47:08.55   & $-$20:45:37.4   &   ~3.65   &     ~~7      \\ 
\hline
\end{tabular} \\
\caption{Column 1: Galaxy name. Column 2 and 3: J2000 right ascension and declination from the R25 apertures of \citet{cook14a}. Column 4: distance in Mpc from \citet{kennicutt08}. Column 5: RC3 Morphological T-type from \citet{kennicutt08}. The full table is available online.}

\label{tab:genprop}
\end{table*}

%##################################################################################
%##################################################################################
\section{Galaxy Regions} \label{sec:galregs}
To study individual regions inside of a galaxy, the extent of a galaxy's flux needs to first be defined. There are three previously published galaxy apertures defined at UV, optical, and IR wavelengths for the LVL sample. The optical apertures are those from the RC3 catalog \citep{rc3} and are defined as isophotal ellipses with surface brightness of 25 mag/arcsec$^2$ in the $B-$band filter. The optical apertures are tabulated in \citet{cook14a}. 

The Spitzer Space Telescope IR elliptical apertures of \citep{dale09} were chosen to encompass the majority of the emission seen at Spitzer IR (3.6$\mu$m--160$\mu$m) wavelengths. In practice, these apertures were usually determined by the extent of the 3.6$\mu$m emission given the superior sensitivity of the 3.6$\mu$m array coupled with the relatively bright emission from older stellar populations at this wavelength. However, in several instances the emission between at 160$\mu$m wavelengths were spatially more extended, and thus images at these wavelengths were used to determine the IR apertures. The resulting median ratio of IR-to-optical semi-major axes is 1.5.

The GALEX UV elliptical apertures of \citet{lee11} were defined as an isophotal ellipse outside of which the photometric error was greater than 0.8 mag or the intensity fell below the sky level. The resulting median ratio of UV-to-optical semi-major axes is 2.3. 

A source of scatter in any relationship between LF slope and global galaxy properties is the crowding of sources due to high galaxy inclination angles. Nearly edge-on (high inclination angle) galaxies will contain star-forming regions which are partially, or fully, obscured by the disk of the galaxy, or blended with sources along the line of sight. Therefore, we make a cut on inclination angle which is calculated via the equation: 

\begin{equation}
  i = \cos^{-1} \Bigg( \Bigg[\dfrac{(b/a) - q_0^2}{1 - q_0^2} \Bigg] \Bigg),
\end{equation}

\noindent where $i$ is the inclination angle, $a$ and $b$ are the semi-major and minor axis of each galaxy, and $q_0$ is the intrinsic axis ratio for an edge-on galaxy set to 0.2 \citep{tully77}. We adopt the UV apertures to define the inclination angles and have chosen to exclude galaxies with $i$ greater than 60$\degr$ \citep[see also,][]{prescott07}. However, as a sanity check we compare the inclination angles based on all three apertures (UV, R25, and IR).

The median difference between the inclination angles based on UV apertures and those of the IR and R25 is $10\fdg3\pm 10\fdg5$ and $0\fdg6\pm 4\fdg1$, respectively. We find 43 galaxies whose inclination angle is cut based on the UV apertures but whose IR or R25 inclination angle is less than 60$\degr$. Visual inspection of these galaxies reveals that 22 of them (DDO210, ESO119-G016, NGC2366, NGC2903, NGC2976, NGC3031, NGC3274, NGC3521, NGC3627, NGC4068, NGC4236, NGC4258, NGC4490, UGC01249, UGC04426, UGC05272, UGC05340, UGC05666, UGC06541, UGC06817, UGC07599, and WLM) have morphologies which are irregular and show little evidence of blended sources (e.g., NGC2366) or have spiral morphologies but the entire disk is clearly identifiable (e.g., NGC3031). We verify that our results are not affected if we exclude these galaxies from the analysis. A total of 77 galaxies are excluded from the analysis due to inclination angle.

%The final galaxy sample used here to study star-forming regions is \#\#, Figure~\ref{fig:sample} shows the general properties of the entire sample and the subset of galaxies in the final sample. CHECK THIS The histograms of Figure~\ref{fig:sample} shows that the final sample spans the same galaxy property range of the full LVL sample, thus probing the full range of Local Universe galaxy properties.

%##################################################################################
%##################################################################################
\section{Star-Forming Regions \label{sec:SFregs}}
In this section we describe the identification, photometry, and extinction correction methods for our star-forming regions as well as the detection limits for each galaxy. We identify star-forming regions based on the GALEX FUV images, which have a FWHM of $\sim 5\arcsec$. This resolution corresponds to $\sim$24 pc for the closest galaxy at $\sim$1~Mpc and $\sim$250 pc for the furthest galaxy at 10.5~Mpc. Due to the large physical scale of FUV sources at the greatest distances of the LVL sample, the FUV regions identified here can include bound star clusters (a few parsecs), loose association (a few 10s of parsecs), or, in some cases, star cluster complexes. However, due to the fractal or scale-free, structure of star formation, we expect that the star-forming regions studied here will exhibit similar distributions as those of the densest parts of the hierarchy, star clusters (see \S~\ref{sec:distsim}).

%##################################################################################
%##################################################################################
\subsection{Contaminating Source Removal}\label{sec:contrem}
To prepare the data for identification and photometry, contaminating sources (e.g., background galaxies and foreground stars) were identified and masked. We utilize the contaminant regions identified by \cite{cook14a}, which are available in all LVL filter bandpasses (i.e., FUV, NUV, UBVR, JHK, IRAC1,2,3,4, MIPS24). A detailed description of the contaminating regions and removal can be found in \cite{cook14a}, but we provide a brief overview here.

The initial contaminant catalogs were generated by \cite{dale09} which were based on Spitzer Space Telescope IR colors and high resolution $HST$ images. These catalogs were then updated by \cite{cook14a} using additional high-resolution HST images via the Hubble Legacy Archive\footnote{http://hla.stsci.edu/hlaview.html}. $HST$ images were visually checked for sources with spiral structure, extended profiles, and  optical colors indicating a background galaxy, and sources with diffraction spikes indicating a foreground star. The new contaminant regions were tailored to each image and added to the modified contamination source catalog of \citet{dale09}. With an updated contaminating source catalog, each contaminant was removed through an interpolation of the surrounding local sky using the IRAF task IMEDIT. 

\subsection{Source Identification \label{sec:id}} 
To ensure that sources in the outer parts of a galaxy are not excluded, we identify star-forming regions inside the aperture with the largest area. In most cases, the UV apertures have the greater area and were subsequently used to define the galaxy's extent. The apertures for each galaxy were visually inspected to verify that no bonafide star-forming regions were found outside the aperture's extent.

The identification of FUV sources was carried out with the publicly available code SExtractor \citep{sex96}. At the GALEX resolution of 5$\arcsec$ most sources inside the the galaxy aperture will be point sources. However, as some sources will be blended or possibly cluster complexes with various morphologies, SExtractor is ideal to identify these regions since the algorithm locates contiguous pixels above a threshold yielding source identification of any morphology.  The SExtractor background input parameters are set to 128 and 6 for BACKSIZE and BACKFILT, respectively. Other SExtractor input parameters are determined from tests designed to optimize the extraction of FUV sources in the LVL sample.

Before the detection of pixels above a threshold is performed, we apply a SExtractor filter (e.g., Gaussian, tophat, mexhat, etc.). To determine the filter best suited to identifying FUV sources in both crowded and uncrowded regions, we run SExtractor on all FUV images using a default (i.e., a pyramidal function), tophat, mexhat, and no filter with the deblending parameters set to their default values.

% and a 2 sigma detection threshold. Although we use a 2 sigma detection threshold here, we find similar results when using a 3 sigma detection threshold. 

\begin{figure*}
  \begin{center}
  \includegraphics[scale=0.5]{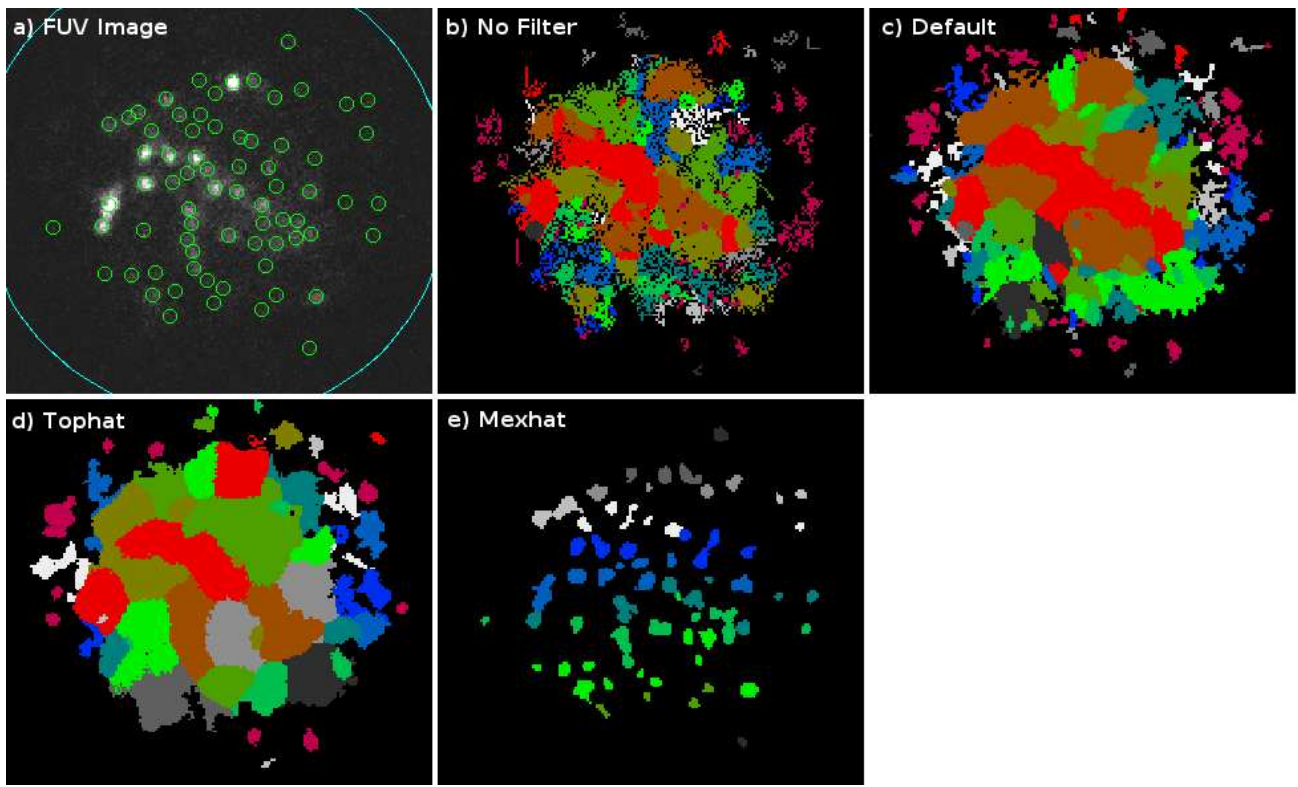}
  \caption{The FUV image and graphical representations of the SExtractor identifications using different filters for the dwarf galaxy UGC07608. Panel (a) is the GALEX FUV image where the cyan ellipse represents the UV galaxy apertures and the green circles represent the SExtractor identifications using a Mexhat filter. Panels (b-e) are the segmentation images using no filter, default (i.e., a pyramidal function), tophat, mexhat filters, respectively. The colors represent a range of integers assigned to each identified source.}
   \label{fig:seg1}
   \end{center}
\end{figure*}  

A graphical representation of SExtractor identifications can be seen in the segmentation images, where the contiguous pixels of different sources are color coded. Figure~\ref{fig:seg1} shows the original FUV image of the dwarf UGC07608 in panel a and the segmentation maps for no filter, default, tophat, mexhat filters in panels (b), (c), (d), and (e), respectively. The cyan ellipse in panel (a) represents the UV galaxy aperture defined by \cite{lee11} and the green circles represent the SExtractor identifications using a mexhat filter. The circular apertures have a radius of 3 pixels, which is the size of the photometric aperture (see \S\ref{sec:phot}).

Visual inspection of the segmentation maps for all galaxies show that using no filter creates contiguous pixels which have boundaries that are ill-defined and have filamentary structure. A typical example is shown in panel (b) of Figure~\ref{fig:seg1} for UGC07608, where many sources in the galaxy and near the edges of the galaxy show irregular morphologies that sometimes extend in filaments away from the source. This filamentary structure indicates that some of these pixels are fluctuations in the background and can cause centering inaccuracies and boundary confusion with nearby sources. 

The default filter segmentation map shown in panel (c) of Figure~\ref{fig:seg1} shows similar results to using no filter, but with somewhat reduced filamentary structure. In addition, the contiguous pixels for many of the sources have been extended due to the filtering of neighboring pixels yielding boundary confusion between neighboring sources similar to using no filter. The tophat filter shown in panel (d) of Figure~\ref{fig:seg1} shows fewer spurious faint sources located near the edges of the galaxy, but shows extended contiguous pixels similar to using both no filter and the default filter.

The mexhat filter shown in panel (e) of Figure~\ref{fig:seg1} shows almost no filamentary structure and well-behaved boundaries (i.e., more round) in the contiguous pixels of most sources. Furthermore, the clear separation in the boundaries between nearby sources will yield more accurate centering information and better deblending of sources in crowded regions. In addition, the mexhat segmentation maps is visually similar to the original FUV image of panel (a) indicating robust identification.

One caveat when using the mexhat filter is that faint sources in the outskirts are sometimes not detected requiring that a 1$\sigma$ threshold be used in order to identify them. This is most likely due to counts being spread into the wings of the PSF when convolved with the filter. Using a 1$\sigma$ detection threshold could result in the identification of some spurious sources, however a 3$\sigma$ detection cut is applied to the FUV source catalog during the photometry process which uses a local background (see \S\ref{sec:phot}).

\begin{figure*}
  \begin{center}
  \includegraphics[scale=0.57]{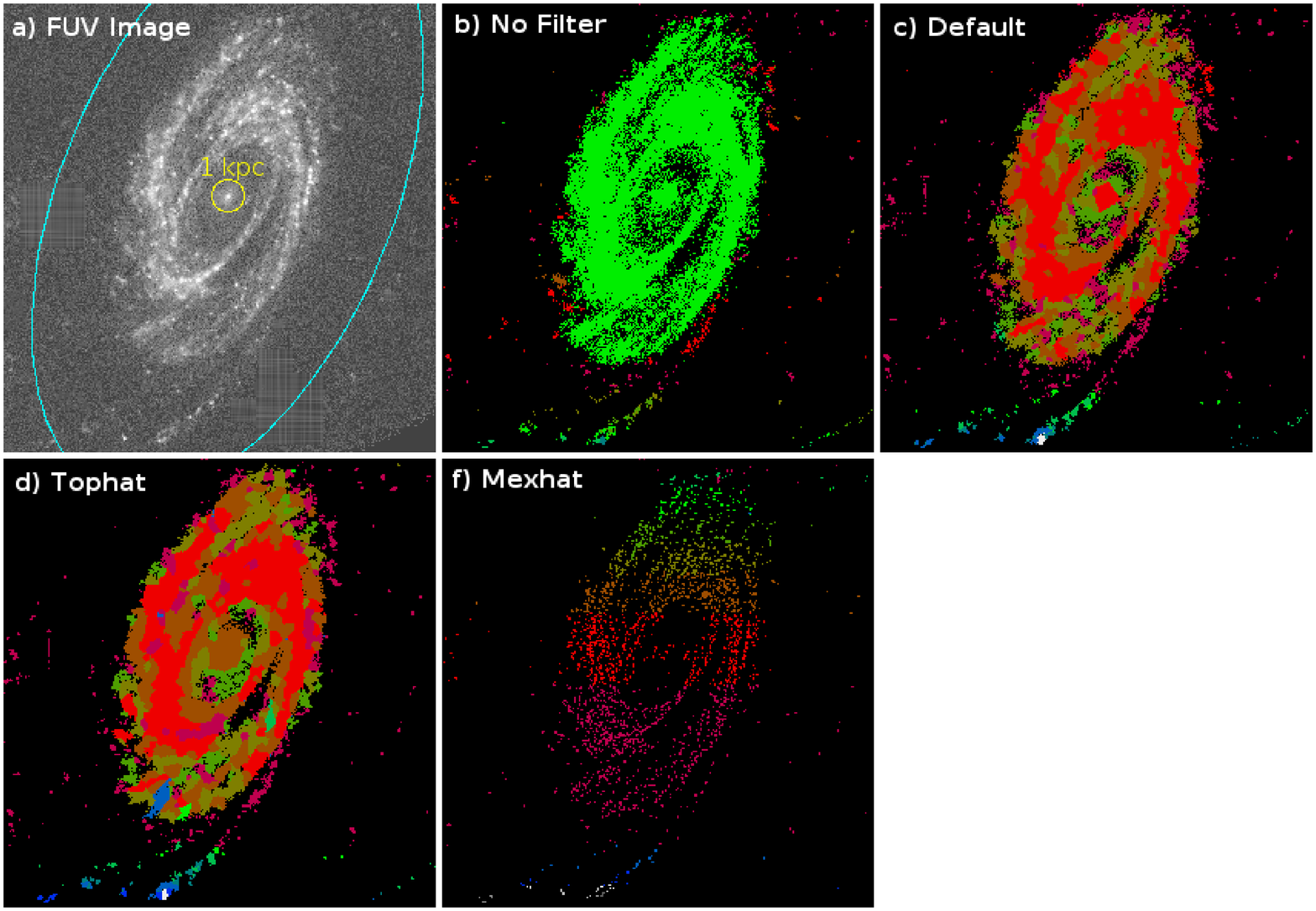}
  \caption{The same as Figure~\ref{fig:seg1}, but for the spiral galaxy NGC3031 (M81). It is interesting to note that the FUV regions identified with the mexhat filter in panel (e) of Figure~\ref{fig:seg2} show clear spiral structure mimicking the structure of the original FUV image of panel (a). }
   \label{fig:seg2}
   \end{center}
\end{figure*}  

Figure~\ref{fig:seg2} is similar to Figure~\ref{fig:seg1}, but for the spiral galaxy NGC3031 (M81). The panels are the same for those of Figure~\ref{fig:seg1}, however, we have removed the circular apertures for visual clarity. We find similar filtering results for this galaxy, where the mexhat filter contiguous pixels of identified sources show well-behaved boundaries and generally good boundary separation between nearby sources. It is interesting to note that the FUV regions identified with the mexhat filter in panel (e) of Figure~\ref{fig:seg2} show clear spiral structure mimicking the structure of the original FUV image of panel (a). We use a mexhat filter with a 1$\sigma$ detection threshold for all FUV galaxies; as noted previously, we apply a 3$\sigma$ detection cut on all sources during the photometry process.

We have also added a yellow region in the center of panel (a) which represents the central 1 kpc region of this galaxy. The nucleus of NGC3031 is clearly separated from the star-forming disk and is likely to be a conglomeration of many star-forming regions. There are similar central regions in the following galaxies: NGC1291, NGC3344, NGC4258, NGC5194, NGC1512, NGC3351, NGC4736, NGC5236, NGC2903, NGC3368, NGC5055, NGC5457, NGC3031, NGC3623, NGC5068, NGC7090. All of the central regions in these galaxies have been visually identified and subsequently removed to avoid contamination of the star-forming region sample \citep[see also,][]{prescott07}.

Blending of nearby sources is likely to occur in our FUV source catalog due to the relatively large PSF of the FUV images. SExtractor utilizes a deblending algorithm which allows the user to chose the degree of deblending through the parameters DEBLEND\_NTHRESH and DEBLEND\_MINCONT. To test which parameters are best suited to the LVL FUV images we analyze the SExtractor outputs using a reasonable range of both DEBLEND\_NTHRESH and DEBLEND\_MINCONT while keeping the background and filtering parameters the same for each deblending parameter. We also note that galaxies at greater distances may be affected more by blending issues; however, we provide a thorough investigation of these effects in \S~\ref{sec:distsim} and find that blending due to distance does not significantly affect our results.

Since the process of deblending of sources will increase the total number of sources identified, the deblending parameters are optimized by examining the total number of identified sources. We find that the total number of all FUV sources identified in the LVL sample increases nearly asymptotically with the DEBLEND\_MINCONT parameter towards a value of $\sim$65,000. The total number plateaus near a DEBLEND\_MINCONT value of 1e-5, which is the value we adopt. For each DEBLEND\_MINCONT value the maximum number of sources occurs for the DEBLEND\_NTHRESH parameter of 64, which is the value we adopt. The total number of sources found in all of our galaxies is $\sim$65,000.

%##################################################################################
%##################################################################################
\subsection{Photometry} \label{sec:phot}
Photometry for FUV sources was carried out via the IRAF task \textsc{PHOT} within an aperture of a 3 pixel (4$\farcs$5) radius, where the centers are from SExtractor. Due to the large PSF of the FUV images, many of the sources have contaminating nearby sources. Visual inspection of crowded regions in the LVL sample reveals that an aperture of 3 pixels (1.8 $\times$ FWHM) minimizes contaminating light from nearby sources while capturing the majority of light for the target region. The sky annuls is set to a radius of 7 pixels and a width of 1 pixel with a mode algorithm to minimize local contaminants. Next, we apply an aperture correction to recover the total flux of our star-forming regions.

The aperture corrections for GALEX FUV point sources have been characterized by previous studies \citep{martin05,morrissey07}, but the star-forming regions studied here may not be point sources even at the 5$\arcsec$ resolution of the GALEX FUV images. In addition, these previous studies have found that the FHWM of GALEX point sources can change from 4$\farcs$5 to 6$\farcs$ from image to image and across the FOV of an image. For these reasons we use a training set of isolated FUV sources located inside the galaxies aperture to investigate an appropriate, empirical aperture correction. The techniques used in this study are similar to those used in star cluster studies of higher resolution \citep[e.g.,][]{chandar10b}

The training set is composed of FUV sources in all galaxies manually selected to have no contaminating flux within a 15 pixel radius. For each training set object we can measure an aperture correction defined as the magnitude difference (i.e., flux ratio) at 15 pixels and 3 pixels radii, where the sky annulus is set to 16 pixels and a width of 1 pixel. A sample-wide aperture correction can be determined by taking the average and standard deviation that can be applied to all sources yielding a correction of -0.86 $\pm$ 0.34 mag. Although an average correction is straight forward, it may not reflect an accurate aperture correction for each source; especially if there is a significant range of PSFs or morphologies (i.e., the extent of the radial profile).

We test the dependence of each object's aperture correction on the radial extent by measuring the concentration index (hereafter CI) of all training set objects, where we define CI as the magnitude difference measured at 1 and 3 pixels. The 1 and 3 pixel radii definition was chosen since smaller than 1 pixel apertures will have increased photometry uncertainties and since crowding may become an issue for apertures larger than 3 pixels. Previous studies of star forming regions (i.e., star clusters) have used CI to quantify the extent of an object's radial profile and have found relationships between the two quantities \citep[e.g.,][]{chandar10b}. The dependence of aperture correction on CI makes sense since a more extended object (i.e., higher CI) will have more light in the wings compared to a point source (i.e., smaller CI) and therefore will have a greater aperture correction.

\begin{figure}
  \begin{center}
  \includegraphics[scale=0.48]{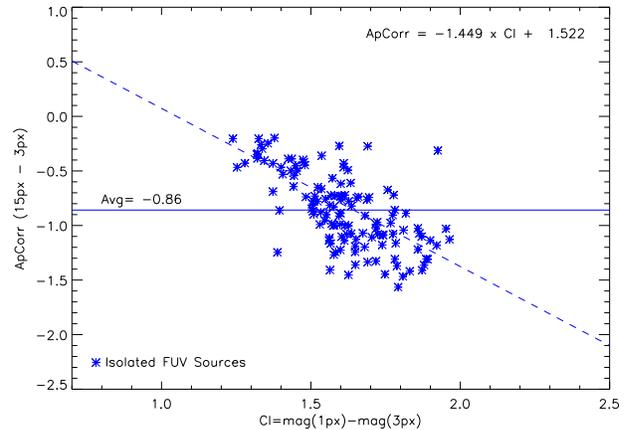}
  \caption{The plot of aperture correction versus concentration index for isolated FUV sources. The aperture correction is defined as the magnitude difference at 15 and 3 pixel radii (the photometry aperture) and the concentration index is defined as the magnitude difference at 1 pixels and 3 pixels. Concentration index is a measure of a source's radial profile extent and the relationship shows that more extended objects require a larger aperture correction.}
   \label{fig:apcorrci_fuv}
   \end{center}
\end{figure}

Figure \ref{fig:apcorrci_fuv} shows a relatively strong correlation between aperture correction and CI for the training set FUV sources, where the solid line represents the average aperture correction to all isolated FUV sources and the dashed line represents the fitted relationship between aperture correction and CI. The fitted relationship between aperture correction and CI is described by the equation:

\begin{equation}
  ApCorr (FUV) = -1.45 * CI + 1.52.
\end{equation}

Although an average aperture correction is suitable for most studies of star-forming regions, Figure~\ref{fig:apcorrci_fuv} shows that at the low and high CI extremes of the training set, the CI-based aperture correction can differ from the average aperture correction by as much as 0.6-0.8 mag. These total flux differences could result in systematic deviations in the LFs of star-forming regions. We use the CI-based aperture corrections to yield the final magnitudes of our star-forming regions and to analyze the relationships between LF slope and galaxy properties. However, we have verified that using an average aperture correction does not significantly change our results; although the LF slope trends with galaxy properties do show some increased scatter compared to the results using the CI-based aperture corrections.

Since the light from an FUV source will be affected by attenuation due to dust we additionally match any 24$\mu m$ sources to our FUV sources, which will be used to account for this attenuated light (see \S\ref{sec:dustcorr}). Photometry is performed on any 24$\mu m$ matched source using the same photometry parameters used for the FUV sources: 3 pixel aperture, 7 pixel sky and 1 pixel width sky annuls. Although the FWHM for 24$\mu m$ and FUV point sources is similar, we derive an empirical aperture correction for 24$\mu m$ sources using the same methods for FUV sources.

We select isolated 24$\mu m$ sources inside each galaxy's aperture and require them to have an FUV counterpart. An aperture correction (mag(15 pixels)-mag(3 pixels)) and CI (mag(1 pixels)-mag(3 pixels)) are measured for each isolated source and plotted in Figure~\ref{fig:apcorrci_fuv}. We find a strong correlation between aperture correction and CI which is described by the equation:

\begin{equation}
  ApCorr (24\mu m) = -1.91 * CI + 2.41.
\end{equation}

\noindent Similar to FUV sources, the difference between the CI-based aperture corrections for 24$\mu m$ can differ from the average aperture correction by as much 0.5-0.7 mag. Since this difference is not insignificant, we use the CI-based aperture corrections to measure the total flux for 24$\mu m$ sources. 

\begin{figure}
  \begin{center}
  \includegraphics[scale=0.48]{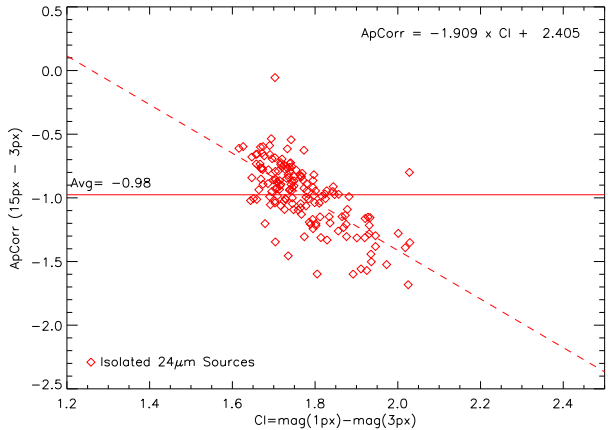}
  \caption{The plot of aperture correction versus concentration index for isolated 24$\mu m$ sources. The relationship shows that more extended objects require a larger aperture correction.}
   \label{fig:apcorrci_mips}
   \end{center}
\end{figure}   

The detection limits for the FUV images are defined here as the 5$\sigma$ detection limit for a point source with a FWHM of 5$\arcsec$. The standard deviation in the sky was measured within equal area regions distributed in an annuli around the galaxy at a distance of 2-4 times the D\_25 radius. The resulting average 5$\sigma$ detection limit for the FUV images is $\log L (erg/s) = 38.0$ with a standard deviation of 0.83, and the corresponding AB magnitude average limit is $21.6 \pm 1.3$ mag. We express the luminosities for individual regions in our LFs as $\rm{d}\nu\rm{L}_{\nu}$ integrated under the filter curve of the FUV filter ($\lambda=1516 \rm{\AA~and}~\rm{d}\lambda=268\rm{\AA}$); note that a monochromatic luminosity (i.e., $\nu\rm{L}_{\nu}$) can be recovered with a 0.75 dex shift in log luminosity.

%##################################################################################
%##################################################################################
\subsection{Dust Correction of Star-Forming Regions} \label{sec:dustcorr}
To account for FUV flux which has been absorbed by dust and re-radiated into IR wavelengths, we apply a dust correction to each FUV identified star-forming region based on the 24$\mu m$/FUV luminosity ratio. There are several previous studies that have derived a relationship between extinction at FUV wavelengths ($A_{\rm{FUV}}$) and an IR(TIR, 24$\mu m$, etc.)/FUV luminosity ratio for both galaxies \citep[e.g.,][]{buat05,hao11} and regions within galaxies \citep[e.g.,][]{leroy08,liu11}. However, we derive our own dust correction prescription from panchromatic data of individual star-forming regions where $A_{\rm{FUV}}$ is based on a direct measure of dust attenuation (i.e., Hydrogen recombination lines).

We derive a dust correction prescription based on the FUV, 24$\mu m$, Pa$\alpha$, and H$\alpha$ luminosities of individual star-forming regions in M51 published by \cite{calzetti05}, where the FUV and 24$\mu m$ luminosities are expressed as $\nu\rm{L}_{\nu}$ and the nebular line luminosities are expressed as $\rm{d}\nu\rm{L}_{\nu}$ integrated under the filter curve \citep[see][]{calzetti05}. We convert our FUV and 24$\mu m$ luminosities to monochromatic luminosities ($\nu\rm{L}_{\nu}$) when calculating dust corrections. Since the Pa$\alpha$/H$\alpha$ luminosity ratio is a direct measure of the flux absorbed by dust, we have derived an empirical relationship between 24$\mu m$/FUV luminosity ratio and $A_{\rm{FUV}}$ based on the Pa$\alpha$/H$\alpha$ luminosity ratio for individual star-forming regions. 

Assuming case B recombination, the E(B-V) value for each region can be derived via the relationship between the nebular emission line color excess and the luminosity ratio of Pa$\alpha$ and H$\alpha$ \citep[see the appendix in][for details]{momcheva13}:

\begin{equation}  \label{eqn:ebv}
  \begin{split}
  E(B-V) = 
  \dfrac{E(H\alpha - Pa\alpha)}{k(\lambda_{H\alpha}) - k(\lambda_{Pa\alpha})} \\
  = \dfrac{2.5}{k(\lambda_{H\alpha}) - k(\lambda_{Pa\alpha})}
  \log_{10}{\Bigg( \dfrac{L(Pa\alpha/H\alpha)_{obs}}{L(Pa\alpha/H\alpha)_{int}}  \Bigg)},
  \end{split}
\end{equation}

\noindent where $k(\lambda)$ is the extinction curve values \citep[$k(\lambda_{H\alpha})=2.535$ and $k(\lambda_{Pa\alpha})=0.45$;][]{cardelli89} evaluated at the given wavelength, $L(Pa\alpha/H\alpha)_{int}$ is the intrinsic luminosity ratio of 0.128 \citep{osterbrock06}, and $L(Pa\alpha/H\alpha)_{obs}$ is the observed luminosity ratio of Pa$\alpha$ and H$\alpha$ in units of ergs/s. $A_{\rm{FUV}}$ can then be calculated given the equation:

\begin{equation}  \label{eqn:afuv}
  A_{\lambda} = k(\lambda) E(B-V)',
\end{equation}

\noindent where we have adopted the \cite{calzetti00} extinction curve value of 10.27 for $k(\lambda_{\rm{FUV}})$. We have also included a factor of 0.44 in E(B-V) to convert the emission line attenuation (i.e., Hydrogen recombination lines) to stellar continuum attenuation \citep{calzetti01}, where the stellar continuum color excess is represented by $E(B-V)'$.

\begin{figure}
  \begin{center}
  \includegraphics[scale=0.48]{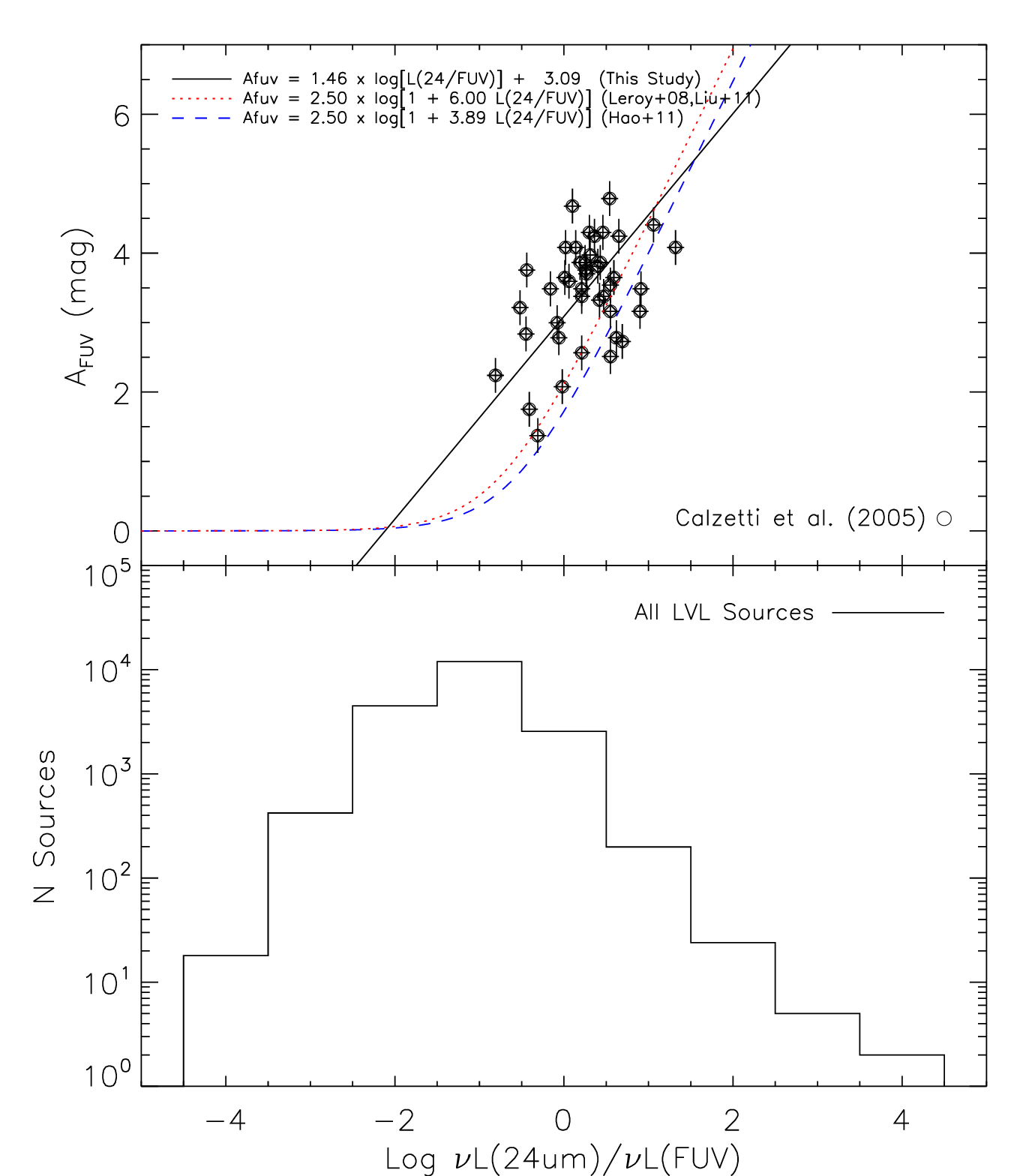}
  \caption{Top panel: the plot of extinction in the FUV bandpass ($A_{\rm{FUV}}$) versus the logarithmic ratio of 24$\mu m$/FUV luminosities for the star-forming regions of \citet{calzetti05} in M51. The solid black line is the bisector fit to the data of \citet{calzetti05}, the blue-dashed line is the relationship prescribed by \citet{hao11} for galaxies, and the red-dashed line is the relationship prescribed by \citet{leroy08,liu11} for individual regions. We use the relationship derived from the star-forming regions of \citet{calzetti05} to correct our FUV sources for flux attenuated by dust. Bottom panel: the histogram of the 24$\mu m$/FUV luminosity ratios for all identified star-forming regions in our study.}
   \label{fig:afuv}
   \end{center}
\end{figure}  

The top panel of Figure~\ref{fig:afuv} shows a correlation between $A_{\rm{FUV}}$ and the 24$\mu m$/FUV luminosity ratios of 42 regions in \cite{calzetti05}. The solid line is the bisector fit to the data which is described by the equation:

\begin{equation}
  A_{\rm{FUV}} = 1.46 \times \log_{10}\Bigg(\dfrac{L(24\mu m)}{L (\rm{FUV})}\Bigg) + 3.09,
\end{equation}

\noindent where the luminosities are in units of erg/s and $A_{\rm{FUV}}$ is in AB magnitude units. 

The blue-dashed line in Figure~\ref{fig:afuv} represents the galaxy-wide dust correction relation of \cite{hao11}. We note that the data for individual regions lies above the galaxy-wide dust correction line of \cite{hao11} indicating that the dust attenuation is greater than those of derived for galaxies. It is not likely that galaxy-wide dust corrections are applicable to individual star-forming regions \citep[][]{calzetti05,boquien15}, and this increased dust attenuation for individual regions has been observed by previous authors \citep[e.g., ][]{calzetti05,calzetti07,leroy08,liu11}. 

The red-dotted line in Figure~\ref{fig:afuv} represents a previously derived dust correction for individual regions \citep{leroy08,liu11}. We find that the previous dust correction prescription for individual regions shows better agreement with the M51 data compared to the galaxy-wide prescription, but also falls below the majority of the M51 data points. The dust correction prescription derived in this study (the bisector fit) shows better agreement with the M51 data. Consequently, we use the bisector fit to correct the FUV luminosities of our star-forming regions. In addition, we verify that all three dust correction prescriptions provide similar trends between luminosity function slope and galaxy properties (see \S~\ref{sec:4way}). 

It could be argued that the functional form of the previous dust correction prescriptions \citep{hao11,leroy08,liu11} might better reflect the relationship between dust attenuation and attenuation indicators (i.e., UV/IR luminosity ratios). We perform a fit with this functional form to the M51 data and obtain similar results to those of \cite{leroy08} and \cite{liu11}, where our fitted slope is slightly higher than those of the previous prescriptions. We also note that there is almost no difference in the trends between LF slope and galaxy properties when using the \cite{leroy08} and \cite{liu11} dust prescription and a dust prescription derived from M51 data using the functional form of previous studies.

The bottom panel of Figure~\ref{fig:afuv} shows the 24$\mu m$/FUV luminosity ratio distribution for all star-forming regions identified in this study. We note that $A_{\rm{FUV}}$ values will become negative for regions with 24$\mu m$/FUV luminosity ratios less than $\sim-$2. To prevent negative dust corrections, we set any negative $A_{\rm{FUV}}$ values to zero.

%##################################################################################
%##################################################################################
\section{Luminosity Functions} \label{sec:lf}
In this section we present the luminosity functions for our star-forming regions. We describe the methods used to generate the luminosity functions and the methods used to derive the fitted power-law slopes. In addition, we provide luminosity function examples representative of the different types of galaxies in our sample. Given the results of our LF binning tests, we define our logarithmic luminosity bins via an equal number of sources in each bin where the bin center is the midpoint of the bins.

\subsection{Fitting \& Binning Methods}
Previous studies of star clusters and \hii regions have found that the LFs are adequately approximated by a power-law ($dN/dL \propto L^{\alpha}$), which can be characterized by the exponent $\alpha$. Our LFs are generated by dividing the luminosities of the FUV sources into luminosity bins and counting the number of sources in each bin. A power-law is represented as a straight line in logarithmic space, and we subsequently use $\chi^2$ minimization to fit a line to our LFs in logarithmic space. The fitted slope of the line is taken as the power-law exponent ($\alpha$) where the errors in each bin represent the Poisson noise of the sources in the bin and the error in the fitted slope is taken as the error in $\alpha$. 

To derive accurate $\alpha$ values, we fit a power-law to luminosity bins which are reasonably complete, in other words, in bins that are not missing star-forming regions due to the depth of the observations. A simple measure of the luminosity at which our sampling is reasonably complete is the peak of the luminosity histogram. Since star-forming regions are represented as a power-law, the number of regions will continue to increase at lower luminosities. If the LF of star-forming regions is accurately described by a power law with a single slope, then a drop in the number of regions for bins at fainter luminosities (i.e., fainter than the peak of the luminosity histogram) will indicate where completeness begins to effect the shape of the LF. We note that the peak of the luminosity histogram is only a rough estimation of the completeness limit for each galaxy, but we believe that it does provide a reasonable limit to the luminosity bins used in our LF power-law fits (see Appendix~\ref{sec:allLFs} for a visual representation of all our LFs where the peak of the LF is the dashed-line in the LF for each galaxy). 

%In addition, the star cluster study of Adamo et al. (2016; submitted) found that the peak of the luminosity histogram matched the 90\% completeness limit derived from artificial cluster tests. For each of our LFs, we fit a power-law to luminosity bins brighter than and equal to the peak of the luminosity histogram.

There are different methods with which to populate sources into each luminosity bin and the choice of the binning method can change the slope of the fitted LF \citep{miaz05}. We quantify the effect of binning by deriving LFs using two common techniques: equal luminosity-size bins where each bin has the same luminosity width and equal-number bins where each bin is populated with an equal number of sources. A comparison of $\alpha$ between these two methods serves as a check on the robustness of our LF results.

The equal luminosity-size bins are most commonly used when fitting a power-law to the LFs of star-forming regions. However, since bright star-forming regions are relatively rare, especially in low-mass galaxies which dominate the LVL sample, we expect that the brightest luminosity bins will suffer from systematic biases due to low-number statistics. \cite{miaz05} showed that derived power-law slopes from equal luminosity-size bins are correlated with the number of sources per bin. Furthermore, \cite{miaz05} showed that variable luminosity-size bins, where each bin is populated with an equal number of sources, show a reduced $\alpha$ dependence on the number of sources per bin. 

\begin{figure}
  \begin{center}
  \includegraphics[scale=0.48]{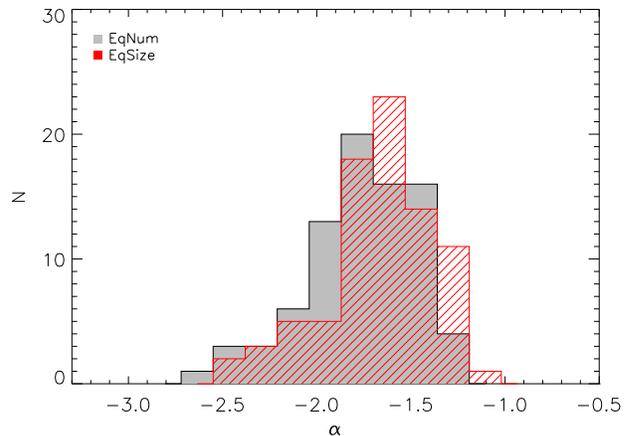}
  \caption{The histogram of luminosity function slopes for two different luminosity binning methods. The grey-shaded histogram represents the LF slopes using equal-number bins and the red diagonal-lined histogram represent slopes using equal luminosity-size bins. The systematic offset of $\alpha$ for equal luminosity-size bins toward flatter slopes is caused by low number statistics per bin.}
   \label{fig:bincomp}
   \end{center}
\end{figure}

Figure~\ref{fig:bincomp} shows the histogram of $\alpha$ values for both binning methods, where the grey shaded histogram represents the LF slopes using equal-number bins and the red diagonal-lined histogram represent slopes using equal luminosity-size bins. The equal luminosity-size binned slopes tend to have flatter slopes compared to the equal-number bins. This is in agreement with the results from \cite{miaz05}, where they found flatter slopes in simulations with lower numbers of sources per bin. We utilize the LF slopes generated with equal-number bins when examining any trends with global galaxy properties. However, we verify that the results of both binning methods yield similar overall trends between LF slope and global galaxy properties (see \S~\ref{sec:4way}).

The difference in LF slopes between equal-number bins and equal luminosity-size bins show an average of \dlfavg (equal-number minus equal luminosity-size bins) with a standard deviation of \dlfstd. The small average $\alpha$ differences compared to the full range of derived $\alpha$ values ($-2.8 < \alpha < -1.0$) provides some measure of confidence in our derived LF slopes. 

In addition to the choice of binning method, we investigate the effect of bin center definition on LF slopes which could also introduce a bias into the LF shape. For example, if the majority of sources in a bin have luminosities that fall on one side of the bin, then an average of the luminosities would result in a bin center that is skewed to one side of the bin compared to the midpoint of the bin. The two bin center definitions we test are: 1) the average of logarithmic luminosities for sources in the bin and 2) the midpoint between the minimum and maximum of logarithmic source luminosities. 

To test the effects of bin center definition, we simulate the LFs for our galaxies based on the real number of sources identified in each galaxy. We assume a universal LF that can be described by a power-law with a slope of $-2$. For each galaxy, we randomly draw star-forming regions equal to the number of sources found in the galaxy and limit these sources to luminosities above the peak in the real luminosity histogram and below the luminosity of the brightest star-forming region found in the entire sample. The lower luminosity limit mimics the sensitivity of the FUV image, and the upper limit allows for the possibility of creating random bright objects (see \S~\ref{sec:simscatter}). The subsequent LF is then fit with the same methods used to fit our real LFs and we iterate 1000 times. 

%We also use the same simulation to test any random scatter in galaxies with low numbers of star-forming regions (see \S~\ref{sec:simscatter}).

\begin{figure}
  \begin{center}
  \includegraphics[scale=0.48]{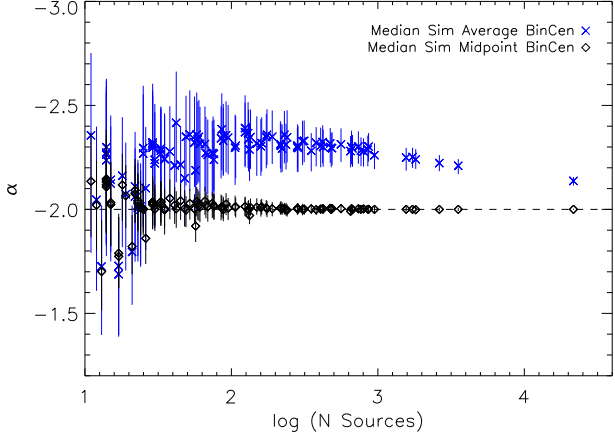}
  \caption{The plot of simulated LF slopes for two bin center definitions, where blue asterisks represent the average of logarithmic luminosities in each bin and the black diamonds represent the midpoint of the logarithmic luminosities in each bin. The average bin center definition results in a bias towards steeper LF slopes compared to the input $-2$ LF slope, while the midpoint bin center definitions recover the input $-2$ LF slope. We use the midpoint bin center definition to construct our LFs.}
   \label{fig:bincen}
   \end{center}
\end{figure}

Figure~\ref{fig:bincen} is a plot of the resulting simulated LF slopes versus the input number of sources for each galaxy. The blue X's represent the median of all simulated $\alpha$ values for average luminosity bin centers, the black diamonds represent the median of all simulated $\alpha$ values for midpoint bin centers, and the error bars represent the 68th and 32nd percentile of the simulations (i.e., the 1$\sigma$ confidence interval).

Visual inspection of Figure~\ref{fig:bincen} shows that the average bin center definition is biased to steeper slopes and have higher errors in each simulation, while the midpoint centering reproduces the input slope of $-2$ with less scatter and lower errors per simulated galaxy. In addition, these simulations also show that both bin centering methods show increased random scatter (above and below the input slope of $-2$) towards lower numbers of sources. We address this increased random scatter in \S~\ref{sec:simscatter}, and ultimately discard the LFs for galaxies with sources less than 30 (see the next section). For the rest of this study we adopt a fiducial LF binning of equal number sources in each bin where the bin center is defined as the midpoint of the bin. 

We also acknowledge that using a maximum likelihood fitting procedure to determine luminosity function slopes would alleviate many of the complicating affects of binning discussed in this section \citep[see][]{whitmore14}. However, \cite{whitmore14} found good agreement for LF slopes when using similar methods to those used here and a maximum likelihood method. We will provide a comparison to a maximum likelihood method in a future work. 

\subsection{Examples}
\begin{figure*}
  \begin{center}
  \includegraphics[scale=1.05]{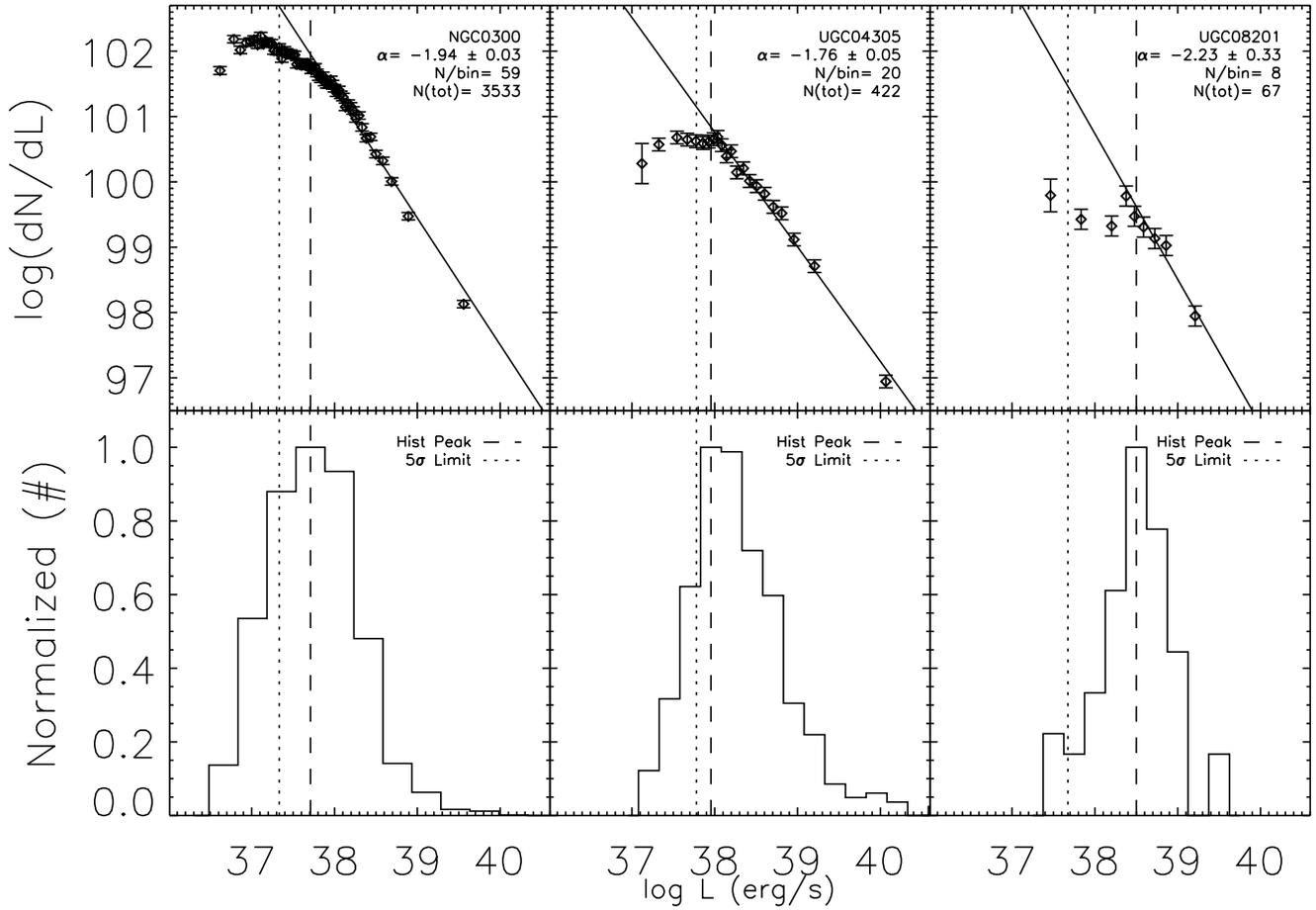}
  \caption{Top panels: Example luminosity functions of galaxies with high, moderate, and low numbers of star forming regions from left-to-right, respectively. The dotted line represents the 5$\sigma$ detection limit for a point source, and the dashed line represents the peak of the luminosity histogram in the bottom panel. The solid line represents the power-law fit to the LF and extends down to the dashed line. The y-axes have been normalized to an arbitrary number. Bottom panels: luminosity histograms for each galaxy in the top panel. The peak in the luminosity histogram accurately traces the LF turnovers in the top panels, and hence, where the effects of incompleteness become significant.}
  \label{fig:histpeak}
  \end{center}
\end{figure*}

Figure~\ref{fig:histpeak} shows the LFs and luminosity histogram examples for three representative galaxies with high ($N\sim1000$s), moderate ($N\sim100$s), and low ($N\sim10$s) total number of sources. The top panels of Figure~\ref{fig:histpeak} show the LF with equal-number bins and the bottom panels show the luminosity histograms for each galaxy. Visual inspection of the top and bottom panel for each galaxy show that the peak of the luminosity histogram accurately traces the LF turnover, and hence, where the effects of incompleteness become significant. We find that the peak of the luminosity histogram accurately traces incompleteness effects for all well-behaved LFs.

%Luminosity histogram bins set to 0.25, however, galaxies with lower numbers sometimes required larger histogram bins to make them look like a Gaussian distribution. Really what I am doing is using equal sized bins to determine the turnover in the equal-number bins...

We define well-behaved LFs as those with log (dN/dL) values which smoothly increase towards lower luminosity bins and either level off or decrease at luminosities fainter than the peak of the luminosity histogram (see Figure~\ref{fig:histpeak}). Conversely, we define ill-behaved LFs as those satisfying any of the following 3 conditions: 1) the log (dN/dL) values do not rise smoothly, or at all, towards lower luminosities; 2) the fit depends on only 2 or 3 luminosity bins and there exists a vertical discontinuity between any of the luminosity bins; and 3) The LF is extremely noisy and the plot is indistinguishable from a scatter plot. There are 11 galaxies whose LFs are not well-behaved (ESO486-G021, NGC3299, NGC4163, NGC4190, NGC5238, UGC06900, UGC07599, UGC07605, UGC07950, UGC09992, UGCA438), and we remove these galaxies from the analysis.

\begin{figure*}
  \begin{center}
  \includegraphics[scale=0.65]{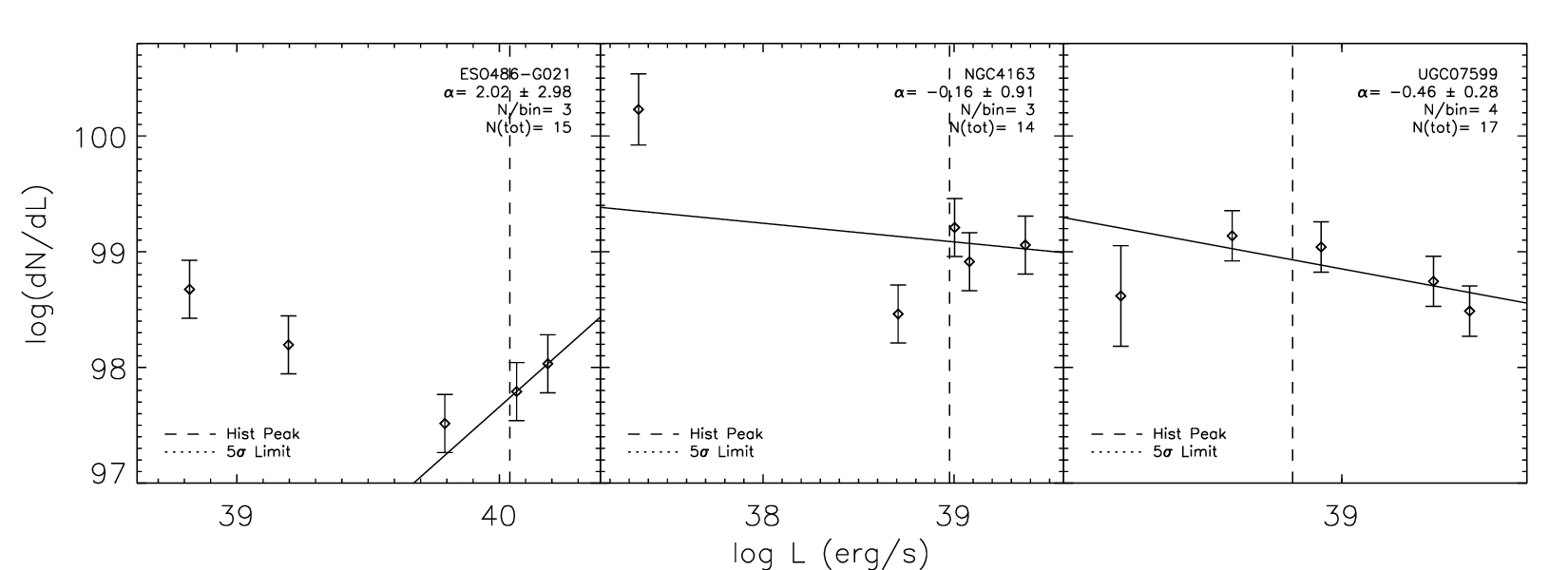}
  \caption{Examples of ill-behaved luminosity functions similar to the top panels of Figure~\ref{fig:histpeak}. From left-to-right, these examples show LFs where: log~(dN/dL) values do not rise continuously towards fainter luminosity bins, the fit depends on only two or three data points, and the data are indistinguishable from a scatter plot. There are 11 galaxies whose LFs are similar to one or more of these examples. The y-axes have been normalized to an arbitrary number.}
  \label{fig:badLFs}
  \end{center}
\end{figure*}

Figure~\ref{fig:badLFs} shows three ill-behaved LFs which exemplify the 3 conditions defined in the previous paragraph. Panel (a) shows a LF whose log~(dN/dL) values do not rise continuously towards fainter luminosity bins; in fact, the log~(dN/dL) values decease for the first 3 luminosity bins. Panel (b) of Figure~\ref{fig:badLFs} shows a LF whose log~(dN/dL) values do increase continuously toward the peak of the luminosity histogram, but the fit depends on only three data points. Furthermore, the large jump in the next two log~(dN/dL) values after the peak (aka discontinuity) indicates that the fit to the brightest luminosity bins are relatively uncertain. Panel (c) of Figure~\ref{fig:badLFs} shows a LF whose log~(dN/dL) values do increase continuously toward the peak of the luminosity histogram, however, the fit depends on only three luminosity bins and the data are indistinguishable from a scatter plot.

All galaxies with ill-defined LFs have similar overall structure to one or more of the LF examples of Figure~\ref{fig:badLFs}. In addition, they all share similar galaxy morphologies and a low total number of sources. The FUV images of these galaxies generally show a ``blobby" morphology where the galaxy has a few central blended sources and a few surrounding faint sources. Furthermore, these galaxies generally have a low total number of sources, where the largest number of sources in a galaxy with an ill-behaved LF is 24 sources. To avoid contaminating our results with unreliable LFs, we exclude any galaxy with a total number of sources $N<30$. 

After removing 99 galaxies $N < 30$ star-forming regions (which include those with ill-behaved LFs) and 77 galaxies with $i \geq 60$ (except those that have been manually added back in; see \S~\ref{sec:galregs}) we have a final galaxy sample of \ntot galaxies. Figures~\ref{fig:allLFs1}$-$\ref{fig:allLFs10} show the LFs for all \ntot galaxies. The LF properties for these galaxies are tabulated in Table~\ref{tab:LFinfo}, where we include the total number of sources, the peak in the luminosity histogram, the luminosity of the brightest source, and the fitted slope $\alpha$ using 4 combinations of 2 aperture correction methods and 2 LF binning methods in each galaxy.

Visual inspection of the LFs in Figures~\ref{fig:allLFs1}$-$\ref{fig:allLFs10} shows a continuous rise in log (dN/dL) values towards fainter luminosity bins and a turnover near the peak in the luminosity histogram where we expect a reduced number of regions due to incompleteness effects. In addition, we find no clear evidence for a break in any individual LF. We use the fitted power-law parameters to investigate any possible relationships between LFs and global galaxy properties. 

\newpage
\onecolumn
\begin{longtable}{lccccccc}
\caption{Luminosity Function Properties}\\
\hline
\hline
Galaxy & N$_{srcs}$ & log (L$_{max}$) & log (Lum Peak) & $\alpha$  & $\alpha$  & $\alpha$  & $\alpha$   \\
 Name & (\#) & (erg/s) & (erg/s) & EqNum-CI & EqNum-Avg & EqSize-CI & EqSize-Avg  \\
 (1) & (2) & (3) & (4) & (5) & (6) & (7) &  (8) \\
\hline
\endfirsthead
\caption{\ref{tab:LFinfo} Continued}\\
\hline
\endhead
\hline
\endfoot
\caption{The properties of the sources identified in each galaxy and the luminosity functions constructed from these sources. Column 1: Galaxy name. Column 2: The number of sources identified in each galaxy. Column 3: The peak of the luminosity histogram for sources in each galaxy. Column 4: The luminosity of the brightest source found in each galaxy. Columns 5-8: The LF slopes ($\alpha$) measured using 4 combinations of 2 binning methods and 2 aperture correction methods: 1) EqNum-CI: CI-based aperture correction and equal number binning; 2) EqNum-Avg: average aperture correction and equal number binning; 3)EqSize-CI: CI-based aperture correction and equal luminosity size binning; 4) EqSize-Avg: average aperture correction and equal luminosity size binning.}\\
\hline
\endlastfoot
  ESO119-G016    &  ~~~30  &  40.09  &  39.33  &  -1.37  $\pm$  0.41  &  -1.45  $\pm$  0.40  &  -1.59  $\pm$  0.60  &  -1.55  $\pm$  0.40  \\
  ESO245-G005    &  ~~130  &  39.95  &  38.56  &  -1.62  $\pm$  0.16  &  -1.71  $\pm$  0.18  &  -1.61  $\pm$  0.17  &  -1.81  $\pm$  0.21  \\
  ESO245-G007    &  ~~~48  &  36.82  &  36.22  &  -1.61  $\pm$  0.40  &  -2.39  $\pm$  0.73  &  -1.27  $\pm$  0.60  &  -2.02  $\pm$  0.64  \\
  IC5152         &  ~~150  &  39.99  &  38.53  &  -1.77  $\pm$  0.16  &  -1.91  $\pm$  0.16  &  -1.74  $\pm$  0.18  &  -1.79  $\pm$  0.17  \\
  IC5332         &  ~~652  &  41.18  &  39.32  &  -1.92  $\pm$  0.07  &  -1.87  $\pm$  0.06  &  -1.88  $\pm$  0.08  &  -1.78  $\pm$  0.06  \\
  NGC0045        &  ~~478  &  40.67  &  38.70  &  -1.55  $\pm$  0.06  &  -1.81  $\pm$  0.09  &  -1.49  $\pm$  0.06  &  -1.75  $\pm$  0.09  \\
  NGC0300        &  ~3533  &  40.08  &  37.71  &  -1.94  $\pm$  0.03  &  -2.21  $\pm$  0.03  &  -2.05  $\pm$  0.03  &  -2.18  $\pm$  0.04  \\
  NGC0598        &  21617  &  40.04  &  36.87  &  -1.75  $\pm$  0.01  &  -1.98  $\pm$  0.01  &  -1.70  $\pm$  0.01  &  -1.93  $\pm$  0.01  \\
  NGC0628        &  ~~795  &  41.65  &  38.78  &  -1.46  $\pm$  0.03  &  -1.50  $\pm$  0.03  &  -1.43  $\pm$  0.03  &  -1.45  $\pm$  0.03  \\
  NGC1291        &  ~~230  &  41.36  &  39.33  &  -2.42  $\pm$  0.12  &  -2.54  $\pm$  0.13  &  -2.23  $\pm$  0.19  &  -2.30  $\pm$  0.18  \\
  NGC1313        &  ~~236  &  41.43  &  39.12  &  -1.60  $\pm$  0.06  &  -1.67  $\pm$  0.07  &  -1.56  $\pm$  0.06  &  -1.62  $\pm$  0.06  \\
  NGC1487        &  ~~~76  &  41.92  &  39.14  &  -1.50  $\pm$  0.10  &  -1.45  $\pm$  0.12  &  -1.35  $\pm$  0.12  &  -1.36  $\pm$  0.13  \\
  NGC1512        &  ~~314  &  41.81  &  39.24  &  -1.93  $\pm$  0.07  &  -1.86  $\pm$  0.06  &  -1.80  $\pm$  0.07  &  -1.79  $\pm$  0.08  \\
  NGC1744        &  ~~218  &  40.76  &  38.74  &  -1.53  $\pm$  0.07  &  -1.54  $\pm$  0.07  &  -1.53  $\pm$  0.08  &  -1.73  $\pm$  0.12  \\
  NGC2366        &  ~~191  &  41.61  &  38.46  &  -1.78  $\pm$  0.07  &  -1.86  $\pm$  0.08  &  -1.58  $\pm$  0.08  &  -1.56  $\pm$  0.11  \\
  NGC2403        &  ~1727  &  41.73  &  38.32  &  -1.64  $\pm$  0.02  &  -1.71  $\pm$  0.03  &  -1.56  $\pm$  0.03  &  -1.63  $\pm$  0.03  \\
  NGC2500        &  ~~~88  &  41.18  &  39.08  &  -1.41  $\pm$  0.10  &  -1.39  $\pm$  0.12  &  -1.38  $\pm$  0.10  &  -1.38  $\pm$  0.12  \\
  NGC2552        &  ~~~69  &  40.75  &  39.41  &  -2.02  $\pm$  0.22  &  -1.96  $\pm$  0.19  &  -1.59  $\pm$  0.28  &  -1.85  $\pm$  0.19  \\
  NGC2903        &  ~~421  &  42.57  &  39.21  &  -1.59  $\pm$  0.04  &  -1.65  $\pm$  0.04  &  -1.47  $\pm$  0.05  &  -1.53  $\pm$  0.05  \\
  NGC2976        &  ~~~59  &  41.37  &  39.25  &  -1.65  $\pm$  0.16  &  -1.85  $\pm$  0.12  &  -1.48  $\pm$  0.14  &  -1.59  $\pm$  0.12  \\
  NGC3031        &  ~1559  &  41.36  &  38.56  &  -1.69  $\pm$  0.03  &  -1.58  $\pm$  0.02  &  -1.63  $\pm$  0.03  &  -1.71  $\pm$  0.04  \\
  NGC3239        &  ~~106  &  41.96  &  39.77  &  -1.86  $\pm$  0.13  &  -1.84  $\pm$  0.11  &  -1.68  $\pm$  0.13  &  -1.58  $\pm$  0.12  \\
  NGC3274        &  ~~~58  &  41.00  &  38.63  &  -1.47  $\pm$  0.11  &  -1.65  $\pm$  0.20  &  -1.37  $\pm$  0.10  &  -1.46  $\pm$  0.19  \\
  NGC3344        &  ~~474  &  41.63  &  39.06  &  -1.75  $\pm$  0.05  &  -1.78  $\pm$  0.06  &  -1.60  $\pm$  0.07  &  -1.58  $\pm$  0.08  \\
  NGC3351        &  ~~351  &  42.48  &  39.30  &  -1.80  $\pm$  0.05  &  -2.00  $\pm$  0.07  &  -1.33  $\pm$  0.05  &  -1.76  $\pm$  0.10  \\
  NGC3368        &  ~~159  &  41.47  &  39.61  &  -1.92  $\pm$  0.12  &  -2.00  $\pm$  0.12  &  -1.74  $\pm$  0.15  &  -1.61  $\pm$  0.20  \\
  NGC3486        &  ~~283  &  41.36  &  39.38  &  -1.58  $\pm$  0.07  &  -1.67  $\pm$  0.08  &  -1.56  $\pm$  0.07  &  -1.61  $\pm$  0.10  \\
  NGC3521        &  ~~285  &  41.35  &  38.88  &  -1.32  $\pm$  0.05  &  -1.36  $\pm$  0.04  &  -1.33  $\pm$  0.04  &  -1.34  $\pm$  0.04  \\
  NGC3627        &  ~~125  &  42.24  &  39.03  &  -1.28  $\pm$  0.06  &  -1.31  $\pm$  0.06  &  -1.22  $\pm$  0.05  &  -1.27  $\pm$  0.05  \\
  NGC4068        &  ~~~41  &  40.39  &  38.79  &  -1.85  $\pm$  0.23  &  -1.94  $\pm$  0.22  &  -1.80  $\pm$  0.28  &  -1.61  $\pm$  0.26  \\
  NGC4214        &  ~~669  &  41.46  &  38.00  &  -1.75  $\pm$  0.04  &  -1.89  $\pm$  0.04  &  -1.62  $\pm$  0.04  &  -1.78  $\pm$  0.05  \\
  NGC4236        &  ~~862  &  41.01  &  38.62  &  -1.89  $\pm$  0.05  &  -1.97  $\pm$  0.05  &  -1.84  $\pm$  0.06  &  -1.88  $\pm$  0.05  \\
  NGC4242        &  ~~~95  &  40.57  &  39.38  &  -1.83  $\pm$  0.20  &  -2.06  $\pm$  0.20  &  -1.81  $\pm$  0.30  &  -2.00  $\pm$  0.20  \\
  NGC4258        &  ~~767  &  42.09  &  38.82  &  -1.72  $\pm$  0.03  &  -1.83  $\pm$  0.04  &  -1.61  $\pm$  0.04  &  -1.74  $\pm$  0.05  \\
  NGC4288        &  ~~~56  &  41.04  &  38.59  &  -1.42  $\pm$  0.11  &  -1.43  $\pm$  0.12  &  -1.31  $\pm$  0.09  &  -1.36  $\pm$  0.10  \\
  NGC4395        &  ~~949  &  41.22  &  38.41  &  -1.82  $\pm$  0.04  &  -2.04  $\pm$  0.05  &  -1.60  $\pm$  0.04  &  -1.91  $\pm$  0.06  \\
  NGC4449        &  ~~239  &  41.85  &  39.00  &  -1.37  $\pm$  0.06  &  -1.37  $\pm$  0.05  &  -1.32  $\pm$  0.06  &  -1.34  $\pm$  0.05  \\
  NGC4490        &  ~~160  &  41.74  &  39.60  &  -1.43  $\pm$  0.09  &  -1.48  $\pm$  0.08  &  -1.30  $\pm$  0.10  &  -1.27  $\pm$  0.09  \\
  NGC4618        &  ~~~86  &  41.24  &  40.16  &  -1.48  $\pm$  0.22  &  -1.18  $\pm$  0.09  &  -1.65  $\pm$  0.28  &  -1.19  $\pm$  0.11  \\
  NGC4625        &  ~~~92  &  41.19  &  38.49  &  -1.42  $\pm$  0.07  &  -1.45  $\pm$  0.08  &  -1.35  $\pm$  0.08  &  -1.43  $\pm$  0.11  \\
  NGC4707        &  ~~~41  &  40.15  &  39.04  &  -1.80  $\pm$  0.33  &  -1.82  $\pm$  0.34  &  -1.86  $\pm$  0.26  &  -1.87  $\pm$  0.54  \\
  NGC4736        &  ~~379  &  41.68  &  38.12  &  -1.46  $\pm$  0.04  &  -1.46  $\pm$  0.03  &  -1.40  $\pm$  0.04  &  -1.40  $\pm$  0.04  \\
  NGC4826        &  ~~~57  &  41.92  &  39.97  &  -2.20  $\pm$  0.23  &  -2.05  $\pm$  0.18  &  -1.55  $\pm$  0.15  &  -1.61  $\pm$  0.16  \\
  NGC5055        &  ~~661  &  41.27  &  38.95  &  -1.48  $\pm$  0.04  &  -1.55  $\pm$  0.04  &  -1.44  $\pm$  0.04  &  -1.52  $\pm$  0.05  \\
  NGC5068        &  ~~221  &  41.39  &  39.61  &  -1.54  $\pm$  0.09  &  -1.41  $\pm$  0.06  &  -1.55  $\pm$  0.11  &  -1.58  $\pm$  0.10  \\
  NGC5194        &  ~~488  &  41.92  &  39.38  &  -1.42  $\pm$  0.04  &  -1.42  $\pm$  0.04  &  -1.41  $\pm$  0.04  &  -1.38  $\pm$  0.04  \\
  NGC5195        &  ~~~33  &  41.20  &  38.81  &  -1.73  $\pm$  0.16  &  -1.85  $\pm$  0.20  &  -1.37  $\pm$  0.14  &  -1.39  $\pm$  0.15  \\
  NGC5204        &  ~~173  &  40.59  &  38.95  &  -1.55  $\pm$  0.14  &  -1.61  $\pm$  0.09  &  -1.49  $\pm$  0.13  &  -1.53  $\pm$  0.09  \\
  NGC5236        &  ~~856  &  43.05  &  38.63  &  -1.50  $\pm$  0.02  &  -1.57  $\pm$  0.03  &  -1.40  $\pm$  0.02  &  -1.44  $\pm$  0.03  \\
  NGC5457        &  ~2624  &  42.29  &  39.05  &  -1.66  $\pm$  0.02  &  -1.68  $\pm$  0.02  &  -1.54  $\pm$  0.02  &  -1.61  $\pm$  0.02  \\
  NGC5474        &  ~~307  &  40.77  &  39.01  &  -1.77  $\pm$  0.09  &  -1.79  $\pm$  0.08  &  -1.76  $\pm$  0.09  &  -1.81  $\pm$  0.09  \\
  NGC5585        &  ~~216  &  40.56  &  38.56  &  -1.56  $\pm$  0.08  &  -1.62  $\pm$  0.08  &  -1.69  $\pm$  0.11  &  -1.64  $\pm$  0.09  \\
  NGC5832        &  ~~~74  &  40.30  &  39.01  &  -1.46  $\pm$  0.16  &  -1.46  $\pm$  0.17  &  -1.47  $\pm$  0.17  &  -1.30  $\pm$  0.17  \\
  NGC7793        &  ~~731  &  40.83  &  38.27  &  -1.33  $\pm$  0.03  &  -1.44  $\pm$  0.04  &  -1.30  $\pm$  0.03  &  -1.40  $\pm$  0.04  \\
  SextansA       &  ~~279  &  38.95  &  37.57  &  -1.80  $\pm$  0.11  &  -1.86  $\pm$  0.12  &  -1.77  $\pm$  0.11  &  -1.61  $\pm$  0.09  \\
  UGC00668       &  ~1813  &  38.80  &  36.69  &  -1.99  $\pm$  0.04  &  -2.16  $\pm$  0.05  &  -1.92  $\pm$  0.04  &  -2.07  $\pm$  0.05  \\
  UGC01176       &  ~~~45  &  40.18  &  39.11  &  -2.02  $\pm$  0.32  &  -1.69  $\pm$  0.31  &  -1.81  $\pm$  0.25  &  -1.61  $\pm$  0.23  \\
  UGC01249       &  ~~~88  &  40.86  &  39.86  &  -2.16  $\pm$  0.39  &  -2.12  $\pm$  0.25  &  -2.08  $\pm$  0.28  &  -2.03  $\pm$  0.36  \\
  UGC04305       &  ~~422  &  40.77  &  37.95  &  -1.76  $\pm$  0.05  &  -1.76  $\pm$  0.05  &  -1.70  $\pm$  0.05  &  -1.68  $\pm$  0.06  \\
  UGC05139       &  ~~~85  &  39.65  &  38.31  &  -1.78  $\pm$  0.21  &  -2.35  $\pm$  0.29  &  -1.83  $\pm$  0.23  &  -1.76  $\pm$  0.27  \\
  UGC05336       &  ~~~49  &  39.07  &  38.31  &  -2.16  $\pm$  0.41  &  -2.07  $\pm$  1.19  &  -2.31  $\pm$  0.46  &  -2.09  $\pm$  1.32  \\
  UGC05340       &  ~~~35  &  39.89  &  38.64  &  -1.31  $\pm$  0.23  &  -1.21  $\pm$  0.20  &  -1.10  $\pm$  0.18  &  -1.06  $\pm$  0.25  \\
  UGC05364       &  ~~107  &  37.90  &  36.83  &  -2.10  $\pm$  0.23  &  -1.78  $\pm$  0.24  &  -1.88  $\pm$  0.22  &  -1.70  $\pm$  0.27  \\
  UGC05373       &  ~~130  &  39.01  &  37.73  &  -2.20  $\pm$  0.20  &  -2.28  $\pm$  0.23  &  -2.06  $\pm$  0.17  &  -2.16  $\pm$  0.27  \\
  UGC05666       &  ~~564  &  40.77  &  38.20  &  -1.65  $\pm$  0.04  &  -1.86  $\pm$  0.06  &  -1.60  $\pm$  0.05  &  -1.79  $\pm$  0.06  \\
  UGC05829       &  ~~133  &  40.65  &  39.03  &  -1.80  $\pm$  0.12  &  -1.78  $\pm$  0.13  &  -1.63  $\pm$  0.12  &  -1.72  $\pm$  0.13  \\
  UGC05889       &  ~~~35  &  40.02  &  38.91  &  -2.22  $\pm$  0.38  &  -2.47  $\pm$  0.53  &  -1.99  $\pm$  0.50  &  -1.87  $\pm$  0.63  \\
  UGC06817       &  ~~~75  &  39.43  &  37.88  &  -2.22  $\pm$  0.19  &  -2.18  $\pm$  0.25  &  -1.64  $\pm$  0.23  &  -1.82  $\pm$  0.28  \\
  UGC07559       &  ~~~52  &  39.78  &  38.37  &  -1.62  $\pm$  0.19  &  -1.60  $\pm$  0.16  &  -1.58  $\pm$  0.15  &  -1.50  $\pm$  0.15  \\
  UGC07577       &  ~~~62  &  39.22  &  38.16  &  -1.97  $\pm$  0.31  &  -2.10  $\pm$  0.30  &  -2.48  $\pm$  0.47  &  -2.15  $\pm$  0.36  \\
  UGC07608       &  ~~~66  &  40.51  &  39.23  &  -1.83  $\pm$  0.24  &  -1.89  $\pm$  0.19  &  -1.63  $\pm$  0.24  &  -1.58  $\pm$  0.19  \\
  UGC07690       &  ~~~30  &  40.95  &  39.44  &  -1.79  $\pm$  0.27  &  -1.80  $\pm$  0.26  &  -1.73  $\pm$  0.26  &  -1.75  $\pm$  0.26  \\
  UGC07698       &  ~~124  &  39.93  &  38.34  &  -1.91  $\pm$  0.12  &  -1.76  $\pm$  0.10  &  -1.72  $\pm$  0.12  &  -1.80  $\pm$  0.18  \\
  UGC07866       &  ~~~76  &  40.09  &  38.54  &  -2.00  $\pm$  0.17  &  -2.07  $\pm$  0.16  &  -1.78  $\pm$  0.19  &  -1.87  $\pm$  0.22  \\
  UGC07949       &  ~~~38  &  40.15  &  39.01  &  -2.06  $\pm$  0.28  &  -3.23  $\pm$  0.56  &  -2.46  $\pm$  0.46  &  -2.84  $\pm$  0.58  \\
  UGC08024       &  ~~~59  &  39.27  &  38.45  &  -1.41  $\pm$  0.29  &  -1.99  $\pm$  0.36  &  -1.30  $\pm$  0.26  &  -1.61  $\pm$  0.33  \\
  UGC08188       &  ~~176  &  39.97  &  38.95  &  -2.46  $\pm$  0.28  &  -2.11  $\pm$  0.18  &  -2.12  $\pm$  0.24  &  -2.00  $\pm$  0.18  \\
  UGC08201       &  ~~~67  &  39.49  &  38.50  &  -2.23  $\pm$  0.33  &  -2.69  $\pm$  0.33  &  -2.17  $\pm$  0.30  &  -1.85  $\pm$  0.27  \\
  UGC08651       &  ~~~33  &  39.52  &  38.19  &  -2.54  $\pm$  0.34  &  -2.55  $\pm$  0.42  &  -2.22  $\pm$  0.39  &  -1.95  $\pm$  0.41  \\
  UGC12613       &  ~~~33  &  38.38  &  37.10  &  -2.60  $\pm$  0.41  &  -2.71  $\pm$  0.37  &  -1.85  $\pm$  0.30  &  -2.40  $\pm$  0.39  \\
  UGCA106        &  ~~135  &  40.90  &  39.45  &  -1.95  $\pm$  0.16  &  -2.07  $\pm$  0.16  &  -1.97  $\pm$  0.19  &  -1.87  $\pm$  0.17  \\
  WLM            &  ~~407  &  39.04  &  37.23  &  -1.95  $\pm$  0.08  &  -2.06  $\pm$  0.09  &  -1.85  $\pm$  0.09  &  -1.99  $\pm$  0.11  \\
\hline
\label{tab:LFinfo}
\end{longtable}

\twocolumn
\newpage

%Galaxies with a possible truncation
%IC5332, NGC300, NGC0598, NGC0628, NGC2403, NGC2903, NGC3031, NGC3239, NGC3351, NGC4214, NGC4068, NGC4449, NGC4490, NGC4826, NGC5055, NGC5194, UGC00668, UGC1291, UGC8188, 

%##################################################################################
%##################################################################################
\section{Global Galaxy Properties}  \label{sec:galprop}
In this section we describe the global galaxy properties of the LVL sample. The properties presented here were derived in \cite{cook14c}, but we provide a brief overview and any relevant caveats.

\subsection{Global Dust Correction}\label{sec:gdustcorr}
The internal dust corrections for the majority of the galaxies are carried out via the prescription of \citet{hao11} where the extinction in the FUV bandpass ($A_{\rm{FUV}}$) is first calculated via the empirical relationship with the 24$\mu m$-to-FUV luminosity ratio ($L_{24}$/$L_{\rm{FUV}}$). All other bandpass extinctions ($A_{\rm{\lambda}}$) are derived from $A_{\rm{FUV}}$ in combination with the dust extinction curve of \citet{draine03}. 

However, not all galaxies have 24$\mu$m and/or FUV detections ($N=47$) from which to derive internal extinctions. For these galaxies we calculate $A_{\rm{H\alpha}}$ based on the empirical scaling relationship from \citet{lee09b} between $A_{\rm{H\alpha}}$ and $M_B$, where the $A_{\rm{H\alpha}}$ values are derived from spectroscopic Balmer decrement measurements. We derive all other bandpass extinctions for these 47 galaxies using the dust reddening curve of \citet{draine03}. \citet{lee09b} quotes a 20\% scatter for galaxies with $-14.7 > M_B > -18$ and only 10\% for galaxies fainter than $-14.7$. Since the majority ($N=36$) of these 47 galaxies are low-luminosity dwarfs ($M_B>-$14.7) and the remaining galaxies are fainter than $M_B > -18$, we do not expect these galaxies to add a significant amount of scatter to the trends presented in \S\ref{sec:trends}.

%\begin{equation}
%  A_{\rm{H\alpha}} =  
%    \begin{cases}
%    0.10 			   &  \textrm{if $M_B > -$14.5}\\
%    1.971 + 0.323M_B + 0.0134M_B^2 &  \textrm{if $M_B \leq -$14.5},
%   \end{cases}
%\end{equation}

%\noindent  

%Quantities in this study which have been corrected for internal extinction are denoted with a subscript ``0" and represent an intrinsic measurement.

\subsection{$M_{\rm{B,0}}$}
The apparent optical fluxes are taken from \citep{cook14a} measured inside the UV apertures of \citet{lee11}, where the optical images have been cleaned of contaminating sources (i.e., background galaxies and foreground stars). These optical fluxes have been converted to absolute magnitudes via the distance moduli of \citet{kennicutt08}, and corrected for internal dust correction (see previous section).

\subsection{Stellar Mass}
We utilize the LVL stellar masses of \cite{cook14c} which are derived from a constant Spitzer 3.6$\mu m$ mass-to-light ratio ($\Upsilon_{\star}^{3.6\mu m}$) of 0.5. In recent years, a consistent $\Upsilon_{\star}^{3.6\mu m}$ of 0.5 has begun to emerge between population synthesis models and the baryonic Tully-Fisher relation measurements \citep{oh08,eskew12,mcgaugh14,meidt14,barnes14,mcgaugh15}. 

%Due to the uniform Spitzer 3.6 $\mu m$ coverage for nearly the entire sample, we do not expect our mass estimate to add any additional scatter in the relationships presented in \S\ref{sec:trends}.

\subsection{SFR, $\Sigma_{\rm{SFR}}$, sSFR }
The SFRs are derived via FUV fluxes taken from the work of \citet{lee11} where we corrected for internal extinction as discussed in \S\ref{sec:gdustcorr}. We use the prescription of \citet{murphy11} to convert the dust-corrected FUV fluxes into a SFR where the uncertainties on the SFR include only the photometric uncertainties. The SFR prescriptions of \citet{murphy11} are an updated version of \citet{kennicutt98} using a Kroupa \citep{kroupa01} IMF. The calibration coefficient of \citet{kennicutt98} is a factor of $\sim$1.6 larger than the \citet{murphy11} coefficient \citep[see][]{calzetti07}.

The $\Sigma_{\rm{SFR}}$ values are derived via normalizing the SFR by the area of the galaxy aperture. We choose to define our areas by the R25 optical apertures instead of the UV apertures since the R25 apertures are more self consistent as these apertures are defined as the isophotal ellipse at which the optical surface brightness equals 25 mag/arcsec$^2$ in the $B-$band. For example, there are many galaxies whose UV aperture extends well beyond the extent of our FUV identified sources, thus leaving a non-negligible fraction of the area as empty space. This occurs for the following reasons: 1) irregular morphology of dwarfs and non-symmetric spiral features in spirals can result in UV apertures which are skewed in one direction relative to the center of the galaxy leaving empty space inside the UV aperture on the opposite side (e.g., NGC2366, NGC5457/M101); and 2) many galaxies have extended, low-surface brightness UV emission at large radii \citep[i.e., XUV galaxies;][]{thilker07} with a few, or often no, identifiable FUV sources above a 3$\sigma$ detection threshold. However, we verify that using either the UV or IR apertures do not significantly affect our results In \S\ref{sec:trends}. The specific SFRs (sSFRs) are derived by normalizing the SFRs by the LVL stellar mass discussed in the previous section.

\subsection{Metallicity}
We utilize the metallicity catalog compiled by \cite{cook14c}. This catalog is an updated version of the \citet{marble10} compilation, where both with ``direct" and ``strong-line" oxygen abundances measured by \citet{berg12} and ``strong-line" oxygen abundances from \citet{moustakas10} were added. \cite{cook14c} replaced all previous ``strong-line" abundances with ``direct" abundances and updated any previous ``strong-line" abundances with newer ``strong-line" abundances. The majority of the ``strong-line" metallicities are on similar calibration scales \citep[i.e., similar to][]{mcgaugh91}. The final metallicity sample is composed of roughly half ``direct" and half ``strong-line" values.

%\subsection{TIR}
%\subsection{PAH/TIR emission}

%##################################################################################
%##################################################################################
\section{Results: Trends with Galaxy Properties} \label{sec:trends}
In this section we examine the relationships between LF slope and global galaxy properties, where LF slope forms the strongest trends with star formation rate properties. We end this section with an evaluation of how our choice of dust correction, aperture correction, and luminosity function binning affect the strength of the relationships between LF slope and galaxy properties. We find that the trends found between LF slope and galaxy environment are not significantly affected by these choices.

\subsection{LF trends}
\begin{figure}
  \begin{center}
  \includegraphics[scale=0.48]{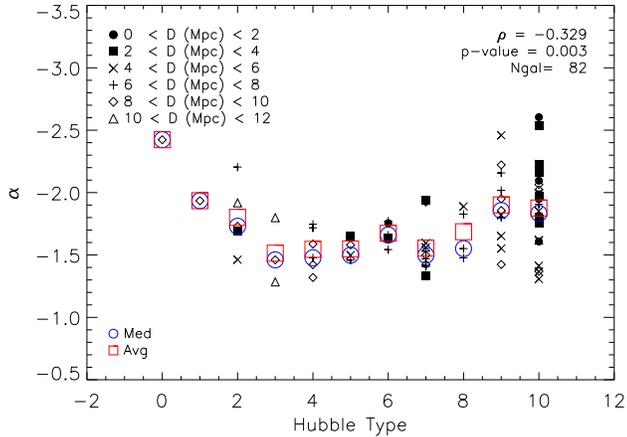}
  \caption{A plot of the LF slope ($\alpha$) versus optical morphology. The symbols are defined in the lower-left corner, where we have binned our galaxies according to their distance. Filled symbols represent closer galaxies and open symbols represent distant galaxies. The number of galaxies and the Spearman correlation coefficient ($\rho$) are indicated in the upper right legend. We find no clear trend between $\alpha$ and morphology, and find increased scatter for later-type galaxies ($T \geq 9$).}
  \label{fig:morph}
  \end{center}
\end{figure}  
  
Figure~\ref{fig:morph} is a plot of the LF slope ($\alpha$) versus optical morphology for the LVL sample. The symbols for individual galaxies are described in the upper-left legend representing distance bins in intervals of 2~Mpc, where filled symbols represent closer galaxies and open symbols represent distant galaxies. The larger open squares and open circles represent the average and median, respectively, of the galaxies in each morphological type. 

We find no discernible trend with galaxy type in Figure~\ref{fig:morph}. This is supported by the small Spearman correlation coefficient ($\rho$) of -0.329. We also find that there is increased scatter for later galaxy types ($T \geq 9$), which are generally the lower-mass, dwarf galaxies. It should also be noted that the majority of the dwarf galaxies tend to be closer (i.e., filled symbols) and have fewer sources.

\begin{figure}
  \begin{center}
  \includegraphics[scale=0.48]{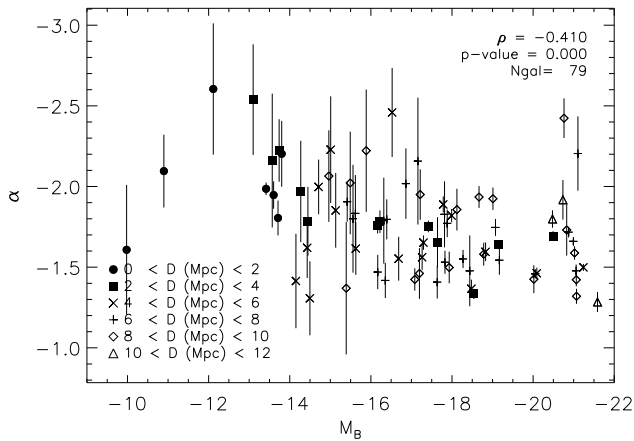}
  \caption{A plot of the LF slope ($\alpha$) versus absolute $B-$band magnitude ($M_{\rm{B}}$) similar to that of Figure~\ref{fig:morph}. We find a weak trend between these two quantities.}
  \label{fig:Mb}
  \end{center}
\end{figure}  
  
Figure~\ref{fig:Mb} is a plot of $\alpha$ versus the absolute $B-$band magnitude ($M_{\rm{B}}$), which has been corrected for internal dust extinction. We find the LF slope shows a weak correlation with $M_{\rm{B}}$, where low-luminosity galaxies tend to have steeper (i.e., more negative) slopes. The weak trend is supported by a correlation coefficient of -0.410. 

\begin{figure}
  \begin{center}
  \includegraphics[scale=0.48]{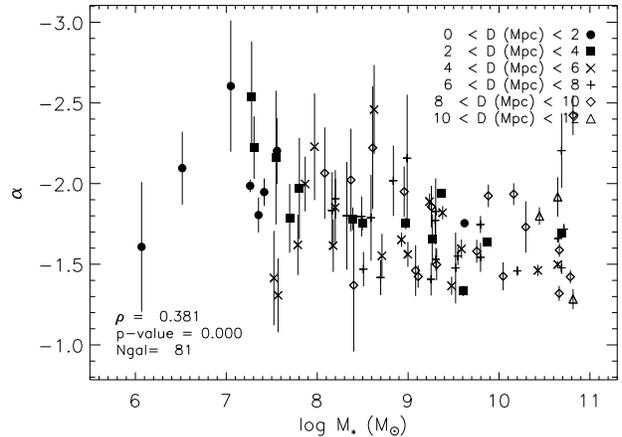}
  \caption{A plot of the LF slope ($\alpha$) versus stellar mass ($M_{\star}$) similar to that of Figure~\ref{fig:morph}. We find a weak trend between these two quantities similar to that found for $M_{\rm{B}}$.}
  \label{fig:mstar}
  \end{center}
\end{figure}  

Figure~\ref{fig:mstar} is a plot $\alpha$ versus the stellar mass ($M_{\star}$). We find the LF slope shows a weak correlation with $M_{\star}$, where low-mass galaxies tend to have steeper slopes. The trend is supported by a correlation coefficient of 0.381. We note that this trend is similar to that of galaxy luminosity ($M_{\rm{B}}$) due to the tight correlation between optical luminosity and $M_{\star}$ \citep[][]{cook14c}.
  
\begin{figure}
  \begin{center}
  \includegraphics[scale=0.48]{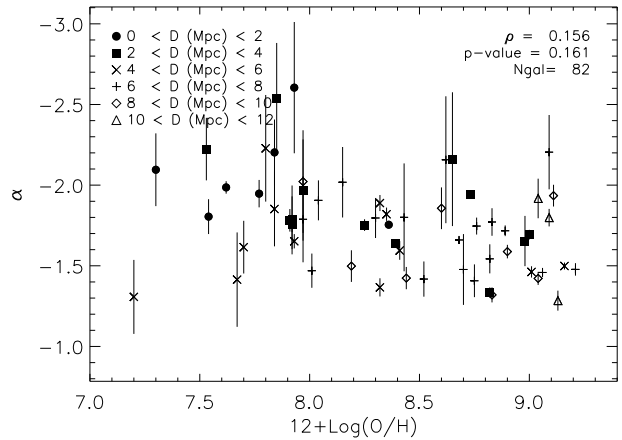}
  \caption{A plot of the LF slope ($\alpha$) versus metallicity similar to that of Figure~\ref{fig:morph}. We find a no trend between these two quantities.}
  \label{fig:metal}
  \end{center}
\end{figure}  
  
Figure~\ref{fig:metal} is a plot of $\alpha$ versus the metallicity (12+log(O/H)) for 82 of the LVL galaxies. The metallicities were taken from the most recent compilation of the LVL galaxies by \cite{cook14c}. We find no trend with a correlation coefficient of $\rho \sim 0.156$. The lack of a trend is somewhat surprising since metallicity forms correlations with both $M_{\rm{B}}$ and $M_{\star}$ \citep[e.g.,][]{hlee03,hlee07,berg12}. 

\begin{figure}
  \begin{center}
  \includegraphics[scale=0.48]{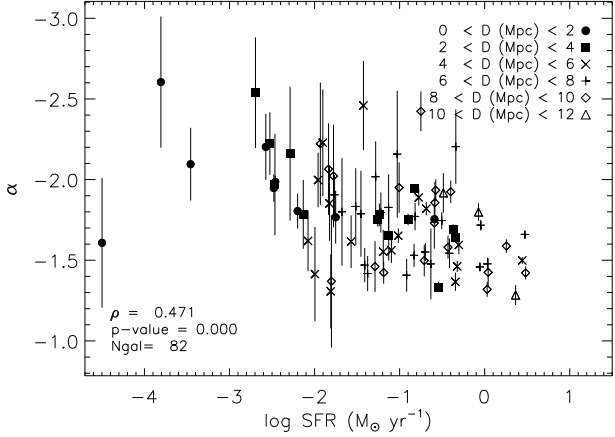}
  \caption{A plot of the LF slope ($\alpha$) versus SFR similar to that of Figure~\ref{fig:morph}. We find a weak-to-moderately strong trend between these two quantities.}
  \label{fig:SFR}
  \end{center}
\end{figure}  
  
Figure~\ref{fig:SFR} is a plot of $\alpha$ versus the FUV-derived SFR which has been corrected for internal dust extinction. We find a weak-to-moderate correlation ($\rho = 0.471 $) between $\alpha$ and SFR, where low-SFR galaxies tend to have steeper slopes. Despite the tight correlations between SFR and $M_{\star}$ and $M_{\rm{B}}$ \citep[e.g.,][]{ellison08,salim14}, the stronger trend between $\alpha$ and SFR suggests that the properties of star-forming regions are more closely connected to the star formation properties of a galaxy.

\begin{figure}
  \begin{center}
  \includegraphics[scale=0.48]{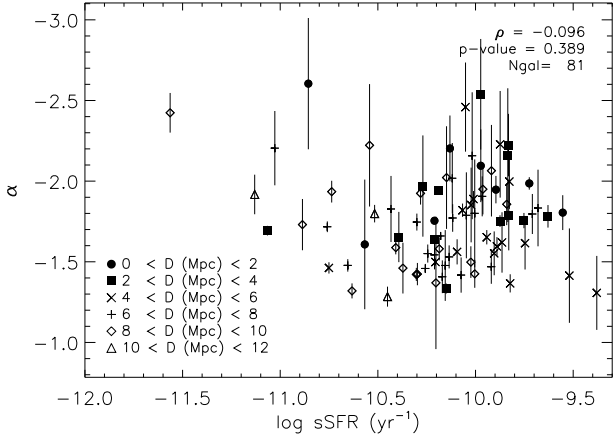}
  \caption{A plot of the LF slope ($\alpha$) versus sSFR similar to that of Figure~\ref{fig:morph}. We find no trend between these two quantities.}
  \label{fig:sSFR}
  \end{center}
\end{figure}  
  
Figure~\ref{fig:sSFR} is a plot of $\alpha$ versus sSFR (sSFR$\equiv$SFR/M$_{\star}$), which is the global SFR of a galaxy normalized by the the total stellar mass. Despite the weak-to-moderate relationship between $\alpha$ and SFR, we find no trend between LF slope and sSFR ($\rho = 0.096$).

\begin{figure}
  \begin{center}
  \includegraphics[scale=0.48]{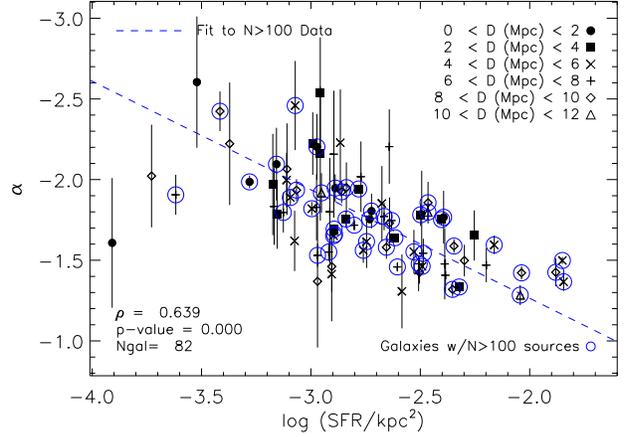}
  \caption{A plot of the LF slope ($\alpha$) versus $\Sigma_{\rm{SFR}}$ similar to that of Figure~\ref{fig:morph}. We find a moderately strong trend between these two quantities. The blue-open circles represent galaxies with a total number of sources greater than 100 where we expect random scatter due to stochastic effects to be minimal. The correlation coefficient for galaxies with $N>100$ is 0.705; stronger than that of all galaxies in the figure. The blue-dashed line is a bisector fit to the $N>100$ data and is described by equation Eqn~\ref{eqn:sfrarea}.}
  \label{fig:sfrarea}
  \end{center}
\end{figure}  

Figure~\ref{fig:sfrarea} is a plot of $\alpha$ versus star formation rate density ($\Sigma_{\rm{SFR}}\equiv$SFR/kpc$^2$), which is the global SFR of a galaxy normalized by the the area of the galaxy. We find that $\Sigma_{\rm{SFR}}$ exhibits a moderately strong correlation ($\rho = 0.639$; the strongest for any global galaxy property) with LF slope, where low $\Sigma_{\rm{SFR}}$ environments tend to have steeper slopes. It is interesting to note that the strength of the trends between two global star formation properties (sSFR and $\Sigma_{\rm{SFR}}$) are drastically different from the trend with SFR when normalizing by $M_{\star}$ and area. The moderate trend between $\alpha$ and $\Sigma_{\rm{SFR}}$ and no trend between $\alpha$ and sSFR suggests that the density of star formation is physically connected to the properties of star-forming regions.

In Figure~\ref{fig:sfrarea} we have also overlaid large blue circles onto the symbols of galaxies with a total number of sources greater than 100, where it is possible that low number statistics may add random scatter to LF slopes (see \S~\ref{sec:simscatter}). We find that correlation between LF slope and $\Sigma_{\rm{SFR}}$ is reproduced when using only galaxies with good number statistics, and that these galaxies span nearly the full range of the entire galaxy sample. Furthermore, we find that the correlation coefficient is stronger for $N>100$ data with a value of 0.705. To quantify the relationship between LF slope and $\Sigma_{\rm{SFR}}$, we perform a bisector fit to the $N>100$ data which is described by the equation: .

\begin{equation}
\alpha = 0.67 \times \log \Sigma_{\rm{SFR}} + 0.08.
  \label{eqn:sfrarea}
\end{equation}

\begin{figure}
  \begin{center}
  \includegraphics[scale=0.48]{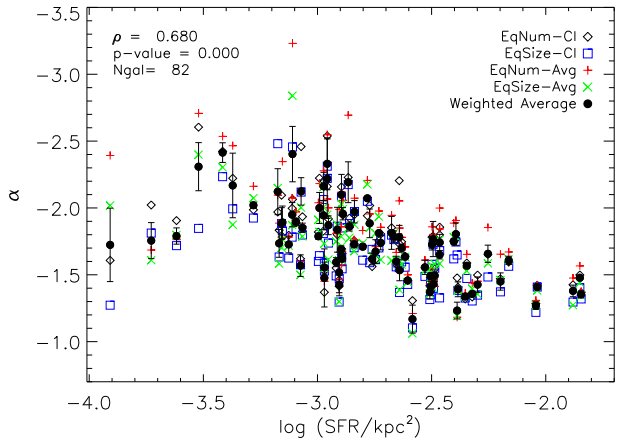}
  \caption{A plot of the LF slope ($\alpha$) versus $\Sigma_{\rm{SFR}}$ where we have combined 2 photometry methods and 2 LF binning methods to produce 4 combinations of the LF slope. This graphically shows the systematics of our methods. We find a strong trend between these $\alpha$ and $\Sigma_{\rm{SFR}}$ when using a weighted average LF slope for the 4 combinations of photometry and binning methods.}
  \label{fig:sfrarea4way}
  \end{center}
\end{figure}  

\subsection{The Effects of Dust Correction, Aperture correction, and LF Binning}\label{sec:4way}
In previous sections, we have made choices on how to obtain the total, unobscured luminosities of our regions and how to bin these luminosities to construct our LFs and consequently measure the LF slopes. In our analysis we have chosen a fiducial set of methods (deriving our own dust correction for individual regions, CI-based aperture correction, equal numbering binning) which represent what we believe to be a reasonable set of methods that minimizes any biases. In this section, we provide an evaluation on our choice of dust correction, aperture correction, and LF binning on the strength of the trends between LF slope and galaxy environment.

To test the effects of our dust correction for individual regions, we evaluate the strength of the trend between LF slope and $\Sigma_{\rm{SFR}}$ (i.e., the strongest trend) when applying no dust correction, the galaxy-wide correction of \cite{hao11}, and previous corrections derived for individual regions \citep{leroy08,liu11}. We find that the plots for no correction, galaxy-wide correction, and the correction derived for regions are very similar in shape and structure to Figure~\ref{fig:sfrarea}, and that the correlation coefficients are $0.402$, $0.523$, and $0.561$, respectively. The persistence of a trend between $\alpha$ and $\Sigma_{\rm{SFR}}$ when using no dust correction and two different dust corrections suggests that our results are not significantly sensitive to the choice of dust correction methods.

%Although the correlation coefficient for no dust correction is somewhat lower than the one derived in our fiducial method, we do not believe that no dust correction is appropriate for our FUV sources since light at this wavelength is easily absorbed by dust and since many of our galaxies have high dust content \citep[i.e., $M_{\rm{B} bright than }$][]{cook14b} 

The effects of aperture correction and LF binning can be evaluated via generating four sets LF slopes by combining the two choices of aperture corrections and the two LF binning methods for each galaxy: 1) EqNum-CI: CI-based aperture correction and equal number binning; 2) EqNum-Avg: average aperture correction and equal number binning; 3)EqSize-CI: CI-based aperture correction and equal luminosity size binning; 4) EqSize-Avg: average aperture correction and equal luminosity size binning. 

Figure~\ref{fig:sfrarea4way} illustrates the variability of our trends when using different methods where the LF slopes of all 4 combinations are plotted against $\Sigma_{\rm{SFR}}$ (i.e., the strongest trend). The symbols with different colors represent the results for each combination of aperture correction and LF binning, and the filled circles represent the weighted average for the 4 combinations of methods. There are a handful of galaxies where the LF slope for one or two methods are outliers from the weighted average (i.e., $\Sigma_{\rm{SFR}} \sim-3.9$ with $-1.3 < \Delta\alpha < -2.4$); however, the majority of galaxies show only small LF slope differences between the 4 combinations of methods. 
 
It is interesting to note that the weighted average of all 4 LF slopes forms a stronger correlation with $\Sigma_{\rm{SFR}}$ compared to any single method. This moderate-to-strong correlation ($\rho \sim 0.680$) between the weighted average $\alpha$ values and $\Sigma_{\rm{SFR}}$ suggests that our results are robust to our choice of photometry and luminosity binning.

%##################################################################################
%##################################################################################
\section{Discussion}\label{sec:disc}
In this section we investigate observational effects that may lead to artificial trends in the LF slope. We ultimately find that these effects cannot fully account for the correlations between LF slope and SFR nor $\Sigma_{\rm{SFR}}$. We then compare our LF slope$-$host galaxy trends to previous studies, and end the section with a discussion of the implications of these trends on the star formation process.

\subsection{Competing Effects}
In this section we provide a careful examination of two competing factors that can change the LF slope: stochastic sampling of the LF in galaxies with low number statistics and blending of sources at lower resolution due to distance. To quantify these effects, we simulate the scatter in LF slope expected given the number of sources in each galaxy, and degrade the resolution of two large, nearby spiral galaxies to greater distances and re-measure the LF slopes.

%##################################################################################
%##################################################################################
\subsubsection{Stochastic Sampling of the LF} \label{sec:simscatter}
The luminosity distributions of star-forming regions follow a power-law, where there are fewer bright regions than faint regions. Sampling a luminosity distribution with low number statistics over many iterations results in the random presence of bright regions that can effect the slope of the distribution. For example, in a distribution where there is expected to be just one region in the brightest luminosity bin, the presence of many bright regions with low probabilities would flatten the LF slope while the presence of no bright regions would steepen the LF slope.  

\begin{figure}
  \begin{center}
  \includegraphics[scale=0.48]{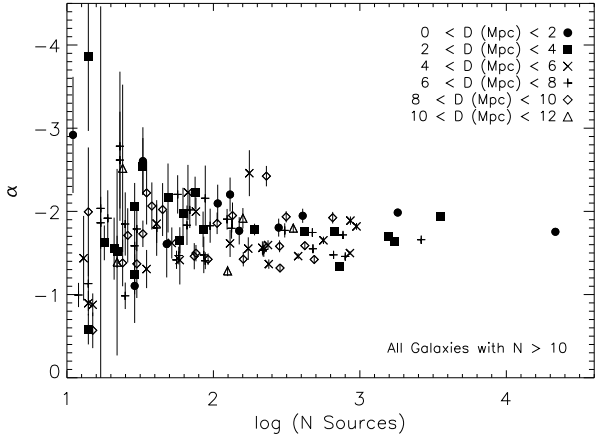}
  \caption{A plot of the LF slope ($\alpha$) versus number of sources similar to that of Figure~\ref{fig:morph}. We find increased scatter towards lower numbers of sources, and decreased scatter towards higher numbers of sources. The average slope of all LFs is $-1.76 \pm 0.30$ (median$=-1.75$).}
  \label{fig:nsrcs}
  \end{center}
\end{figure}

It is not clear a priori if stochastic scatter in $\alpha$ introduces a bias (i.e., if the LF slope is systematically steeper or flatter for galaxies with low number statistics) nor is the amplitude of this scatter constrained. A first order check of both the potential systematic bias and quantification of this scatter can be seen in Figure~\ref{fig:nsrcs}, where $\alpha$ is plotted against the total number of sources in each galaxy. Visual inspection of Figure~\ref{fig:nsrcs} reveals that $\alpha$ shows increased scatter for galaxies with lower numbers of star-forming regions ($N\lesssim 100$), and that $\alpha$ converges to a value slightly flatter than $-2$. The median of all good LFs is $-1.75 \pm 0.30$ (the average is $-1.76$). 

The random scatter of $\alpha$ values at lower number statistics suggests that stochasticity does not systematically bias the LF slope to either steeper nor flatter slopes. Also, it is not clear why our LFs converge to flatter slopes than those found in star cluster and \ha studies of local spiral and dwarf galaxies; $-2$. However, a recent star cluster study using multi-band ($NUV,U,B,V,R$) HST imaging has found a systematic trend for LF slopes derived from the same star clusters at different wavelengths, where the LF slopes derived from shorter wavelengths tend to have flatter slopes (Adamo et al. 2016; submitted). It is plausible that the flatter FUV-derived LF slopes measured here are an extension of this trend with shorter wavelengths.

We can also make a first order check on the effects that low number statistics might have on the trends found in \S~\ref{sec:trends} between LF slope and galaxy properties via examination of these trends when only using galaxies with a total number of sources greater 100. The blue circles of Figure~\ref{fig:sfrarea} represent galaxies with $N>100$ and form a stronger correlation between LF slope and $\Sigma_{\rm{SFR}}$. This stronger relationship is supported by the Spearman correlation coefficient of $\rho=0.705$ for $N>100$ galaxies. In addition, the properties of these galaxies span nearly the full range of both LF slope and $\Sigma_{\rm{SFR}}$ as the entire galaxy sample. The persistence of the trend between LF slope and $\Sigma_{\rm{SFR}}$ for galaxies with good number statistics suggests that stochasticity is not responsible for the this trend. 

To quantify this scatter, we simulate the expected change in $\alpha$ given the number of sources in each galaxy. For our simulations, we assume a universal LF that can be described by a power-law with a slope of $-2$. For each galaxy, we randomly draw star-forming regions equal to the number of sources found in the galaxy and limit these sources to luminosities above the peak in the real luminosity histogram and below the luminosity of the brightest star-forming region found in the entire sample. The lower luminosity limit mimics the sensitivity of the FUV image, and the upper limit allows for the possibility of creating random bright objects. The subsequent LF is then fit with the same methods used to fit our real LFs and we iterate 1000 times. 

\begin{figure}
  \begin{center}
  \includegraphics[scale=0.48]{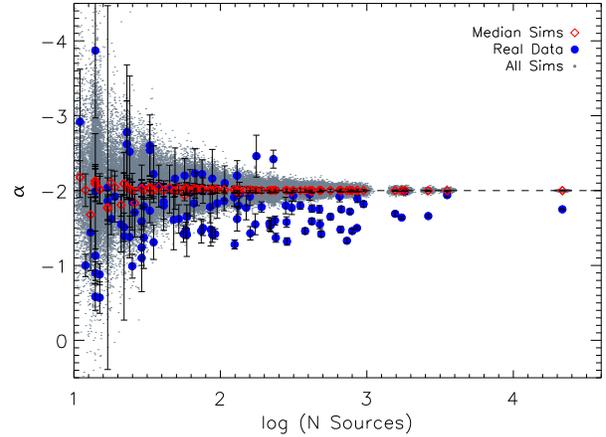}
  \caption{A plot of simulated LF slope ($\alpha$) versus the number of sources, where the red diamonds represent the median of all simulations, the median error bars represent the 68th and 32nd percentile of the simulations (i.e., the 1$\sigma$ confidence interval), and the grey dots represent all simulated realizations of the LF slope. In addition, the filled-blue circles represent the real data of our final sample. We find that the both the real and simulated LF slopes randomly scatter above and below the input $-2$ slope. Thus stochastic scatter is not responsible for the trends found between LF slope and galaxy properties.}
  \label{fig:nsrcs_sim}
  \end{center}
\end{figure}  

Figure~\ref{fig:nsrcs_sim} shows the simulated $\alpha$ values plotted against the total number of sources found in each galaxy, where the red diamonds represent the median of all simulations, the median error bars represent the 68th and 32nd percentile of the simulations (i.e., the 1$\sigma$ confidence interval), and the grey dots represent all simulated realizations of the LF slope. For comparison purposes, we have overplotted the real data points measured for each galaxy represented by filled-blue circles.

Visual inspection of Figure~\ref{fig:nsrcs_sim} shows that the median of all simulations randomly scatter around a $-2$ slope for galaxies with both high and low number statistics, but find that the scatter in the simulations increases symmetrically around a slope of $-2$ towards galaxies with low number statistics. When comparing the real LF slopes to the simulated slopes, we find that galaxies with high number statistics ($N>100$) tend to have flatter slopes than the simulations. However, for galaxies with low number statistics, we find similar distributions of scatter between our simulations and real data, where the scatter is not systematic, but rather randomly distributed both above and below the input $-2$ LF slope. The random scatter seen in both the observed and simulated LF slopes in galaxies with low number statistics (i.e., where we expect stochasticity to be prominent) suggests that stochastic scatter is not responsible for the trends found between LF slope and galaxy properties. 

\subsubsection{Resolution Effects}\label{sec:distsim}
Changes in the physical resolution of an image (i.e., due to galaxy distance) can change the degree to which distinct sources can be distinguished from one another. Consequently resolution can affect the slope of the LF by changing the number of sources in various luminosity bins. Greater galaxy distances  (i.e., the FWHM is probing larger physical scales) can blend nearby sources into a single brighter source, thus flattening the LF slope by decreasing the number of faint sources and increasing the number of bright sources.

%For example, at the GALEX resolution of 5$\arcsec$, our FUV images will probe a physical scale of 24~pc at a distance of 1~Mpc, and will probe 240~pc at 10~Mpc. 

%The amplitude of the change in $\alpha$ due to resolution is somewhat uncertain since the severity of deblending can depend on the degree of crowding in a galaxy, and since the shape of the LF may not be significantly affected if the majority of faint sources lost due to deblending have luminosities fainter than the peak of the luminosity histogram. 

In a study of \ha sources (i.e., star-forming \hii regions), \cite{kennicutt89} degraded the FWHM of H$\alpha$ images of two nearby galaxies at evenly spaced intervals from 30~pc to 900~pc scales, and found that the shape of the LF showed no significant change below 200 pc scales \citep[for star clusters, see also:][]{randriam13}. At physical scales between 200 and 400~pc, the resolution experiments of \cite{kennicutt89} showed a loss of sources at lower luminosities, but no significant change in the bright end of the LF slope. Conversely, \cite{scoville01} found that the high resolution HST \ha LF of M51 had a steeper LF slope by 0.2-0.5 in $\alpha$ when compared to various ground-based LF slopes, where the physical scales probed by HST for M51 is $\sim$10~pc and is 50-100~pc for the different ground-based studies \citep[see also;][]{pleuss00}.

Another factor to consider when comparing the LF slopes based on imaging which probe different physical scales is the fractal picture of star formation. Many studies on the clustering of young stars and star-forming regions have found similar luminosity distributions when probing a wide range of physical scales (i.e., from a few to 100s of parsecs). \cite{bastian07} tested the change in the LF slope in M33 for different physical scales, and found no significant differences when probing 10-100~pc scales. If star-formation truly follows a fractal, or scale-free, distribution, then the LF slope should remain constant (at least when probing below a reasonably moderate physical scale). The results of \cite{kennicutt89} and \cite{bastian07} suggest that the LF slope should be not be significantly impacted by distance if probing a physical scale less than a few hundred parsecs.

\begin{figure}
  \begin{center}
  \includegraphics[scale=0.48]{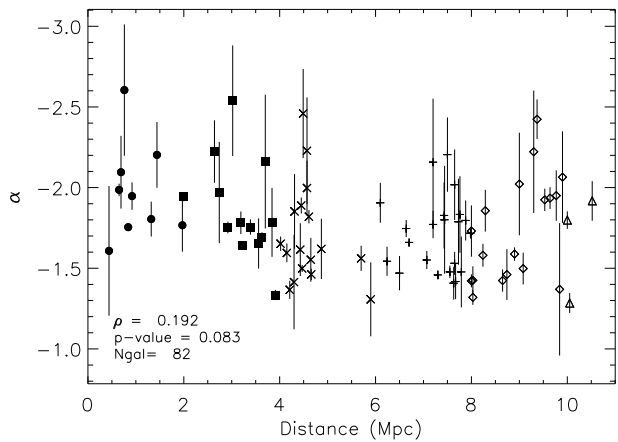}
  \caption{A plot of the LF slope ($\alpha$) versus distance similar to that of Figure~\ref{fig:morph}. We find no trend between these two quantities indicating that resolution effects due to distance has little or no effect on our LF slope-galaxy property relationships.}
  \label{fig:dist}
  \end{center}
\end{figure}  
  
The maximum distance of the \ntot galaxies with clean LFs studied in this work is $\sim$10.5~Mpc, where the FWHM of the FUV image probes a physical scale of $\sim$250~pc. Thus we do not anticipate that the blending of sources due to distance will significantly impact our results. However, we perform several experiments to test the significance of distance/resolution may have on our LF slopes. As a first-order check, we examine the relationship between LF slope ($\alpha$) and distance in Figure~\ref{fig:dist}. We find no correlation between $\alpha$ and distance, which is supported by the small Spearman correlation coefficient ($\rho$=0.192). The lack of correlation between $\alpha$ and distance indicates that distance has a small or no effect on $\alpha$, but does not rule out the effect entirely since the magnitude of this effect may be small.

Another check on the effect of resolution is to examine the distribution of distances in the trends between $\alpha$ and the two properties that show the strongest correlations: SFR and $\Sigma_{\rm{SFR}}$. A visual inspection of both Figure~\ref{fig:SFR} and Figure~\ref{fig:sfrarea} show that the filled symbols, which represent the closest galaxies, show a similar trend to that of the entire sample and span nearly the entire range of properties in both $\alpha$ and galaxy property. The persistence of these trends for the closest galaxies also suggests that resolution effects due to distance cannot be entirely responsible for the trends found between $\alpha$ and the galaxy properties of SFR and $\Sigma_{\rm{SFR}}$.

Despite no clear trend between $\alpha$ and distance and the similar trend between $\alpha$ and star formation properties for the closest galaxies, it is possible that the strength of these trends may be exaggerated by resolution effects due to distance. To quantify this effect, we simulate the LF slopes of NGC0598 (M33; distance$\sim0.8$Mpc) and NGC0300 (distance$\sim2.0$Mpc) at increased distances by smoothing the GALEX FUV FWHM to larger pixel values, thus probing larger physical scales. The 5$\arcsec$ native resolution of the FUV images corresponds to a physical scale of 24~pc and 48~pc for NGC0598 and NGC0300, respectively. In addition, the FUV native resolution at 10.5~Mpc (the maximum distance of our sample) corresponds to a 250~pc physical scale. We degrade the native resolution GALEX FUV image to equally spaced 30~pc physical scales out to 270~pc to simulate the effects of distance on LF slope. 

Our smoothing procedure takes into account the smoothing of FUV images to a larger FWHM values, the decrease of flux from sources due to an increased distance, and the increase of Poisson noise due to the decrease in flux for sources. For each physical scale we perform the following operations: 1) convert the native resolution image's pixel units into counts, 2) convolve the native resolution FUV image with the appropriate Gaussian to produce a smoothed image with a FWHM equal to the desired physical scale, 3) scale down the image counts due to distance, 4) rebin the smoothed image (i.e., increase the size of the pixels) to recover the 5$\arcsec$ FWHM of the native FUV image, 5) draw a Poisson deviate per pixel to increase the Poisson noise, and 5) convert the resultant image back into the original pixel units. This procedure effectively yields GALEX FUV observations as if they were located at increased distances.

With the distance simulated images constructed, we then use identical methods to those used on the native resolution image to identify FUV regions, perform photometry, and generate LFs for each simulated image. We visually verify that all obvious sources are identified in our simulated images. In addition, since our original analysis required a dust correction for individual sources via the $L_{24}$/$L_{\rm{FUV}}$ luminosity ratio, we have also smoothed the Spitzer 24$\mu m$ image to the corresponding physical scales using the same procedure to that of the FUV images.  

Rebinning the smoothed images recovers the native resolution in pixels, thus, rebinning facilitates using the same photometry parameters (in pixels), the same aperture correction relationships derived in \S\ref{sec:phot}, and the same dust correction relationship derived in \S\ref{sec:dustcorr} to measure the total flux for each source in the smoothed images. 

\begin{figure*}
  \begin{center}
  \includegraphics[scale=0.48]{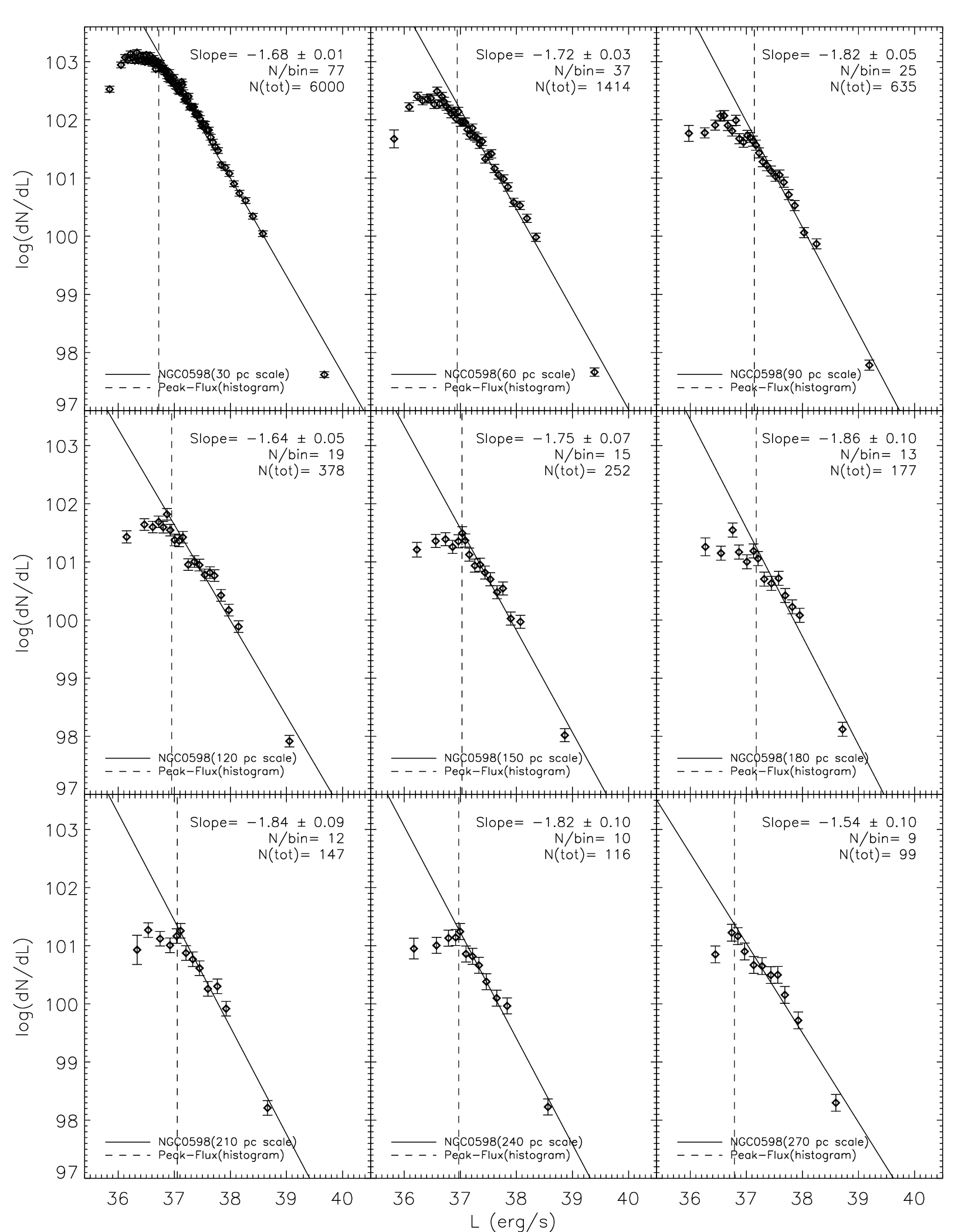}
  \caption{The simulated LFs for NGC0598 at increased distances, where each panel presents the results for increasing distances from left-to-right and from top-to-bottom (30-270~pc physical scales), and the dashed line represents the peak of the luminosity histogram for the sources found in each smoothed image. There is a general trend of decreasing numbers of sources in the faint luminosity bins for increasing distances. However, we find that each simulated LF slope are consistent with the native resolution LF slope within the errors. }
  \label{fig:distsim_all_ngc0598}
  \end{center}
\end{figure*}  

\begin{figure*}
  \begin{center}
  \includegraphics[scale=0.48]{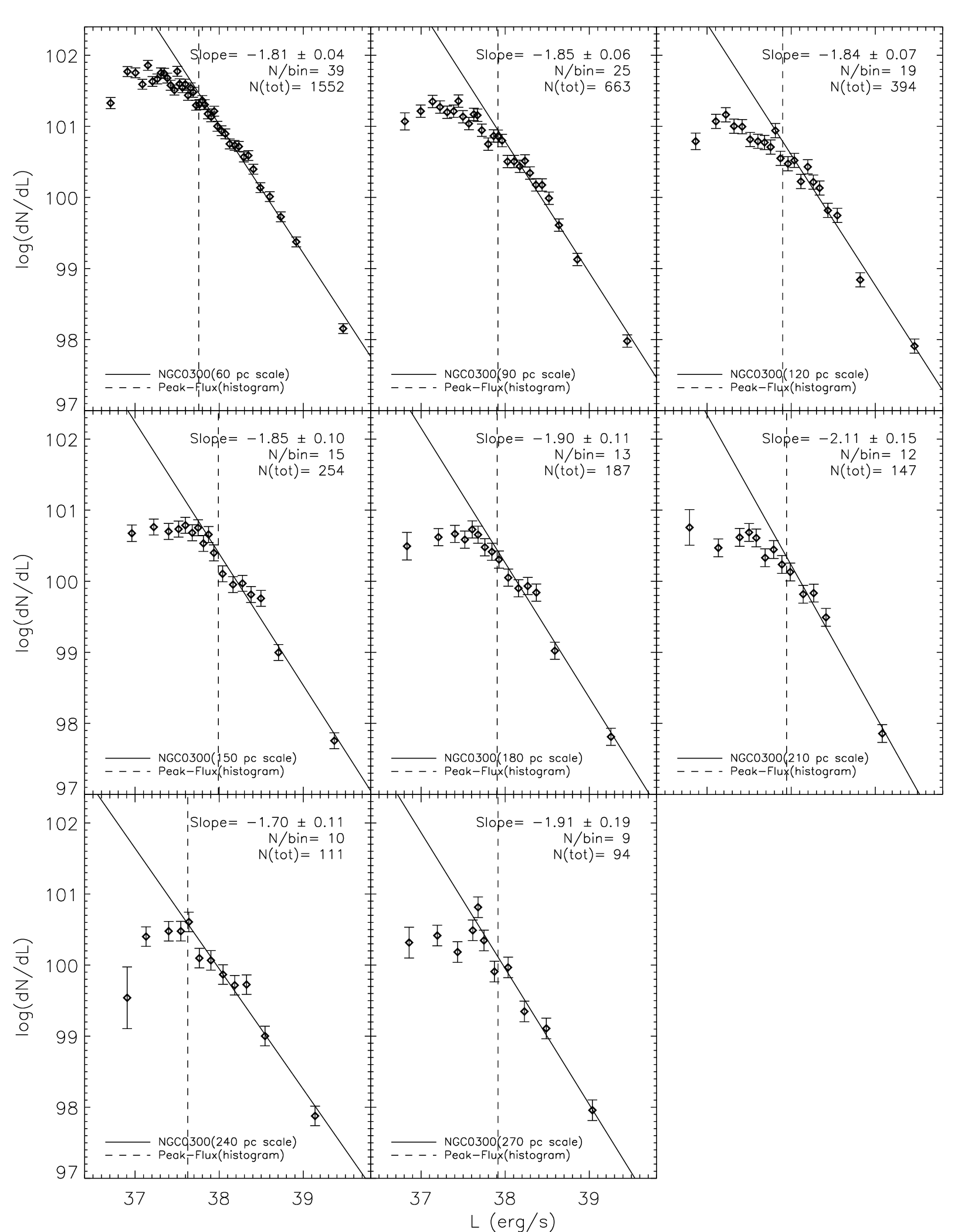}
  \caption{The simulated LFs for NGC0300 at increased distances, where each panel presents the results for increasing distances from left-to-right and from top-to-bottom (60-270~pc physical scales), and the dashed line represents the peak of the luminosity histogram for the sources found in each smoothed image. There is a general trend of decreasing numbers of sources in the faint luminosity bins for increasing distances. However, we find that each simulated LF slope are consistent with the native resolution LF slope within the errors.}
  \label{fig:distsim_all_ngc0300}
  \end{center}
\end{figure*}  

We then fit a power-law to the LF of simulated sources using the same binning and fitting methods as those for the native resolution images. Figure~\ref{fig:distsim_all_ngc0598} and Figure~\ref{fig:distsim_all_ngc0300} show the resultant LFs for NGC0598 and NGC0300, respectively, where each panel represents the LF for different physical scales (i.e., physical scale increases from left-to-right and from top-to-bottom). For each galaxy, there is a general trend of lower numbers of sources in the lower luminosity bins at larger physical scales (i.e., greater distances). It is likely that the reduced numbers of faint objects is largely due to a combination of the reduced counts for sources and the subsequent increased Poisson noise for increasing simulated distances. These two effects would act to hide faint sources in the noise of the simulated images. 

It is reasonable to anticipate that fewer numbers of faint sources would result in flatter LF slopes. However we find that the slope of the brighter luminosity bins do not change significantly across all simulated distance. These results are consistent with the resolution experiments performed by \cite{kennicutt89}, where they found no significant change in LF slope below physical scales of 200~pc and found fewer faint sources at physical scales between 200 and 400~pc but found no change on the bright end of the LF shape. 

We also note that the reduced number of fainter sources indicate that the completeness (and detection) limits of our simulated images is increased for greater simulated distances. However, our fitting method takes into account varying completeness limits by fitting a power law to luminosity bins at and above the peak of the luminosity histogram, where incompleteness is likely to become prominent. Thus, our fitted LF slopes at greater simulated distances will continue to accurately describe the shape of the bright end of the luminosity function. The increased completeness limits for our simulated images can seen in both Figure~\ref{fig:distsim_all_ngc0598} and Figure~\ref{fig:distsim_all_ngc0300} where the peak of the luminosity histogram (vertical-dashed lines) show a rough, overall trend toward higher luminosities at greater simulated distances. For instance the top 3 panels of Figure~\ref{fig:distsim_all_ngc0598} show a peak histogram luminosity of 36.7, 37.0, and 37.2 erg/s for physical scales of 30, 60, and 90~pc, respectively.

\begin{figure}
  \begin{center}
  \includegraphics[scale=0.48]{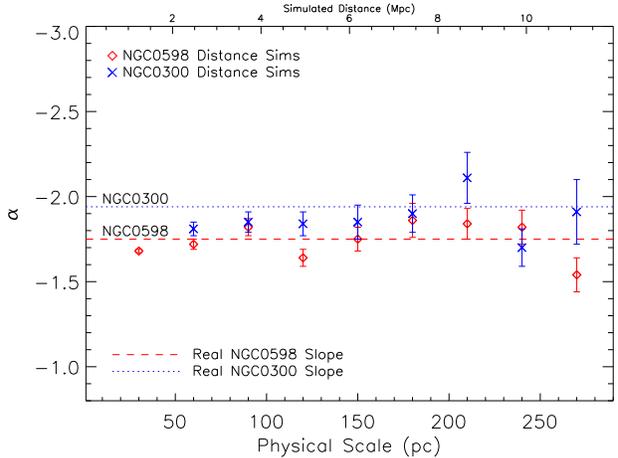}
  \caption{The combined LF slope results for our distance simulations where the red diamonds represent the simulated LF slopes for NGC0598, the blue-dotted line represents the LF slope measured on the native resolution image of NGC0598, the blue asterisks represent the simulated LF slopes for NGC0300, and the red-dashed line represents the LF slope measured on the native resolution image of NGC0598. We find no clear trend between the simulated LF slopes with distance. In addition, we find that the simulated LF slopes are consistent with the native resolution LF slope within the errors.}
  \label{fig:distsim}
  \end{center}
\end{figure}

The power-law fits to each of the simulated LFs are well behaved and yield similar slopes to the native resolution values within the fitted errors. This can be seen in a plot of the LF slope for each distance simulation versus the physical scale probed by each distance in Figure~\ref{fig:distsim}, where the red diamonds and blue asterisks represent the simulated LF slope for NGC0598 and NGC0300, respectively. The horizontal dashed-red and dotted-blue lines represent the native resolution LF slopes for NGC0598 and NGC0300, respectively. 

There is no clear trend between LF slope and the simulated distance for either galaxy, and we find that the simulated LF slopes are consistent with the native resolution LF slopes ($-1.75$ and $-1.94$ for NGC0598 and NGC0300, respectively) within the error bars of the simulated slopes. The median simulated slopes are $-1.75$ and $-1.85$ with a standard deviation of $0.11$ and $0.12$ for NGC0598 and NGC0300, respectively. The lack of a trend between simulated LF slope and distance suggests that the blending of sources due to distance does not significantly affect the trends found between LF slope and star formation rate properties. 

We note that a lack of a change in the LF slope with increasing distance is only valid when measuring the bright end of the LF and so long as the lower limit of the fit is allowed to adjust to an appropriate luminosity given the completeness limit. Although we find no significant change in the bright end of the LF slope at greater distances, we do find that the faint end does appear to flatten due to the loss and/or blending of faint sources. However, since our LF slopes are derived for the bright end of the LF and our LF fits allow for a variable lower limit based on the completeness limit, we conclude that our trends between LF slope and galaxy properties are robust to blending due to distance.

A caveat of this analysis is the effect of blending due to the filling factor (or number density) of star-forming regions in galaxies with smaller diameters. When comparing the LF slopes of two galaxies at the same distance with the same number of regions but with different effective areas can conceivably blend sources, thus possibly flattening the LF slope of the galaxy with the smaller diameter (i.e., higher SFR per area). As a first order attempt to quantify this effect we have examined any trends between LF slope and area (kpc$^2$) and the number of regions per area (\#/kpc$^2$). We find no trend between alpha and these two quantities ($\rho\sim0.22~\rm{and} \sim0.11$ for area and the number of regions per area, respectively); for brevity we do not show these plots. However, the plot of LF slope versus $N$ per area may not be a definitive test as the true number of sources cannot be known without HST resolution. We plan to simulate this effect in an upcoming paper looking at the LFs of star clusters with HST NUV imaging.

\subsection{Comparison to Previous Trends}
There are only a handful of studies that have previously examined trends between the LF slope and global galaxy properties. Of these previous studies, the majority have used \hii regions \citep[i.e., H$\alpha$ identified sources;][]{kennicutt89,elmegreen1999,youngblood99,vanzee00,thilker02}, one using Pa$\alpha$ identified \hii regions \citep{liu13a}, and one using HST F814W ($\sim$I-band) identified star clusters \citep{whitmore14}. These previous studies examined between 10-35 galaxies, where only the galaxy samples of \cite{kennicutt89,elmegreen1999,whitmore14} contained a range of morphology (i.e., early-type to irregular galaxies). Unfortunately, the other previous studies examined either only irregular galaxies \citep{youngblood99,vanzee00} or only later-type spirals \citep{thilker02,liu13a}. For each of these studies the authors purport that any trends between LF slope and galaxy property are inconclusive for their galaxy sample. 

As a first step toward comparing our trends to those of previous trends, we provide a brief discussion on the effects of LFs derived from different wavelengths. Our star-forming regions are derived from GALEX FUV sources which probe different ages compared to previous studies. FUV wavelengths probe ages less than 100~Myr, while regions identified at \ha wavelengths probe ages less than 10~Myr and star clusters identified at HST F814W wavelengths probe ages $\sim$a few Gyr.

%A possible caveat to address when comparing FUV regions to older star clusters is that star cluster populations will have lost $\sim$ 10\% of clusters per decade in logarithmic age \citep{}. However, it is unclear if the clusters that are disrupted over time

Previous studies have examined the effects of age on the shape of LFs and found that populations of older star-forming regions tend to have steeper LF slopes \citep{oey98}. This steepening is attributed to the dimming of bright star-forming regions due stellar evolution, where the luminous star-forming regions contain relatively more massive stars which have short main sequence lifetimes. We can test this hypothesis by comparing the average LF slopes between \ha, FUV, and F814W LFs. The studies of \cite{kennicutt89} and \cite{elmegreen1999} report an average LF slope of $-2\pm0.5$ for \ha regions, \cite{whitmore14} reports an average LF slope of $-2.37\pm 0.18$ for F814W star clusters, and we report an average $\alpha$ of $-1.76\pm0.30$ for FUV regions. The average LF slope for the oldest population, F814W star clusters, is steeper than both the \ha and FUV averages. However, we find that our FUV (100~Myr) average $\alpha$ value is consistent with \ha (10~Myr) studies. 

Another caveat to address is the effect of incomplete sampling of the stellar IMF in low-mass (similarly low-luminosity) star-forming regions. \cite{hermanowicz13} found that the FUV luminosities of \hii regions scales with the \ha luminosities, but that this scaling breaks down at low luminosities (Log$\sim 37$ erg/s) due to incomplete sampling of the stellar IMF. Since our regions are identified at FUV wavelengths they are not likely to be affected by IMF sampling issues, but this does not rule out the possibility that previous \ha studies are affected by incomplete sampling of the IMF. However, we note that none of our LFs are have fitted luminosity bins below Log$\sim 37$ erg/s, thus we believe our comparison to previous \ha studies is valid.

We also note that making LF slope comparisons from different studies for individual galaxies is not straight forward. There are multiple factors that can change the shape of the LF (i.e., wavelength, resolution, LF identification methods, binning methods, and LF fitting methods) which may have competing effects on the LF slope.  However, we believe that any trends found between LF slope and galaxy properties are robust if the regions were uniformly identified and the LFs were generated with uniform methods for a sample of galaxies. Consequently, we provide only a qualitative comparison of our LF slope trends with galaxy properties to those of previous studies which have galaxy samples with a range in galaxy properties and whose LFs have been generated via uniform methods.

\subsubsection{Galaxy Type}
%The studies of \cite{kennicutt89} and \cite{elmegreen1999} found a correlation between $\alpha$ and galaxy type \citep[see also]{vanzee00}, and \cite{whitmore14} found correlations between $\alpha$ and several galaxy properties: galaxy type, SFR, $M_{\rm{B}}$. However, both \cite{thilker02} and \cite{youngblood99} found no convincing evidence for a trend between LF slope and galaxy type.

\cite{kennicutt89} and \cite{elmegreen1999} found a correlation between the LF slope and galaxy type, where earlier type galaxies (Sa-Sab) exhibit steeper slopes than later-type galaxies (Sbc-Sd) and irregulars (Sm-Irr). \cite{whitmore14} found a similar correlation between $\alpha$ and galaxy type when ignoring the earliest type galaxies ($T<4$), where later-type galaxies and irregulars showed flatter slopes.  

Our trend of $\alpha$ versus galaxy type is shown in Figure~\ref{fig:morph}, where we find no clear trend between LF slope and galaxy type. We find increased scatter in $\alpha$ for later-type galaxies, especially in the irregular galaxies ($T \geq 9$) where the LF slopes span the entire range of $\alpha$ values for whole galaxy sample. 

%The results of our stochastic simulations in \S\ref{sec:simscatter} reveal that part of this scatter can be attributed to random sampling of the LF in environments with low number statistics. 

The disagreement between the results presented in this study and those from previous studies may be due to underlying galaxy properties and the number of galaxies in the latest galaxy type bins ($T \geq 9$). Our final sample is dominated by these later-type galaxies where 42 galaxies have $T \geq 9$, thus our sample reflects the properties of typical irregular galaxies in the local universe. However, the significance of trends in the previous studies heavily rely on the flatter slopes of only a few irregular type galaxies in each study: 3 galaxies (LMC, SMC, and NGC4449) in \cite{kennicutt89}, 1 galaxy (the Antennae ;NGC 4038/39) in \cite{whitmore14}, and 3 galaxies (NGC1156, NGC1569, and NGC2366) in \cite{elmegreen1999}. 

The Antennae is a starbursting merger remnant whose properties do not reflect normal star-forming galaxies, thus may be an outlier in this parameter space. Furthermore, \cite{whitmore14} reported that the significance of the trend between $\alpha$ and galaxy type is reduced by a factor of two when removing the Antennae galaxy. In addition, \cite{kennicutt89} reported that the dispersion in $\alpha$ among spirals of the same galaxy type is similar to the magnitude of the trend between $\alpha$ and galaxy type, leaving 3 irregular galaxies to support the trend. Finally, \cite{elmegreen1999} studied only 11 galaxies in total which contain only 3 irregular galaxies where the range of $\alpha$ among each galaxy type spans more than half the total range of the entire sample (see Figure 8 of \cite{elmegreen1999}). The irregular galaxies in each of the previous studies have enhanced SFRs compared to typical local universe irregulars and may indicate the star formation properties of these galaxies are responsible for the flatter LF slopes.

\subsubsection{$M_{\rm{B}}$}
\cite{whitmore14} found a weak correlation between $\alpha$ and the absolute $B-$band magnitude ($M_{\rm{B}}$), where brighter galaxies tend to have steeper slopes. Figure~\ref{fig:Mb} shows a weak trend between $\alpha$ and $M_{\rm{B}}$, where fainter galaxies tend to have steeper slopes; this is the opposite trend that was found by \cite{whitmore14}. The discrepancy between these two trends is unclear. However, it is possible that the range covered in $M_{\rm{B}}$ by \cite{whitmore14} is not large enough accurately characterize this relationship where the range in $M_{\rm{B}}$ covered by \cite{whitmore14} is from $-17.5$ to $-21.5$~mag compared to our range of $-10$ to $-22$~mag. It is also possible there exists no trend between $\alpha$ and $M_{\rm{B}}$ since both our trend and \cite{whitmore14} trend are only weakly correlated. 

\subsubsection{SFR}
\cite{whitmore14} found a correlation between $\alpha$ and global SFR, where lower SFR galaxies tend to have steeper slopes. We find a similar trend between these two quantities which can be seen in Figure~\ref{fig:SFR}. We find that, in our analysis, the trend between $\alpha$ and SFR is weak-to-moderate, second only to $\Sigma_{\rm{SFR}}$. We note that our trend spans a larger range compared to \cite{whitmore14}, where our data have a Log(SFR) range of $-5$ to $1~\rm{M_{\odot} yr^{-1}}$ and the data of \cite{whitmore14} span a range of $-2$ to $0.5~\rm{M_{\odot} yr^{-1}}$.

\cite{whitmore14} attributes this trend to a ``size-of-sample effect" where higher SFRs create more star-forming regions, and in turn, are more likely to create brighter star-forming regions. This can affect the slope of the LF since low-SFR galaxies are less likely to form bright regions, thus creating a steeper slope.

\begin{figure}
  \begin{center}
  \includegraphics[scale=0.48]{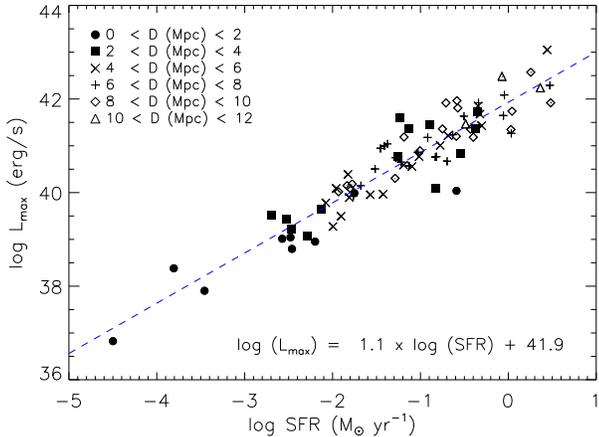}
  \caption{A plot of the luminosity of the brightest star-forming region ($L_{\rm{\rm{max}}}$) and SFR. We find a tight correlation with a correlation coefficient ($\rho$) of 0.904. Since the global SFR of a galaxy scales with the total number of sources, this tight trend indicates that a ``size-of-sample" effect is present in our sample. However, both the real and simulated LFs of galaxies with low number statistics show random scatter both above and below the canonical $-2$ LF slope suggesting that the ``size-of-sample" effect is not responsible for the trends between LF slope and galaxy properties.}
  \label{fig:lmaxsfr}
  \end{center}
\end{figure}  

The ``size-of-sample effect" has been examined by many studies of star clusters \citep[][]{larsen02,weidner04,bastian08,cook12,whitmore14}, and is statistical in nature. For example, randomly drawing star-forming regions from a power-law with a slope of $-2$ from distributions with more regions yields a larger probability to draw a brighter region, and conversely you are less likely to draw a bright region from a distribution with small number statistics. This results in a correlation between the sample size (i.e., the total number of regions and subsequently SFR) and the luminosity of the brightest star-forming region ($L_{\rm{max}}$). We find such a trend between SFR and $L_{\rm{max}}$ in Figure~\ref{fig:lmaxsfr} with a strong correlation ($\rho = 0.904$). We fit a line to the relationship which is described the equation:

\begin{equation}
  \log_{10} (L_{\rm{max}}) = 1.1 \times log_{10} (SFR) + 41.9.
\end{equation}

\noindent The tight correlation in Figure~\ref{fig:lmaxsfr} suggests that the ``size-of-sample" effect is present in our sample, but our stochastic simulations of \S\ref{sec:simscatter} shows that low numbers of sources causes random scatter both above and below the $-2$ slope, not systematically steeper slopes. Thus, we conclude that the steeper slopes in galaxies with low number statistics are not due to a ``size-of-sample" effect.

It is interesting to note that previous star cluster studies have found a relationship between the LF slope ($\alpha$), number of clusters detected \citep[which is proportional to SFR;][]{whitmore14}, and the maximum mass (which scales with luminosity) of star clusters \citep[e.g.,][]{hunter03,larsen02,bastian08}. This relationship is derived in \cite{hunter03} and is described via: 

\begin{equation}
M_{\rm{max}} \propto N^{(1/\alpha - 1)}.
\label{eqn:mmaxn}
\end{equation}

\noindent Since $M_{\rm{max}}$ is proportional to $L_{\rm{max}}$ and the number of regions ($N$) is proportional to SFR, we can use this relationship to estimate the combined LF slope of our star-forming regions by measuring the slope of the $L_{\rm{max}}$-SFR trend in Figure~\ref{fig:lmaxsfr}. In log-log space, the exponent of equation~\ref{eqn:mmaxn} becomes the slope of Figure~\ref{fig:lmaxsfr} which we measure to be $\sim$unity for our sample. This translates into a combined LF slope of $-2$ (it is actually $+2$ but \cite{hunter03} defined $\alpha$ such that the negative sign is already included in the equation).

The combined $-2$ LF slope matches the canonical LF slope, but is different than the median LF slope of our galaxy sample of $-1.75$ (average is $-1.76$). However, when we combine all star-forming regions into one composite sample and fit a LF, we find a bright-end slope of $-1.93$; very near the $-2$ value derived from the $L_{\rm{max}}$-SFR trend (see \S~\ref{sec:lfbreak}).

%##################################################################################
%##################################################################################
\subsection{Implications on the Star Formation Process}
In the previous section we compared our LF slope trends to those of previous studies, where we found disagreement for galaxy type and $M_{\rm{B}}$, however, found agreement for the trend between $\alpha$ and SFR. We also simulated other competing effects and found that the trends between $\alpha$ and star formation properties are not significantly affected. We conclude that the trends between $\alpha$ and star formation environment (SFR and $\Sigma_{\rm{SFR}}$) are real, and we discuss the implications of these trends on the star formation process in this section.

%Figures~\ref{fig:sfrarea} and \ref{fig:sfrarea_sim} reveal that $\Sigma_{\rm{SFR}}$ formed the strongest correlation ($\rho=0.636$) with $\alpha$, and that the stochastic scatter could not account for the range of $\alpha$ values observed in the trend. Furthermore, we found that the $\alpha-\Sigma_{\rm{SFR}}$ trend is reproduced by only those data with $\alpha$ values found to lie $>3\sigma$ outside of all simulated $\alpha$ values. For these reasons, we conclude that the star formation environment is, at least partially, responsible for the correlation between $\alpha$ and $\Sigma_{\rm{SFR}}$.

The star formation process is ultimately governed by the gaseous starting material from which stars are made, and thus dependent on the host galaxy's interstellar medium (ISM) properties: gas density ($\Sigma_{\rm{gas}} \equiv M_{\rm{gas}}$/kpc$^2$) and star formation efficiency ($M_{\star}/M_{\rm{gas}}$). The Kennicutt-Schmidt law characterizes a tight relationship between $\Sigma_{\rm{gas}}$ and $\Sigma_{\rm{SFR}}$, where galaxies with higher $\Sigma_{\rm{gas}}$ tend to have higher $\Sigma_{\rm{SFR}}$ and the vertical normalization of this relationship represents the star formation efficiency \citep{kennicutt98}. Furthermore, \cite{kruijssen12} formulated a theoretical framework of star formation where higher $\Sigma_{\rm{SFR}}$ and $\Sigma_{\rm{gas}}$ environments can achieve higher star formation efficiencies, thus enhancing clustered star formation efficiency ($\Gamma \equiv CFR/SFR$), which is supported by observational evidence \citep[][]{goddard10,adamo11,cook12}.

$\Gamma$ represents the mass fraction of stars located in star clusters compared to those located in the field, thus $\Gamma$ probes the total mass (and similarly the luminosity) of star-forming regions in a galaxy. In this framework, as the $\Sigma_{\rm{SFR}}$ of a galaxy increases there is an increased fraction of stars in the densest, brightest parts of the star formation hierarchy (i.e., star-forming regions). We hypothesize that the flatter LF slopes seen in higher $\Sigma_{\rm{SFR}}$ galaxies in our study are the result of higher $\Sigma_{\rm{gas}}$ and higher star formation efficiencies creating relatively more bright star-forming regions, thus flattening the LF slope. 

A simple test of this hypothesis is to examine any relationship between $\alpha$ and $\Gamma$. This is one of the goals for the second paper in this series where we will derive the masses of our star-forming regions and measure $\Gamma$ for our final galaxy sample.

%!!!PUT IN A DISSCUSSION OF KRUIJSSEN15...HE SAYS THAT SIGMASFR IS MORE RELATED TO THE ISM PROPERTIES OF A GALAXY AND NOT ABSOLUTE QUANTITIES SUCH AS SFR AND MSTAR...In addition, Meidt+15 also connects locat star formation to galaxy properties via gas pressure (how galactic environment regulates star formation).
%##################################################################################
%##################################################################################
\subsection{LF Break}\label{sec:lfbreak}
Visual inspection of all LFs in Appendix~\ref{sec:allLFs} reveal no clear break in the LFs of any galaxy. However, 10 galaxies show reduced numbers in the brightest L bin compared to the extrapolated power-law fit (i.e., the data point of the brightest luminosity bin falls $>2\sigma$ below the fit). A reduced number of sources in the brightest luminsoity bin has been seen in both star cluster studies \citep[e.g.,][]{gieles06a,bastian12a,adamo15} and the studies of giant molecular clouds \citep[][]{rosolowsky07} suggesting that the luminosity and mass functions of star-forming regions may have a high luminosity and mass function turnover. In this section we investigate if there exists a break in the LFs of star-forming regions. 

Although we find no clear break in any individual galaxy's LF, we create a composite LF comprised of all sources from many galaxies to increase the number statistics of bright star-forming regions. The composite LF includes the star-forming regions of any galaxy with a measured inclination angle less than 60$\degr$ (plus those manually put back into the analysis; see \S\ref{sec:galregs}). Consequently, the composite sample includes galaxies with poor LF fits and those with fewer than 30 total sources in addition to galaxies in our final sample. 

%If the hierarchical picture of star formation is correct, then any star-forming regions drawn from the same LF (i.e., a power-law with slope equal to $-2$) will not change the shape of the LF, but rather statistically reconstruct the canonical LF.  

\begin{figure}
  \begin{center}
  \includegraphics[scale=0.48]{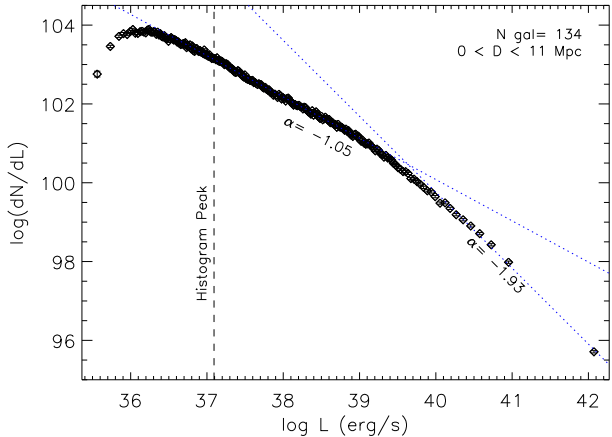}
  \caption{The composite LF where we have combined the star-forming regions of 134 galaxies to increase the number statistics of bright star-forming regions. The y-axis is normalized to the same arbitrary number as those of the LFs of individual galaxies. We find a clear break at log $(L)\sim 39.5$ erg/s. However, we show that this break is due to different luminosity limits of more distant galaxies in the composite sample.}
  \label{fig:LFcomp_all}
  \end{center}
\end{figure}  

\begin{figure}
  \begin{center}
  \includegraphics[scale=0.48]{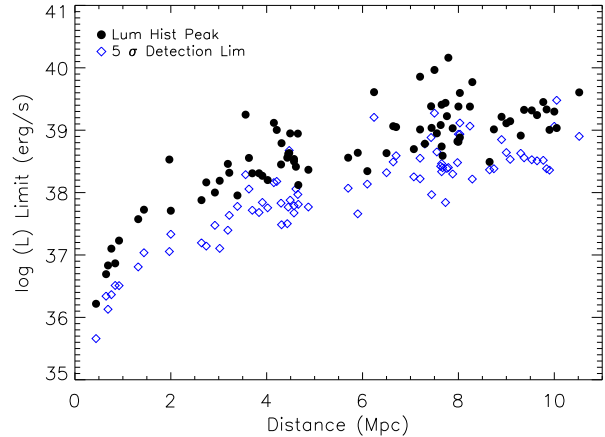}
  \caption{A plot of the peak of the luminosity histogram (filled, black circles) and the 5$\sigma$ detection limit (open, blue diamonds) for each galaxy's FUV image versus the galaxy's distance. We find a relationship between these two luminosity limits and distance, where the limiting luminosity increases with distance and levels off near $\sim$39.5 erg/s. The break in Figure~\ref{fig:LFcomp_all} is due to reduced numbers of faint composite sources of galaxies at distance further away than 7~Mpc.}
  \label{fig:detlim}
  \end{center}
\end{figure}

Figure~\ref{fig:LFcomp_all} shows the composite LF of 134 galaxies where there is a clear break near a luminosity of $\sim$39.5 erg/s. The bright end of the composite LF shows a slope of $\sim-1.93$; however, we find that the flatter slope at luminosities fainter than Log($L$)$\sim39.5$~erg/s is due to incompleteness of more distant galaxies. Figure~\ref{fig:detlim} is a plot of the peak of the luminosity histogram (filled, black circles) and the 5$\sigma$ detection limit (open, blue diamonds) for each galaxy's FUV image versus the galaxy's distance. We find a relationship between these two luminosity limits and distance, where the limiting luminosity increases with distance and levels off near $\sim$39.5 erg/s. 

%The effects of increasing luminosity limits with distance on the LF truncation stem from reduced numbers of faint composite sources for more distant galaxies. 

To understand how the break in the composite LF is related to the completeness limit of galaxies at different distances, we focus on the peak of the luminosity histogram in Figure~\ref{fig:detlim} since each galaxy's LF shows reduced numbers fainter than this luminosity due to incompleteness. The peak of the luminosity histogram levels off at 39.5 erg/s, hence there will be little-to-no reduced numbers of composite sources brighter than 39.5 erg/s due to incompleteness. However, there will be reduced numbers of composite sources below 39.5 erg/s for all galaxies beyond a distance of 7~Mpc, where the peak in the luminosity histogram is, on average, near $\sim$39.5~erg/s. In other words, the reduced numbers of sources in the composite LF fainter than 39.5 erg/s is caused by the incomplete luminosity bins of galaxies more distant than 7~Mpc. This suggests that the break seen in the composite LF of Figure~\ref{fig:LFcomp_all} likely does not reflect the true LF shape. We cannot confirm the existence of a break in the LF of star-forming regions with the data presented in this study. Given these results, we advocate for greater care of the completeness limits of galaxies in the samples of future composite LF studies.

%##################################################################################
%##################################################################################
\section{Summary}
We have used the GALEX FUV images of 258 nearby galaxies to identify tens-of-thousands of star-forming regions and measured each galaxy's luminosity function. Since our star-forming regions consist of point sources and extended sources with different morphologies, we measure the total flux of each source via an aperture correction based on concentration index, which is a measure of the extent of a source's radial profile. In addition, we use panchromatic data of star-forming regions in M51 from \cite{calzetti05} to derive a relationship between 24$\mu m$/FUV luminosity ratio and dust extinction. We apply this dust correction to the FUV luminosities of our star-forming regions to account for the flux reprocessed by dust. 

A comparison between two luminosity binning methods reveals that an average differences of $-$0.21 for luminosity function slopes with equal luminosity-size bins and equal number bins. This difference is relatively small compared to the full range of luminosity slopes (-2.8 to -1.0), which indicates that our slopes are robust. In our final galaxy sample, we exclude any galaxy with an inclination angle less than 60$\degr$, $N<30$ total sources, and ill-behaved luminosity functions leaving \ntot galaxies in the final sample.

We find weak-to-moderate correlations between luminosity function slope and several global galaxy properties; the strongest of which are star formation rate and star formation rate density ($\Sigma_{\rm{SFR}}$). We also find no correlation between luminosity function slope and galaxy type. There is increased scatter for later-type galaxies ($T \geq 9$) where we expect stochastic scatter to be present due to low number statistics in low-SFR, dwarf galaxies. In an examination of the effects of our choice of dust correction, aperture correction, the luminosity function binning method choices on our trends between luminosity function slope and galaxy properties we find the following: 1) our trends persist when using no dust correction and when assuming 2 different dust correction prescriptions; 2) we find that the trends persist when using 2 different aperture correction definitions and 2 luminosity binning methods; and 3) a stronger correlation between luminosity slope and ($\Sigma_{\rm{SFR}}$) when using a weighted average of all 4 combinations of aperture correction and binning methods. 

A careful examination of competing effects that can change the luminosity function slope reveals that neither stochastic sampling of the luminosity function in galaxies with low-number statistics nor the effects of blending due to distance can fully account for these relationships. We find that stochastic scatter due to low number statistics results in random scatter to both flatter and steeper slopes in galaxies with fewer star-forming regions. Since this scatter is not systematic, we conclude that our trends between luminosity function slope and galaxy properties are not due to stochastic scatter. Examination of distance effects reveals no trend between the measured luminosity function slopes and distance. Furthermore, we find that the closest galaxies ($D < 2~\rm{Mpc}$) reproduce the trends between $\alpha$ and the galaxy properties of SFR and $\Sigma_{\rm{SFR}}$ for the entire sample. We also perform distance simulations on two nearby galaxies NGC0300 and NGC0598 (M33) located at a distance of 2.0 and 0.8~Mpc, respectively. In these simulations we smooth and rebin the native resolution FUV images to increasing distances, re-extract sources, and re-measure the LF slopes. The resulting simulated slopes are consistent with the measured native resolution slopes within the errors and show no clear trend between simulated LF slope and the physical scales probed (FWHM range corresponding to 24-270~pc). These results suggest that the effects of resolution due to distance do not significantly affect the correlations seen between luminosity function slope and global star formation properties.

We hypothesize that the trends found between luminosity function slope and star formation rate properties may be connected to the star formation process via gas density and star formation efficiency since star formation is ultimately governed by the gaseous starting material. Support for this hypothesis can be found in previous studies where higher star formation rate densities correlate with 1) higher gas densities and 2) larger fractions of stars forming in star clusters ($\Gamma$). This supporting observational evidence suggests that higher star formation rate density environments obtain higher star formation efficiencies, produce a larger fraction of stars in star clusters, produce relatively more bright star-forming regions, and thus flatten the luminosity function slope. A direct test of this hypothesis is to measure $\Gamma$ for this galaxy sample and examine any relationship between luminosity function slope and $\Gamma$. This is left for future work.

Finally, we test if a break in the LF accurately describes the LF shape by creating a composite LF of many galaxies to increase the number statistics of bright star-forming regions. We find a clear break in the composite luminosity function near a luminosity of $\sim$39.5 erg/s; however, we find that this break is not real. We attribute the break in the power-law to a dearth faint sources where the most distant galaxies are affected by incompleteness below $L\sim$39.5 erg/s.

%##################################################################################
%##################################################################################
\bibliographystyle{bibliographies/mn2e.bst}   
\bibliography{bibliographies/all.bib}  

\begin{thebibliography}{82}
\expandafter\ifx\csname natexlab\endcsname\relax\def\natexlab#1{#1}\fi

\bibitem[{{Adamo} {et~al}\mbox{.}(2015){Adamo}, {Kruijssen}, {Bastian},
  {Silva-Villa}, \& {Ryon}}]{adamo15}
{Adamo} A., {Kruijssen} J.~M.~D., {Bastian} N., {Silva-Villa} E., {Ryon} J.,
  2015, \mnras, 452, 246

\bibitem[{{Adamo}, {{\"O}stlin} \& {Zackrisson}(2011){Adamo}, {{\"O}stlin}, \&
  {Zackrisson}}]{adamo11}
{Adamo} A., {{\"O}stlin} G., {Zackrisson} E., 2011, \mnras, 417, 1904

\bibitem[{{Barnes} {et~al}\mbox{.}(2014){Barnes}, {van Zee}, {Dale},
  {Staudaher}, {Bullock}, {Calzetti}, {Chandar}, \& {Dalcanton}}]{barnes14}
{Barnes} K.~L., {van Zee} L., {Dale} D.~A., {Staudaher} S., {Bullock} J.~S.,
  {Calzetti} D., {Chandar} R., {Dalcanton} J.~J., 2014, \apj, 789, 126

\bibitem[{{Bastian}(2008)}]{bastian08}
{Bastian} N., 2008, MNRAS, 390, 759

\bibitem[{{Bastian} {et~al}\mbox{.}(2012){Bastian}, {Adamo}, {Gieles},
  {Silva-Villa}, {Lamers}, {Larsen}, {Smith}, {Konstantopoulos}, \&
  {Zackrisson}}]{bastian12a}
{Bastian} N. {et~al.}, 2012, \mnras, 419, 2606

\bibitem[{{Bastian} {et~al}\mbox{.}(2007){Bastian}, {Ercolano}, {Gieles},
  {Rosolowsky}, {Scheepmaker}, {Gutermuth}, \& {Efremov}}]{bastian07}
{Bastian} N., {Ercolano} B., {Gieles} M., {Rosolowsky} E., {Scheepmaker} R.~A.,
  {Gutermuth} R., {Efremov} Y., 2007, \mnras, 379, 1302

\bibitem[{Beckman {et~al}\mbox{.}(2000)Beckman, Rozas, Zurita, Watson, \&
  Knapen}]{beckman00}
Beckman J.~E., Rozas M., Zurita A., Watson R.~A., Knapen J.~H., 2000, The
  Astronomical Journal, 119, 2728

\bibitem[{{Berg} {et~al}\mbox{.}(2012){Berg}, {Skillman}, {Marble}, {van Zee},
  {Engelbracht}, {Lee}, {Kennicutt}, {Calzetti}, {Dale}, \& {Johnson}}]{berg12}
{Berg} D.~A. {et~al.}, 2012, \apj, 754, 98

\bibitem[{{Bertin} \& {Arnouts}(1996)}]{sex96}
{Bertin} E., {Arnouts} S., 1996, \aaps, 117, 393

\bibitem[{{Bik} {et~al}\mbox{.}(2003){Bik}, {Lamers}, {Bastian}, {Panagia}, \&
  {Romaniello}}]{bik03}
{Bik} A., {Lamers} H.~J.~G.~L.~M., {Bastian} N., {Panagia} N., {Romaniello} M.,
  2003, \aap, 397, 473

\bibitem[{{Boquien} {et~al}\mbox{.}(2015){Boquien}, {Calzetti}, {Aalto},
  {Boselli}, {Braine}, {Buat}, {Combes}, {Israel}, {Kramer}, {Lord},
  {Rela{\~n}o}, {Rosolowsky}, {Stacey}, {Tabatabaei}, {van der Tak}, {van der
  Werf}, {Verley}, \& {Xilouris}}]{boquien15}
{Boquien} M. {et~al.}, 2015, \aap, 578, A8

\bibitem[{{Bradley} {et~al}\mbox{.}(2006){Bradley}, {Knapen}, {Beckman}, \&
  {Folkes}}]{bradley06}
{Bradley} T.~R., {Knapen} J.~H., {Beckman} J.~E., {Folkes} S.~L., 2006, \aap,
  459, L13

\bibitem[{{Buat} {et~al}\mbox{.}(2005){Buat}, {Iglesias-P{\'a}ramo}, {Seibert},
  {Burgarella}, {Charlot}, {Martin}, {Xu}, {Heckman}, {Boissier}, {Boselli},
  {Barlow}, {Bianchi}, {Byun}, {Donas}, {Forster}, {Friedman}, {Jelinski},
  {Lee}, {Madore}, {Malina}, {Milliard}, {Morissey}, {Neff}, {Rich},
  {Schiminovitch}, {Siegmund}, {Small}, {Szalay}, {Welsh}, \& {Wyder}}]{buat05}
{Buat} V. {et~al.}, 2005, \apjl, 619, L51

\bibitem[{{Calzetti}(2001)}]{calzetti01}
{Calzetti} D., 2001, \pasp, 113, 1449

\bibitem[{{Calzetti} {et~al}\mbox{.}(2000){Calzetti}, {Armus}, {Bohlin},
  {Kinney}, {Koornneef}, \& {Storchi-Bergmann}}]{calzetti00}
{Calzetti} D., {Armus} L., {Bohlin} R.~C., {Kinney} A.~L., {Koornneef} J.,
  {Storchi-Bergmann} T., 2000, \apj, 533, 682

\bibitem[{{Calzetti} {et~al}\mbox{.}(2007){Calzetti}, {Kennicutt},
  {Engelbracht}, {Leitherer}, {Draine}, {Kewley}, {Moustakas}, {Sosey}, {Dale},
  {Gordon}, {Helou}, {Hollenbach}, {Armus}, {Bendo}, {Bot}, {Buckalew},
  {Jarrett}, {Li}, {Meyer}, {Murphy}, {Prescott}, {Regan}, {Rieke}, {Roussel},
  {Sheth}, {Smith}, {Thornley}, \& {Walter}}]{calzetti07}
{Calzetti} D. {et~al.}, 2007, \apj, 666, 870

\bibitem[{{Calzetti} {et~al}\mbox{.}(2005){Calzetti}, {Kennicutt}, {Bianchi},
  {Thilker}, {Dale}, {Engelbracht}, {Leitherer}, {Meyer}, {Sosey}, {Mutchler},
  {Regan}, {Thornley}, {Armus}, {Bendo}, {Boissier}, {Boselli}, {Draine},
  {Gordon}, {Helou}, {Hollenbach}, {Kewley}, {Madore}, {Martin}, {Murphy},
  {Rieke}, {Rieke}, {Roussel}, {Sheth}, {Smith}, {Walter}, {White}, {Yi},
  {Scoville}, {Polletta}, \& {Lindler}}]{calzetti05}
{Calzetti} D. {et~al.}, 2005, \apj, 633, 871

\bibitem[{{Cardelli}, {Clayton} \& {Mathis}(1989){Cardelli}, {Clayton}, \&
  {Mathis}}]{cardelli89}
{Cardelli} J.~A., {Clayton} G.~C., {Mathis} J.~S., 1989, \apj, 345, 245

\bibitem[{{Chandar}, {Fall} \& {Whitmore}(2015){Chandar}, {Fall}, \&
  {Whitmore}}]{chandar15}
{Chandar} R., {Fall} S.~M., {Whitmore} B.~C., 2015, \apj, 810, 1

\bibitem[{{Chandar} {et~al}\mbox{.}(2010){Chandar}, {Whitmore}, {Kim},
  {Kaleida}, {Mutchler}, {Calzetti}, {Saha}, {O'Connell}, {Balick}, {Bond},
  {Carollo}, {Disney}, {Dopita}, {Frogel}, {Hall}, {Holtzman}, {Kimble},
  {McCarthy}, {Paresce}, {Silk}, {Trauger}, {Walker}, {Windhorst}, \&
  {Young}}]{chandar10b}
{Chandar} R. {et~al.}, 2010, \apj, 719, 966

\bibitem[{{Cook} {et~al}\mbox{.}(2014{\natexlab{a}}){Cook}, {Dale}, {Johnson},
  {Van Zee}, {Lee}, {Kennicutt}, {Calzetti}, {Staudaher}, \&
  {Engelbracht}}]{cook14a}
{Cook} D.~O. {et~al.}, 2014{\natexlab{a}}, \mnras, 445, 881

\bibitem[{{Cook} {et~al}\mbox{.}(2014{\natexlab{b}}){Cook}, {Dale}, {Johnson},
  {Van Zee}, {Lee}, {Kennicutt}, {Calzetti}, {Staudaher}, \&
  {Engelbracht}}]{cook14c}
{Cook} D.~O. {et~al.}, 2014{\natexlab{b}}, \mnras, 445, 899

\bibitem[{{Cook} {et~al}\mbox{.}(2012){Cook}, {Seth}, {Dale}, {Johnson},
  {Weisz}, {Fouesneau}, {Olsen}, {Engelbracht}, \& {Dalcanton}}]{cook12}
{Cook} D.~O. {et~al.}, 2012, \apj, 751, 100

\bibitem[{{Dale} {et~al}\mbox{.}(2009){Dale}, {Cohen}, {Johnson}, {Schuster},
  {Calzetti}, {Engelbracht}, {Gil de Paz}, {Kennicutt}, {Lee}, {Begum},
  {Block}, {Dalcanton}, {Funes}, {Gordon}, {Johnson}, {Marble}, {Sakai},
  {Skillman}, {van Zee}, {Walter}, {Weisz}, {Williams}, {Wu}, \& {Wu}}]{dale09}
{Dale} D.~A. {et~al.}, 2009, \apj, 703, 517

\bibitem[{{de Grijs} {et~al}\mbox{.}(2003){de Grijs}, {Anders}, {Bastian},
  {Lynds}, {Lamers}, \& {O'Neil}}]{degrijs03b}
{de Grijs} R., {Anders} P., {Bastian} N., {Lynds} R., {Lamers} H.~J.~G.~L.~M.,
  {O'Neil} E.~J., 2003, \mnras, 343, 1285

\bibitem[{{de Vaucouleurs} {et~al}\mbox{.}(1991){de Vaucouleurs}, {de
  Vaucouleurs}, {Corwin}, {Buta}, {Paturel}, \& {Fouqu{\'e}}}]{rc3}
{de Vaucouleurs} G., {de Vaucouleurs} A., {Corwin}, Jr. H.~G., {Buta} R.~J.,
  {Paturel} G., {Fouqu{\'e}} P., 1991, {Third Reference Catalogue of Bright
  Galaxies. Volume I: Explanations and references. Volume II: Data for galaxies
  between 0$^{h}$ and 12$^{h}$. Volume III: Data for galaxies between 12$^{h}$
  and 24$^{h}$.}

\bibitem[{{Draine}(2003)}]{draine03}
{Draine} B.~T., 2003, \araa, 41, 241

\bibitem[{{Ellison} {et~al}\mbox{.}(2008){Ellison}, {Patton}, {Simard}, \&
  {McConnachie}}]{ellison08}
{Ellison} S.~L., {Patton} D.~R., {Simard} L., {McConnachie} A.~W., 2008, \apjl,
  672, L107

\bibitem[{{Elmegreen}(2006)}]{elmegreen06b}
{Elmegreen} B.~G., 2006, \apj, 648, 572

\bibitem[{{Elmegreen}(2010)}]{elmegreen10}
{Elmegreen} B.~G., 2010, in IAU Symposium, Vol. 266, IAU Symposium, {de Grijs}
  R., {L{\'e}pine} J.~R.~D., eds., pp. 3--13

\bibitem[{{Elmegreen} \& {Salzer}(1999)}]{elmegreen1999}
{Elmegreen} D.~M., {Salzer} J.~J., 1999, \aj, 117, 764

\bibitem[{{Eskew}, {Zaritsky} \& {Meidt}(2012){Eskew}, {Zaritsky}, \&
  {Meidt}}]{eskew12}
{Eskew} M., {Zaritsky} D., {Meidt} S., 2012, \aj, 143, 139

\bibitem[{{Gieles} {et~al}\mbox{.}(2006){Gieles}, {Larsen}, {Bastian}, \&
  {Stein}}]{gieles06a}
{Gieles} M., {Larsen} S.~S., {Bastian} N., {Stein} I.~T., 2006, \aap, 450, 129

\bibitem[{{Goddard}, {Bastian} \& {Kennicutt}(2010){Goddard}, {Bastian}, \&
  {Kennicutt}}]{goddard10}
{Goddard} Q.~E., {Bastian} N., {Kennicutt} R.~C., 2010, \mnras, 405, 857

\bibitem[{{Gouliermis} {et~al}\mbox{.}(2010){Gouliermis}, {Schmeja}, {Klessen},
  {de Blok}, \& {Walter}}]{gouliermis10}
{Gouliermis} D.~A., {Schmeja} S., {Klessen} R.~S., {de Blok} W.~J.~G., {Walter}
  F., 2010, \apj, 725, 1717

\bibitem[{{Gouliermis} {et~al}\mbox{.}(2015){Gouliermis}, {Thilker},
  {Elmegreen}, {Elmegreen}, {Calzetti}, {Lee}, {Adamo}, {Aloisi}, {Cignoni},
  {Cook}, {Dale}, {Gallagher}, {Grasha}, {Grebel}, {Herrero Davo}, {Hunter},
  {Johnson}, {Kim}, {Nair}, {Nota}, {Pellerin}, {Ryon}, {Sabbi}, {Sacchi},
  {Smith}, {Tosi}, {Ubeda}, \& {Whitmore}}]{gouliermis15}
{Gouliermis} D.~A. {et~al.}, 2015, ArXiv e-prints

\bibitem[{{Hao} {et~al}\mbox{.}(2011){Hao}, {Kennicutt}, {Johnson}, {Calzetti},
  {Dale}, \& {Moustakas}}]{hao11}
{Hao} C.-N., {Kennicutt} R.~C., {Johnson} B.~D., {Calzetti} D., {Dale} D.~A.,
  {Moustakas} J., 2011, \apj, 741, 124

\bibitem[{{Hermanowicz}, {Kennicutt} \& {Eldridge}(2013){Hermanowicz},
  {Kennicutt}, \& {Eldridge}}]{hermanowicz13}
{Hermanowicz} M.~T., {Kennicutt} R.~C., {Eldridge} J.~J., 2013, \mnras, 432,
  3097

\bibitem[{{Hunter} {et~al}\mbox{.}(2003){Hunter}, {Elmegreen}, {Dupuy}, \&
  {Mortonson}}]{hunter03}
{Hunter} D.~A., {Elmegreen} B.~G., {Dupuy} T.~J., {Mortonson} M., 2003, \aj,
  126, 1836

\bibitem[{{Kennicutt}(1998)}]{kennicutt98}
{Kennicutt}, Jr. R.~C., 1998, \araa, 36, 189

\bibitem[{{Kennicutt}, {Edgar} \& {Hodge}(1989){Kennicutt}, {Edgar}, \&
  {Hodge}}]{kennicutt89}
{Kennicutt}, Jr. R.~C., {Edgar} B.~K., {Hodge} P.~W., 1989, \apj, 337, 761

\bibitem[{{Kennicutt} {et~al}\mbox{.}(2008){Kennicutt}, {Lee}, {Funes},
  {Sakai}, \& {Akiyama}}]{kennicutt08}
{Kennicutt}, Jr. R.~C., {Lee} J.~C., {Funes}, Jos{\'e}~G. S.~J., {Sakai} S.,
  {Akiyama} S., 2008, \apjs, 178, 247

\bibitem[{{Kroupa}(2001)}]{kroupa01}
{Kroupa} P., 2001, \mnras, 322, 231

\bibitem[{{Kruijssen}(2012)}]{kruijssen12}
{Kruijssen} J.~M.~D., 2012, \mnras, 426, 3008

\bibitem[{{Lada} \& {Lada}(2003)}]{lada03}
{Lada} C.~J., {Lada} E.~A., 2003, ARA\&A, 41, 57

\bibitem[{{Larsen}(2002)}]{larsen02}
{Larsen} S.~S., 2002, \aj, 124, 1393

\bibitem[{{Lee}, {Grebel} \& {Hodge}(2003){Lee}, {Grebel}, \& {Hodge}}]{hlee03}
{Lee} H., {Grebel} E.~K., {Hodge} P.~W., 2003, \aap, 401, 141

\bibitem[{{Lee}, {Zucker} \& {Grebel}(2007){Lee}, {Zucker}, \&
  {Grebel}}]{hlee07}
{Lee} H., {Zucker} D.~B., {Grebel} E.~K., 2007, \mnras, 376, 820

\bibitem[{{Lee} {et~al}\mbox{.}(2011){Lee}, {Gil de Paz}, {Kennicutt},
  {Bothwell}, {Dalcanton}, {Jos{\'e} G.~Funes S.}, {Johnson}, {Sakai},
  {Skillman}, {Tremonti}, \& {van Zee}}]{lee11}
{Lee} J.~C. {et~al.}, 2011, \apjs, 192, 6

\bibitem[{{Lee} {et~al}\mbox{.}(2009){Lee}, {Gil de Paz}, {Tremonti},
  {Kennicutt}, {Salim}, {Bothwell}, {Calzetti}, {Dalcanton}, {Dale},
  {Engelbracht}, {Funes}, {Johnson}, {Sakai}, {Skillman}, {van Zee}, {Walter},
  \& {Weisz}}]{lee09b}
{Lee} J.~C. {et~al.}, 2009, \apj, 706, 599

\bibitem[{{Leroy} {et~al}\mbox{.}(2008){Leroy}, {Walter}, {Brinks}, {Bigiel},
  {de Blok}, {Madore}, \& {Thornley}}]{leroy08}
{Leroy} A.~K., {Walter} F., {Brinks} E., {Bigiel} F., {de Blok} W.~J.~G.,
  {Madore} B., {Thornley} M.~D., 2008, \aj, 136, 2782

\bibitem[{{Liu} {et~al}\mbox{.}(2013){Liu}, {Calzetti}, {Kennicutt},
  {Schinnerer}, {Sofue}, {Komugi}, {Egusa}, \& {Scoville}}]{liu13a}
{Liu} G., {Calzetti} D., {Kennicutt}, Jr. R.~C., {Schinnerer} E., {Sofue} Y.,
  {Komugi} S., {Egusa} F., {Scoville} N.~Z., 2013, \apj, 772, 27

\bibitem[{{Liu} {et~al}\mbox{.}(2011){Liu}, {Koda}, {Calzetti}, {Fukuhara}, \&
  {Momose}}]{liu11}
{Liu} G., {Koda} J., {Calzetti} D., {Fukuhara} M., {Momose} R., 2011, \apj,
  735, 63

\bibitem[{{Ma{\'{\i}}z Apell{\'a}niz} \& {{\'U}beda}(2005)}]{miaz05}
{Ma{\'{\i}}z Apell{\'a}niz} J., {{\'U}beda} L., 2005, \apj, 629, 873

\bibitem[{{Marble} {et~al}\mbox{.}(2010){Marble}, {Engelbracht}, {van Zee},
  {Dale}, {Smith}, {Gordon}, {Wu}, {Lee}, {Kennicutt}, {Skillman}, {Johnson},
  {Block}, {Calzetti}, {Cohen}, {Lee}, \& {Schuster}}]{marble10}
{Marble} A.~R. {et~al.}, 2010, \apj, 715, 506

\bibitem[{{Martin} {et~al}\mbox{.}(2005){Martin}, {Fanson}, {Schiminovich},
  {Morrissey}, {Friedman}, {Barlow}, {Conrow}, {Grange}, {Jelinsky},
  {Milliard}, {Siegmund}, {Bianchi}, {Byun}, {Donas}, {Forster}, {Heckman},
  {Lee}, {Madore}, {Malina}, {Neff}, {Rich}, {Small}, {Surber}, {Szalay},
  {Welsh}, \& {Wyder}}]{martin05}
{Martin} D.~C. {et~al.}, 2005, \apjl, 619, L1

\bibitem[{{McCrady} \& {Graham}(2007)}]{mccrady07}
{McCrady} N., {Graham} J.~R., 2007, \apj, 663, 844

\bibitem[{{McGaugh}(1991)}]{mcgaugh91}
{McGaugh} S.~S., 1991, \apj, 380, 140

\bibitem[{{McGaugh} \& {Schombert}(2014)}]{mcgaugh14}
{McGaugh} S.~S., {Schombert} J.~M., 2014, \aj, 148, 77

\bibitem[{{McGaugh} \& {Schombert}(2015)}]{mcgaugh15}
{McGaugh} S.~S., {Schombert} J.~M., 2015, \apj, 802, 18

\bibitem[{{Meidt} {et~al}\mbox{.}(2014){Meidt}, {Schinnerer}, {van de Ven},
  {Zaritsky}, {Peletier}, {Knapen}, {Sheth}, {Regan}, {Querejeta},
  {Mu{\~n}oz-Mateos}, {Kim}, {Hinz}, {Gil de Paz}, {Athanassoula}, {Bosma},
  {Buta}, {Cisternas}, {Ho}, {Holwerda}, {Skibba}, {Laurikainen}, {Salo},
  {Gadotti}, {Laine}, {Erroz-Ferrer}, {Comer{\'o}n}, {Men{\'e}ndez-Delmestre},
  {Seibert}, \& {Mizusawa}}]{meidt14}
{Meidt} S.~E. {et~al.}, 2014, \apj, 788, 144

\bibitem[{{Momcheva} {et~al}\mbox{.}(2013){Momcheva}, {Lee}, {Ly}, {Salim},
  {Dale}, {Ouchi}, {Finn}, \& {Ono}}]{momcheva13}
{Momcheva} I.~G., {Lee} J.~C., {Ly} C., {Salim} S., {Dale} D.~A., {Ouchi} M.,
  {Finn} R., {Ono} Y., 2013, \aj, 145, 47

\bibitem[{{Morrissey} {et~al}\mbox{.}(2007){Morrissey}, {Conrow}, {Barlow},
  {Small}, {Seibert}, {Wyder}, {Budav{\'a}ri}, {Arnouts}, {Friedman},
  {Forster}, {Martin}, {Neff}, {Schiminovich}, {Bianchi}, {Donas}, {Heckman},
  {Lee}, {Madore}, {Milliard}, {Rich}, {Szalay}, {Welsh}, \&
  {Yi}}]{morrissey07}
{Morrissey} P. {et~al.}, 2007, \apjs, 173, 682

\bibitem[{{Moustakas} {et~al}\mbox{.}(2010){Moustakas}, {Kennicutt},
  {Tremonti}, {Dale}, {Smith}, \& {Calzetti}}]{moustakas10}
{Moustakas} J., {Kennicutt}, Jr. R.~C., {Tremonti} C.~A., {Dale} D.~A., {Smith}
  J.-D.~T., {Calzetti} D., 2010, \apjs, 190, 233

\bibitem[{{Murphy} {et~al}\mbox{.}(2011){Murphy}, {Condon}, {Schinnerer},
  {Kennicutt}, {Calzetti}, {Armus}, {Helou}, {Turner}, {Aniano}, {Beir{\~a}o},
  {Bolatto}, {Brandl}, {Croxall}, {Dale}, {Donovan Meyer}, {Draine},
  {Engelbracht}, {Hunt}, {Hao}, {Koda}, {Roussel}, {Skibba}, \&
  {Smith}}]{murphy11}
{Murphy} E.~J. {et~al.}, 2011, \apj, 737, 67

\bibitem[{{Oey} \& {Clarke}(1998)}]{oey98}
{Oey} M.~S., {Clarke} C.~J., 1998, \aj, 115, 1543

\bibitem[{{Oh} {et~al}\mbox{.}(2008){Oh}, {de Blok}, {Walter}, {Brinks}, \&
  {Kennicutt}}]{oh08}
{Oh} S.-H., {de Blok} W.~J.~G., {Walter} F., {Brinks} E., {Kennicutt}, Jr.
  R.~C., 2008, \aj, 136, 2761

\bibitem[{{Osterbrock} \& {Ferland}(2006)}]{osterbrock06}
{Osterbrock} D.~E., {Ferland} G.~J., 2006, in {Book Review: Astrophysics of
  Gaseous Nebulae and Active Galactic Nuclei (2ND Edition) / University Science
  Books, 2005}

\bibitem[{{Pleuss}, {Heller} \& {Fricke}(2000){Pleuss}, {Heller}, \&
  {Fricke}}]{pleuss00}
{Pleuss} P.~O., {Heller} C.~H., {Fricke} K.~J., 2000, \aap, 361, 913

\bibitem[{{Prescott} {et~al}\mbox{.}(2007){Prescott}, {Kennicutt}, {Bendo},
  {Buckalew}, {Calzetti}, {Engelbracht}, {Gordon}, {Hollenbach}, {Lee},
  {Moustakas}, {Dale}, {Helou}, {Jarrett}, {Murphy}, {Smith}, {Akiyama}, \&
  {Sosey}}]{prescott07}
{Prescott} M.~K.~M. {et~al.}, 2007, \apj, 668, 182

\bibitem[{Randriamanakoto {et~al}\mbox{.}(2013)Randriamanakoto, Vaisanen,
  Ryder, Kankare, Kotilainen, \& Mattila}]{randriam13}
Randriamanakoto Z., Vaisanen P., Ryder S., Kankare E., Kotilainen J.~K.,
  Mattila S., 2013, arXiv, 1302, 768

\bibitem[{{Rosolowsky} {et~al}\mbox{.}(2007){Rosolowsky}, {Keto}, {Matsushita},
  \& {Willner}}]{rosolowsky07}
{Rosolowsky} E., {Keto} E., {Matsushita} S., {Willner} S.~P., 2007, \apj, 661,
  830

\bibitem[{{Salim} {et~al}\mbox{.}(2014){Salim}, {Lee}, {Ly}, {Brinchmann},
  {Dav{\'e}}, {Dickinson}, {Salzer}, \& {Charlot}}]{salim14}
{Salim} S., {Lee} J.~C., {Ly} C., {Brinchmann} J., {Dav{\'e}} R., {Dickinson}
  M., {Salzer} J.~J., {Charlot} S., 2014, \apj, 797, 126

\bibitem[{{Scoville} {et~al}\mbox{.}(2001){Scoville}, {Polletta}, {Ewald},
  {Stolovy}, {Thompson}, \& {Rieke}}]{scoville01}
{Scoville} N.~Z., {Polletta} M., {Ewald} S., {Stolovy} S.~R., {Thompson} R.,
  {Rieke} M., 2001, \aj, 122, 3017

\bibitem[{{Thilker} {et~al}\mbox{.}(2007){Thilker}, {Bianchi}, {Meurer}, {Gil
  de Paz}, {Boissier}, {Madore}, {Boselli}, {Ferguson}, {Mu{\~n}oz-Mateos},
  {Madsen}, {Hameed}, {Overzier}, {Forster}, {Friedman}, {Martin}, {Morrissey},
  {Neff}, {Schiminovich}, {Seibert}, {Small}, {Wyder}, {Donas}, {Heckman},
  {Lee}, {Milliard}, {Rich}, {Szalay}, {Welsh}, \& {Yi}}]{thilker07}
{Thilker} D.~A. {et~al.}, 2007, \apjs, 173, 538

\bibitem[{Thilker {et~al}\mbox{.}(2002)Thilker, Walterbos, Braun, \&
  Hoopes}]{thilker02}
Thilker D.~A., Walterbos R. A.~M., Braun R., Hoopes C.~G., 2002, The
  Astronomical Journal, 124, 3118

\bibitem[{{Tully} \& {Fisher}(1977)}]{tully77}
{Tully} R.~B., {Fisher} J.~R., 1977, \aap, 54, 661

\bibitem[{van Zee(2000)}]{vanzee00}
van Zee L., 2000, The Astronomical Journal, 119, 2757

\bibitem[{{Weidner}, {Kroupa} \& {Larsen}(2004){Weidner}, {Kroupa}, \&
  {Larsen}}]{weidner04}
{Weidner} C., {Kroupa} P., {Larsen} S.~S., 2004, \mnras, 350, 1503

\bibitem[{{Whitmore} {et~al}\mbox{.}(2014){Whitmore}, {Brogan}, {Chandar},
  {Evans}, {Hibbard}, {Johnson}, {Leroy}, {Privon}, {Remijan}, \&
  {Sheth}}]{whitmore14}
{Whitmore} B.~C. {et~al.}, 2014, \apj, 795, 156

\bibitem[{Youngblood \& Hunter(1999)}]{youngblood99}
Youngblood A.~J., Hunter D.~A., 1999, The Astrophysical Journal, 519, 55

\bibitem[{{Zhang} \& {Fall}(1999)}]{zhang99}
{Zhang} Q., {Fall} S.~M., 1999, \apjl, 527, L81

\end{thebibliography}

\newpage
\appendix
\section{All Luminosity Function} \label{sec:allLFs}

\begin{figure*}
  \begin{center}
  \includegraphics[scale=0.65]{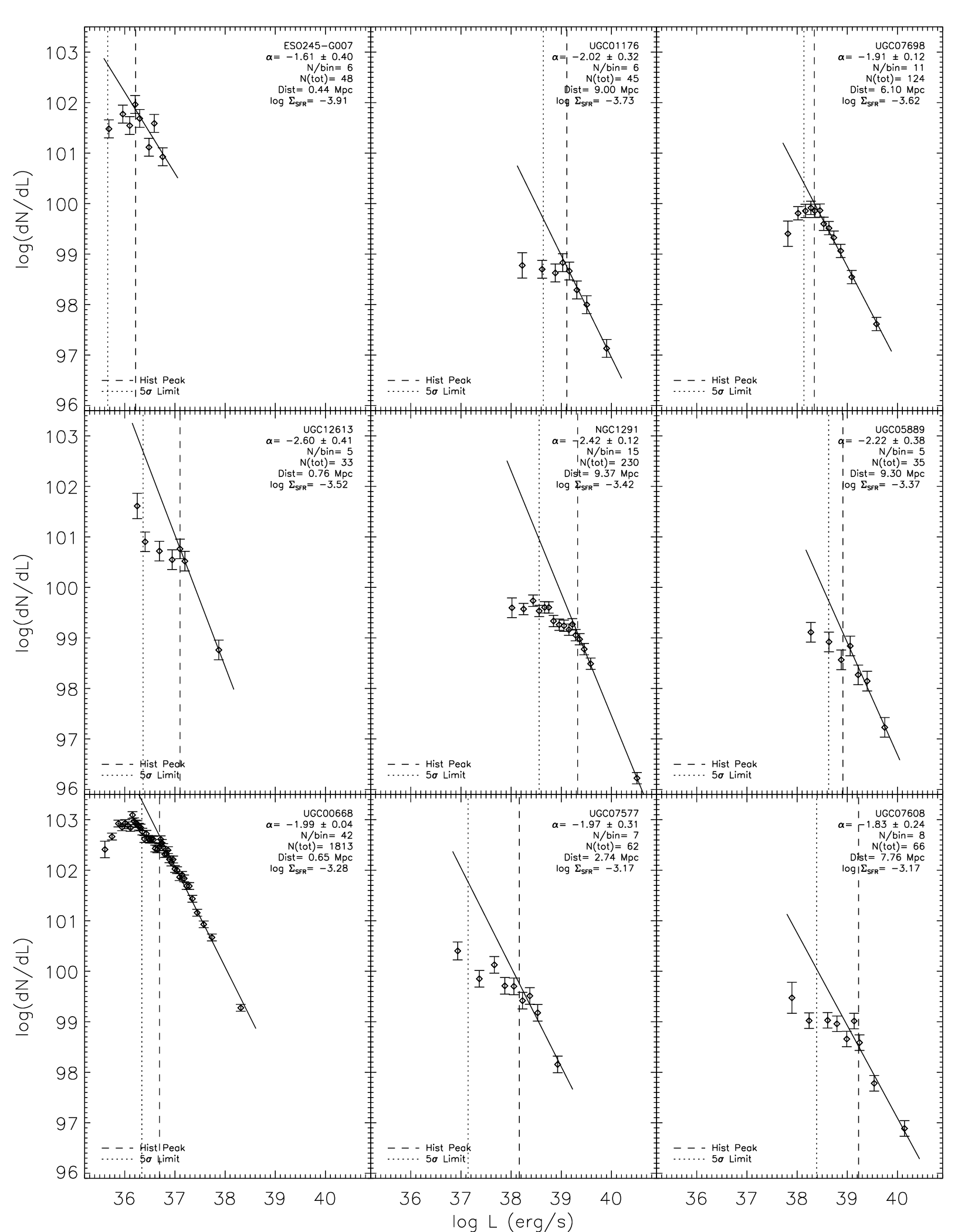}
  \caption{Figures~\ref{fig:allLFs1}-~\ref{fig:allLFs10} show the LFs for all \ntot galaxies in our final sample, where the panels have been sorted by $\Sigma_{\rm{SFR}}$. The y-axes for all LFs in all figures have been normalized to the same arbitrary number to facilitate comparisons of all luminosity functions.}
  \label{fig:allLFs1}
  \end{center}
\end{figure*}  
  
\begin{figure*}
  \begin{center}
  \includegraphics[scale=0.65]{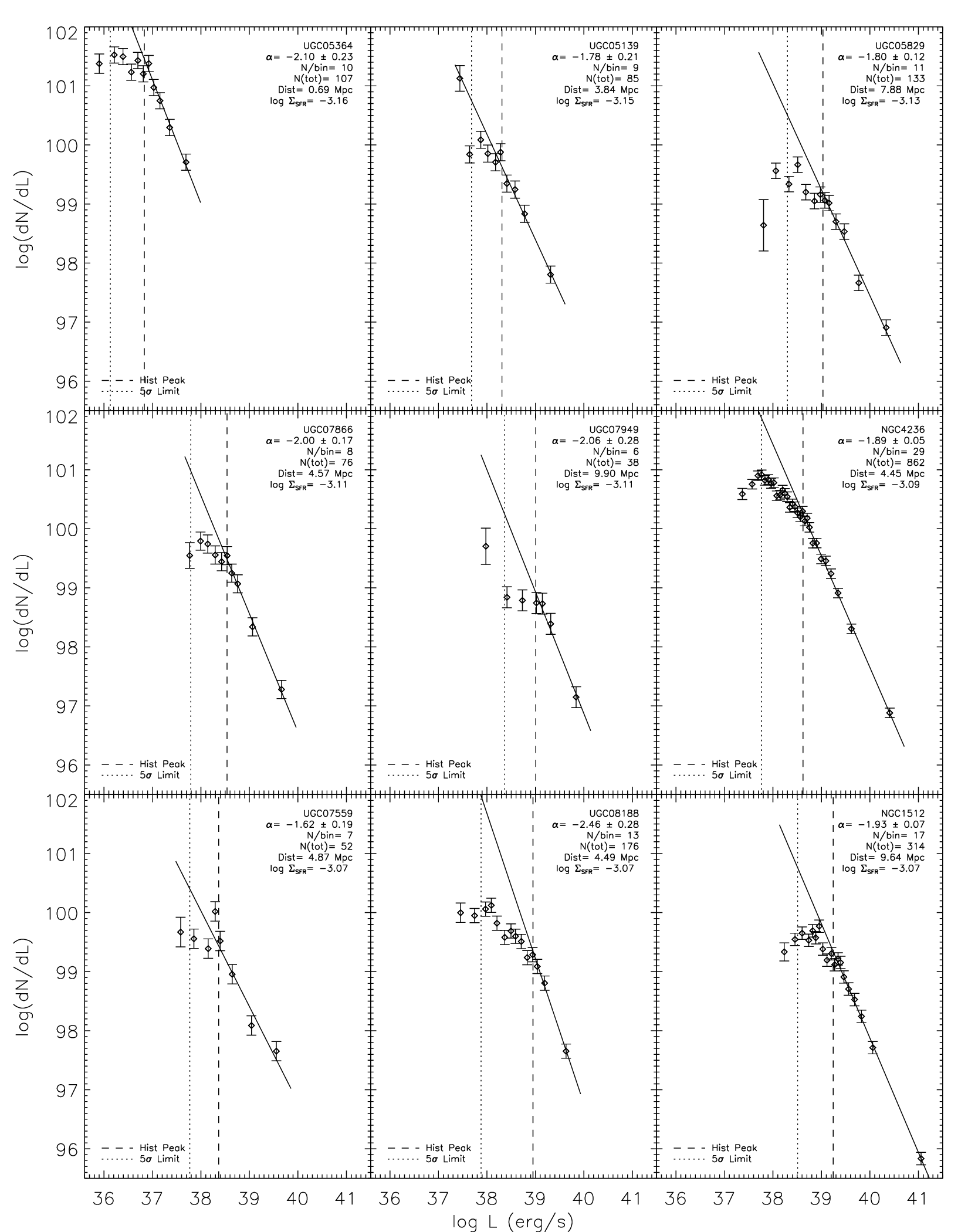}
  \caption{}
  \end{center}
\end{figure*}  
  
\begin{figure*}
  \begin{center}
  \includegraphics[scale=0.65]{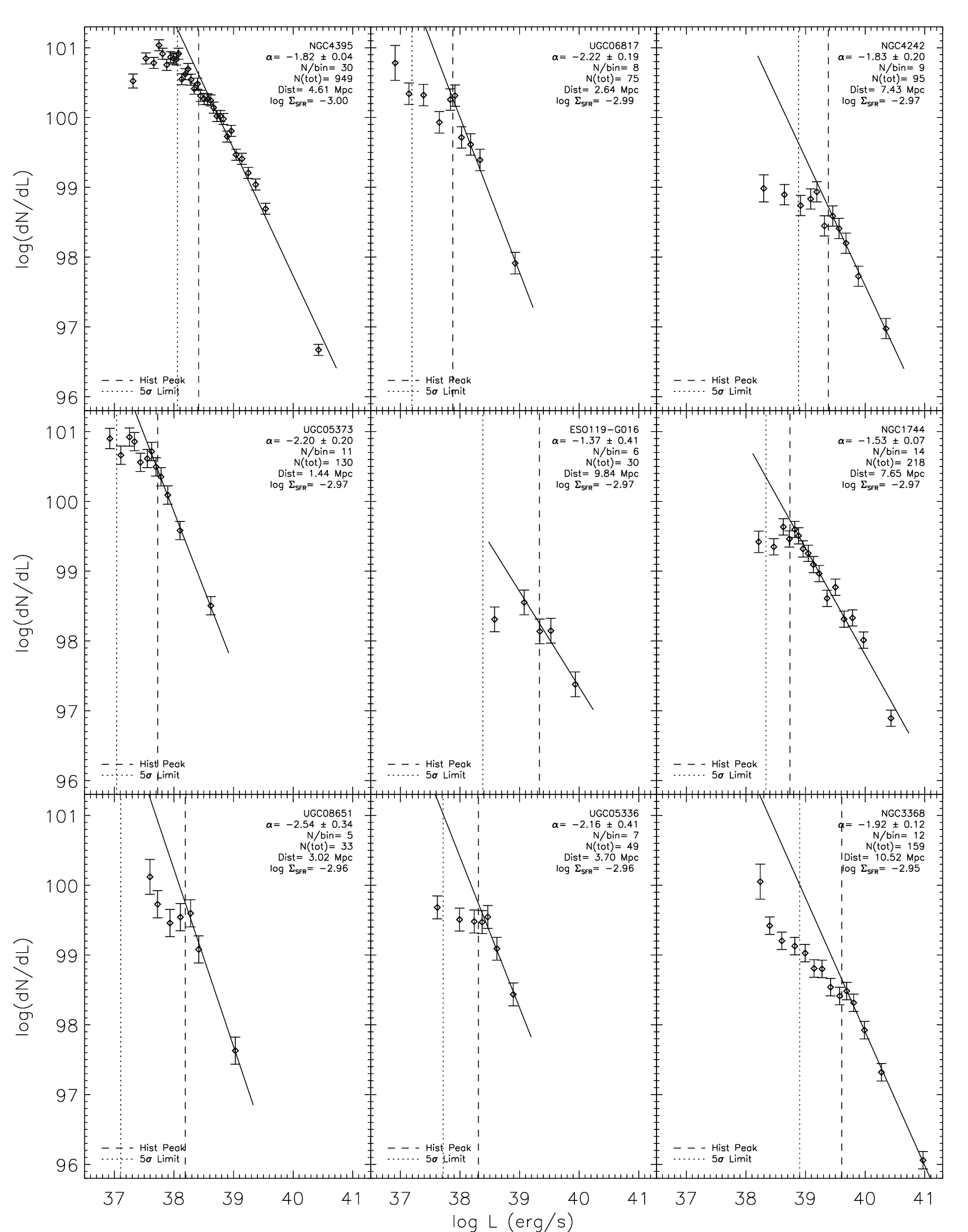}
  \caption{}
  \end{center}
\end{figure*}  
  
\begin{figure*}
  \begin{center}
  \includegraphics[scale=0.65]{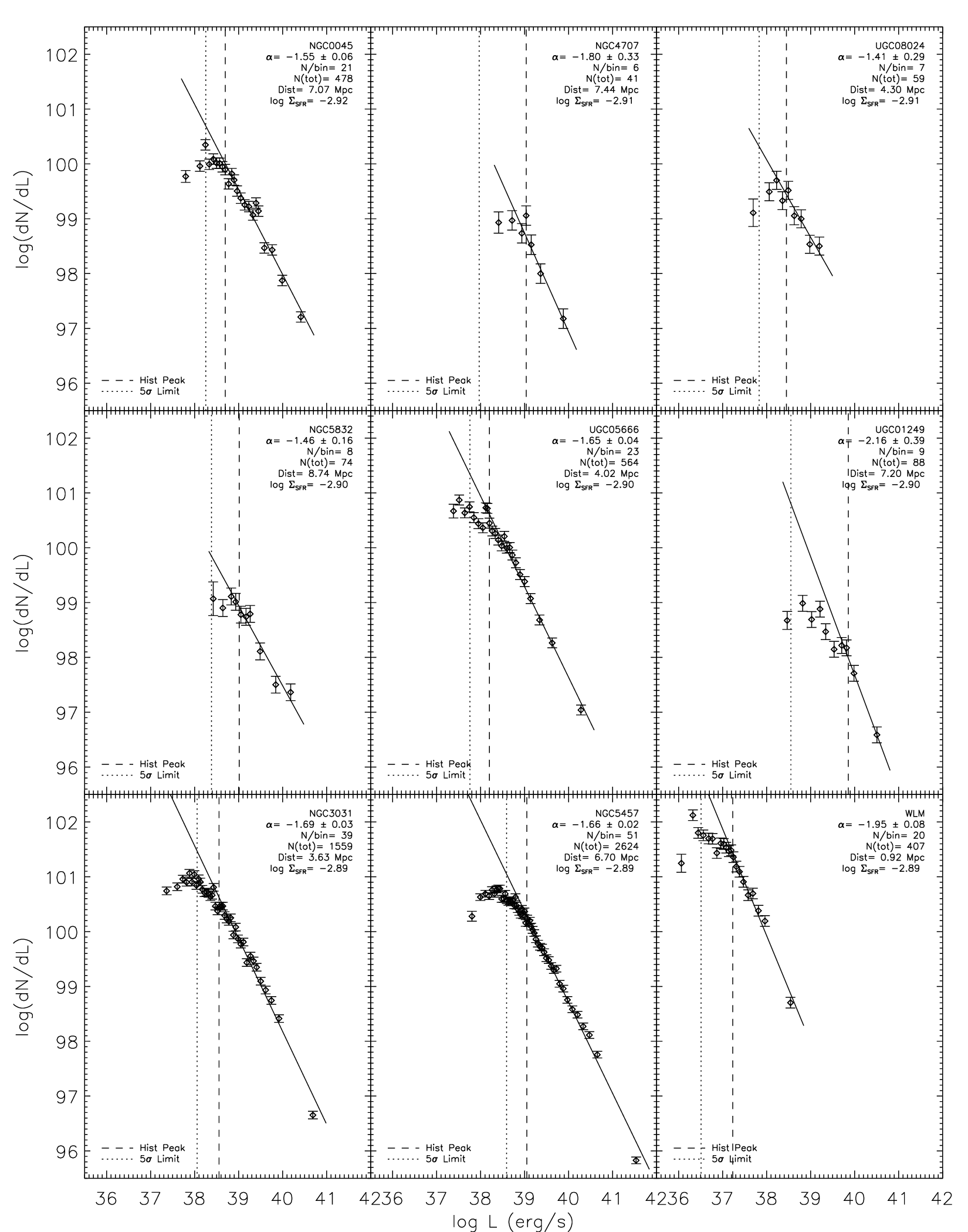}
  \caption{}
  \end{center}
\end{figure*}  
  
\begin{figure*}
  \begin{center}
  \includegraphics[scale=0.65]{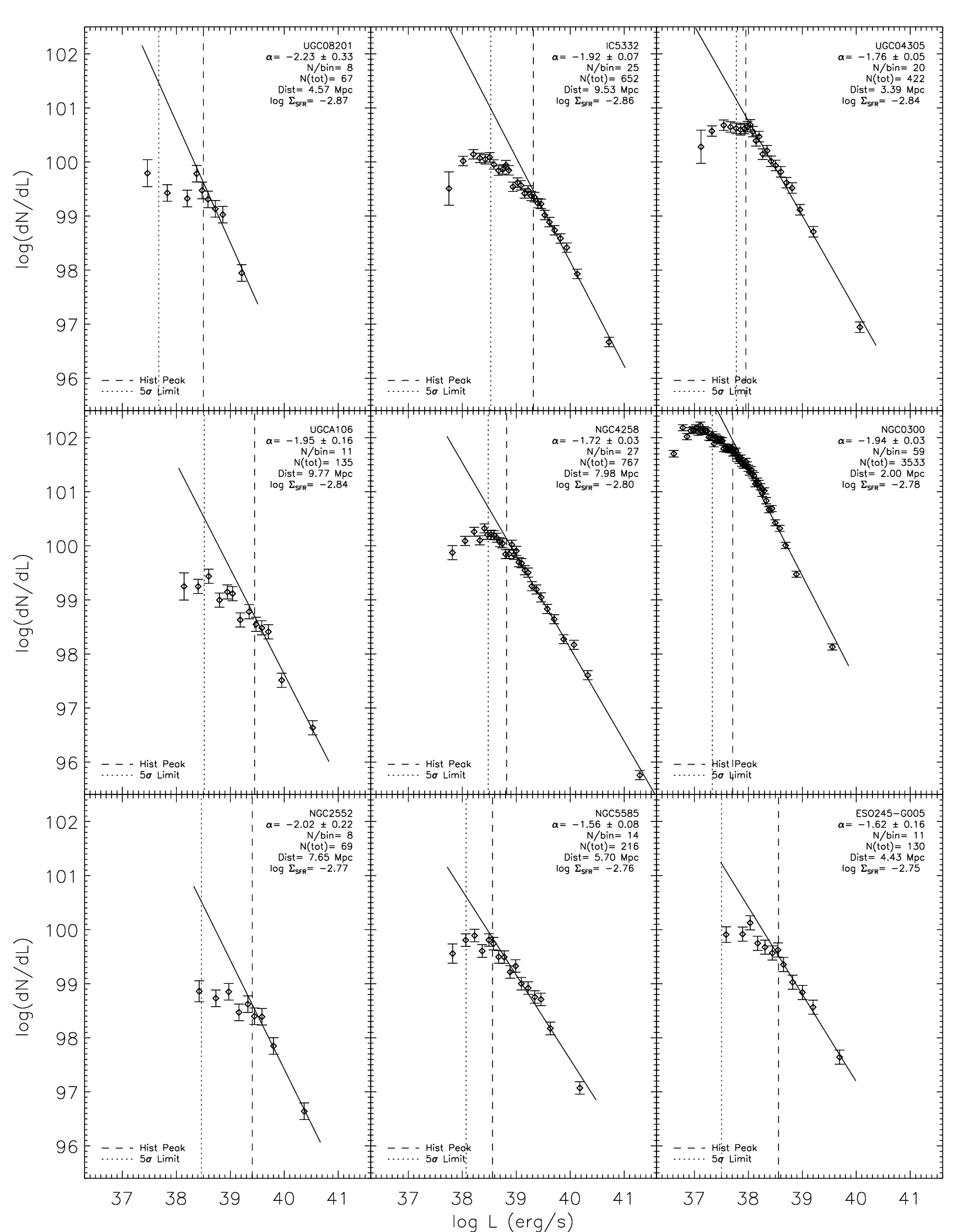}
  \caption{}
  \end{center}
\end{figure*}  
  
\begin{figure*}
  \begin{center}
  \includegraphics[scale=0.65]{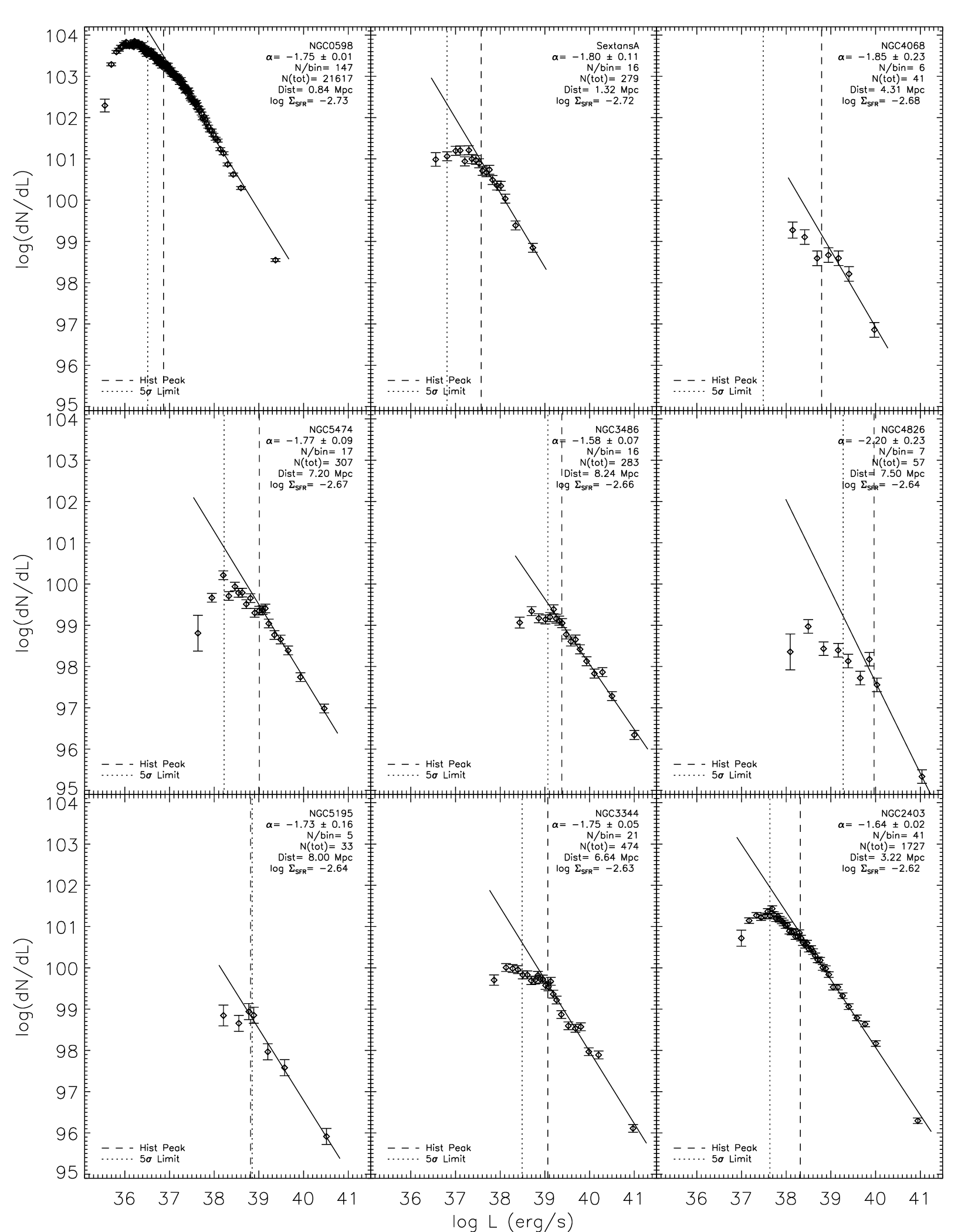}
  \caption{}
  \end{center}
\end{figure*}  
  
\begin{figure*}
  \begin{center}
  \includegraphics[scale=0.65]{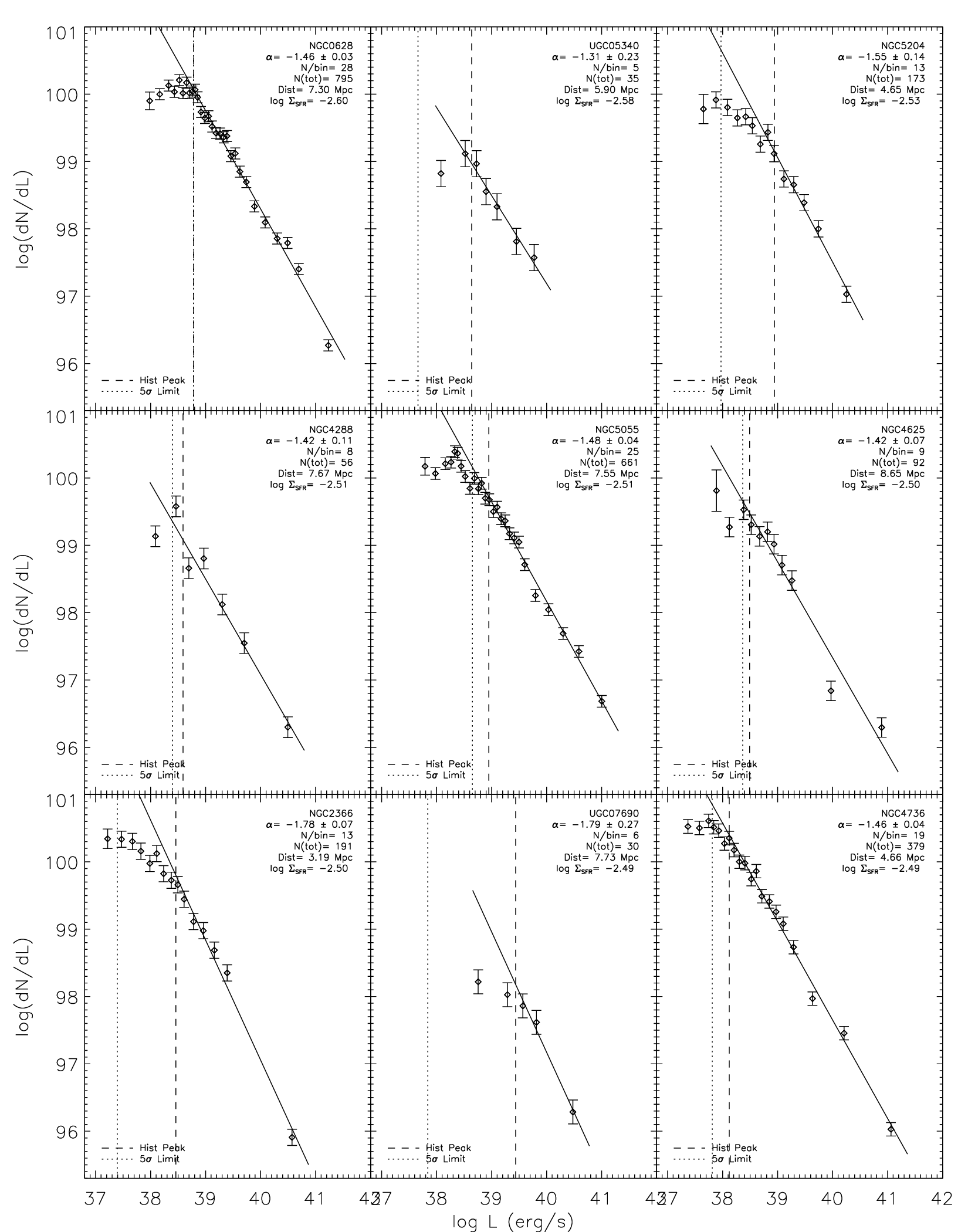}
  \caption{}
  \end{center}
\end{figure*}  
  
\begin{figure*}
  \begin{center}
  \includegraphics[scale=0.65]{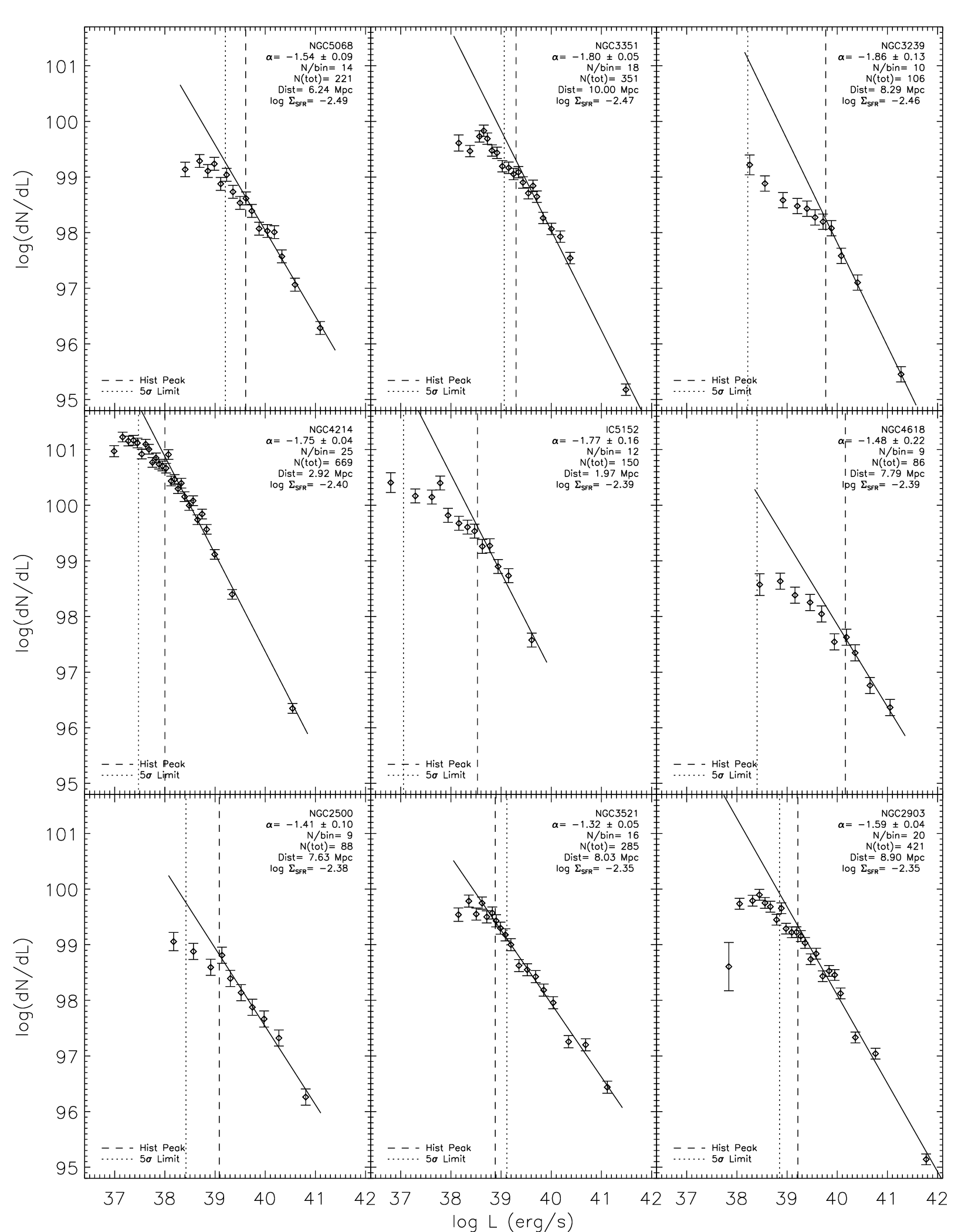}
  \caption{}
  \end{center}
\end{figure*}  
  
\begin{figure*}
  \begin{center}
  \includegraphics[scale=0.65]{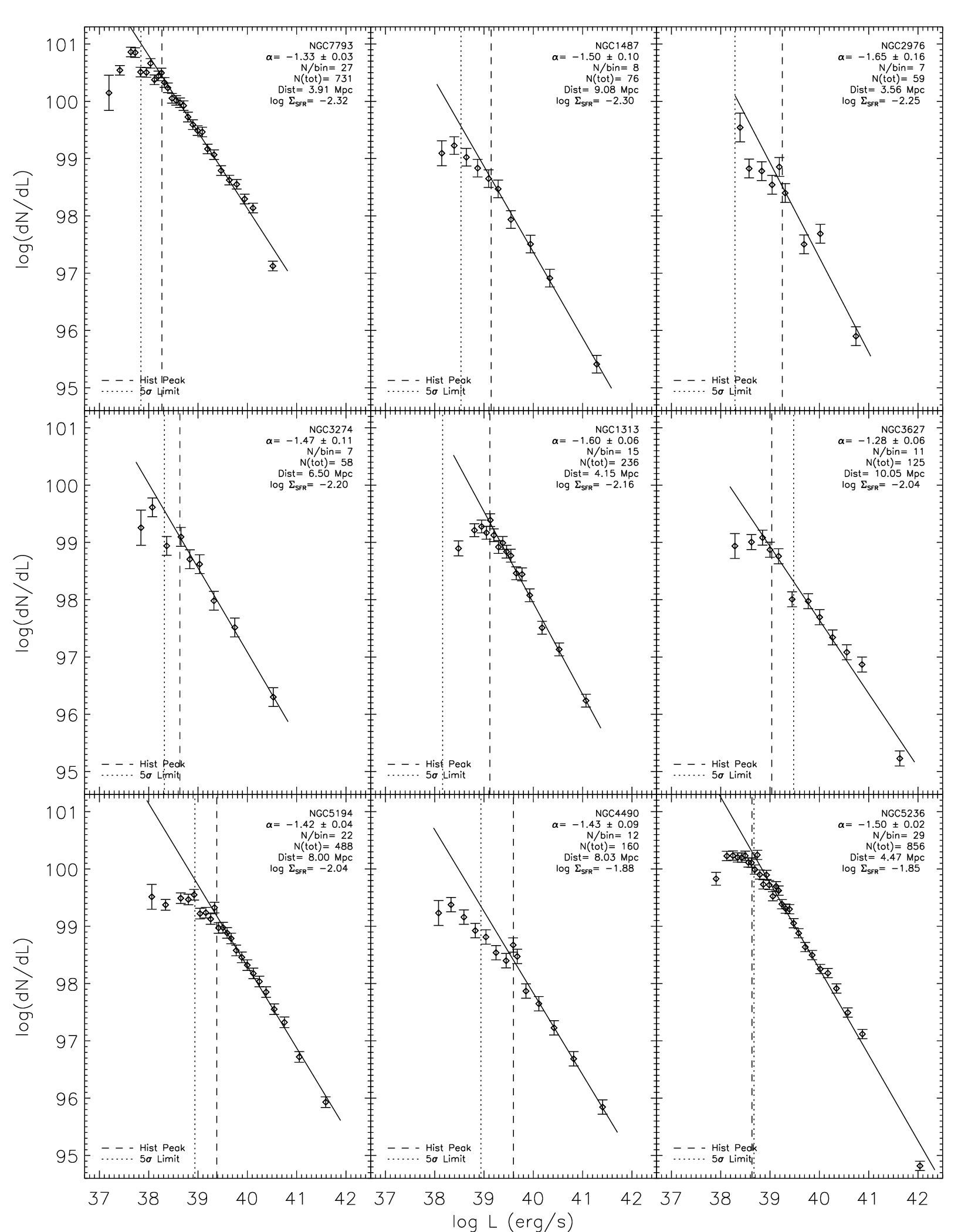}
  \caption{}
  \label{fig:allLFs9}
  \end{center}
\end{figure*}  
  
\begin{figure*}
  \begin{center}
  \includegraphics[scale=0.65]{SortedAllLFs_pg10.ps}
  \caption{}
  \label{fig:allLFs10}
  \end{center}
\end{figure*}

\end{document}